\documentclass[journal]{IEEEtran}
\usepackage{cite}  
\usepackage{graphicx}
\usepackage{algorithm}  
\usepackage{algpseudocode}  
\usepackage{latexsym}
\usepackage{amsmath,amssymb,amsfonts}  
\usepackage{bm}  
\usepackage{xcolor}
\usepackage[acronym,shortcuts]{glossaries}
\usepackage[font=footnotesize]{subcaption}
\usepackage[font=footnotesize]{caption}
\usepackage{makecell}

\definecolor{orange}{rgb}{1.0, 0.5, 0.0}
\definecolor{calpolypomonagreen}{rgb}{0.0, 0.5, 0.0}

\newcommand{\Kabuto}[1]{\textcolor{black}{#1}}
\newcommand{\KabutoSecond}[1]{\textcolor{black}{#1}}
\newacronym{CS}{CS}{compressed sensing}
\newacronym{UE}{UE}{user equipment}
\newacronym{BS}{BS}{base station}
\newacronym{THz}{THz}{sub-terahertz}
\newacronym{NMSE}{NMSE}{normalized mean-squared error}
\newacronym{AWGN}{AWGN}{additive white Gaussian noise}
\newacronym{CSI}{CSI}{channel state information}
\newacronym{RF}{RF}{radio-frequency}
\newacronym{mmWave}{mmWave}{millimeter-wave}
\newacronym{sub-THz}{sub-THz}{sub-terahertz}
\newacronym{OFDM}{OFDM}{orthogonal frequency division multiplexing}
\newacronym{MIMO}{MIMO}{multiple-input-multiple-output}
\newacronym{ULA}{ULA}{uniform linear array}
\newacronym{DFT}{DFT}{discrete Fourier transform}
\newacronym{LS}{LS}{least square}
\newacronym{LMMSE}{LMMSE}{linear minimum mean square error}
\newacronym{ML}{ML}{maximum likelihood}
\newacronym{SOMP}{SOMP}{simultaneous orthogonal matching pursuit}
\newacronym{SIGW}{SIGW}{simultaneous iterative gridless weighted}
\newacronym{CoDL}{CoDL}{combined dictionary learning}
\newacronym{DLHWBS}{DLHWBS}{dictionary learning for hardware impairments under beam squint}
\newacronym{DA-OMP-BS}{DA-OMP-BS}{dictionary adaptive OMP under beam squint}
\newacronym{QPSK}{QPSK}{quadrature phase-shift keying}
\newacronym{SNR}{SNR}{signal-to-noise ratio}
\newacronym{PDF}{PDF}{probability density function}
\newacronym{DCS-SOMP}{DCS-SOMP}{distributed compressed sensing-simultaneous orthogonal matching pursuit}
\newacronym{SE}{SE}{spectral efficiency}
\newacronym{FLOPs}{FLOPs}{floating operations}
\newacronym{MA}{MA}{movable antenna}
\newacronym{FA}{FA}{fluid antenna}
\newacronym{DoF}{DoF}{degrees of freedom}

\newacronym{AoA}{AoA}{azimuth angles of arrival}
\newacronym{AoD}{AoD}{azimuth angles of departure}
\newacronym{ZoD}{ZoD}{zenith angles of departure}
\newacronym{ZoA}{ZoA}{zenith angles of arrival}
\newacronym{UPA}{UPA}{uniform planar array}
\newacronym{CP}{CP}{cyclic prefix}

\begin{document}

\title{Channel Estimation for Hybrid MIMO Systems With Array Model Errors and Beam Squint Effects}

\author{Kabuto Arai,~\IEEEmembership{Graduate Student Member, IEEE} and  Koji Ishibashi,~\IEEEmembership{Senior Member, IEEE}
%
\thanks{
This work was supported in part by Japan Science and Technology Agency (JST) Advanced International Collaborative Research Program (AdCORP) under Grant JPMJKB2307.
An earlier version of this paper was presented in part at the 2023 IEEE 98th Vehicular Technology Conference (VTC2023-Fall) [DOI: 10.1109/VTC2023-Fall60731.2023.10333598.]~\cite{2023Arai_vtc}.
 

K. Arai and K. Ishibashi are with the Advanced Wireless and Communication Research Center (AWCC), The University of Electro-Communications, Tokyo 182-8285, Japan (e-mail: k.arai@awcc.uec.ac.jp, koji@ieee.org)}}

\markboth{Journal of \LaTeX\ Class Files,~Vol.~14, No.~8, August~2021}%
{Shell \MakeLowercase{\textit{et al.}}: Sample article using IEEEtran.cls for IEEE Journals}


\maketitle

\begin{abstract}
    \Kabuto{
    This paper proposes a channel estimation method for wideband hybrid \acf{MIMO} systems operating in high-frequency bands, including \acf{mmWave} and \acf{sub-THz}.
    The proposed method accounts for beam squint effects and array errors, which arise from hardware impairments and time-varying environmental factors, such as thermal effects and dynamic motion of the array.}
    Although conventional channel estimation methods calibrate array errors through offline operation with large training pilots, the calibration errors remain due to time-varying array errors.
    Therefore, the proposed channel estimation method calibrates array errors online with small pilot overhead.
    In the proposed method, array response matrices are explicitly decomposed into a small number of physical parameters, including path gains, angles, \Kabuto{delays}, and array errors, which are iteratively estimated by alternating optimization based on a \acf{ML} criterion.
    To enhance the convergence performance, we introduce a switching mechanism from an on-grid algorithm to an off-grid algorithm depending on the estimation accuracy of the array error during algorithmic iterations.
    Furthermore, we introduce an approximate mutual coupling model to reduce the number of parameters.
    The reduction of parameters not only lowers computational complexity but also mitigates overfitting to noisy observations.
    Numerical simulations demonstrate that the proposed method works effectively online in the presence of array errors even with small pilot overhead.
\end{abstract}

\glsresetall


\IEEEpeerreviewmaketitle

\glsresetall

\section{Introduction}
    \label{sec:intro}

    As the demand for broadband communication grows, high carrier frequencies, such as \ac{mmWave} and \ac{sub-THz}, are becoming increasingly important for future communication systems~\cite{2021Tataria_6G_syrvey}. 
    However, these high-frequency bands suffer from significant attenuation due to weak diffraction, blockages, and molecular absorption~\cite{2021Tarboush_TeraMIMO}. 
    To mitigate these challenges, beamforming techniques are extensively employed with massive antenna arrays by leveraging the short wavelengths of these high-frequency bands. 
    In terms of cost and power consumption, hybrid beamforming architectures, which utilize both analog and digital precoders/combiners, are widely adopted instead of full-digital architectures~\cite{2016Heath_overview_sp}.
    To design the optimal beamformer, \ac{CSI} is essential~\cite{2016Yu_beam_AO}. 
    However, channel estimation in a hybrid beamforming architecture requires a large number of pilots during the training beam search process because the number of \ac{RF} chains is 
    much smaller than the number of antennas~\cite{2018Rodriguez_SOMP}.
    
    To address the pilot overhead in wideband hybrid \ac{MIMO} systems, channel estimation methods based on \ac{CS} techniques have been proposed~\cite{2018Rodriguez_SOMP, 2024Uchimura_tracking, 2020Gonzalez_SIGW, 2022Cui_PSOMP}. 
    These methods can significantly reduce pilot overhead by exploiting channel sparsity in the angle domain. 
    To effectively leverage this sparsity, the virtual channel representation is widely utilized with a dictionary matrix, which is an array response matrix over a set of quantized angle grid points covering the entire angle domain. 
    Since the dictionary matrix is based on the ideal array manifold model without accounting for modeling errors, the estimation performance of \ac{CS}-based methods is adversely affected by modeling errors caused by 
    array errors~\cite{2016Eberhardt_investigation_error} in practical wideband hybrid \ac{MIMO} systems~\cite{2020Xie_CoDL, 2022Maity_CoDL_THz, 2023Xie_DADL, 2023Maity_CoDL_RIS}.

    The array errors arise from the following three factors~\cite{2016Eberhardt_investigation_error, 2020Xie_CoDL, 2022Maity_CoDL_THz, 2023Xie_DADL, 2023Maity_CoDL_RIS}: 
    (1) mutual coupling between antenna elements due to closely implemented massive antenna elements, causing unwanted energy interactions between the elements, 
    (2) gain/phase errors in antenna elements, which result from the nonuniform electrical characteristics of individual devices, and 
    (3) antenna positioning errors stemming from the limited accuracy of manufacturing, especially in high-frequency bands because the wavelength is on the millimeter or micrometer scale in mmWave and sub-THz.
    These array errors introduce modeling errors which severely degrade channel estimation performance because the CS-based channel estimation heavily relies on the ideal array model.
    Therefore, it is necessary to calibrate the array errors for accurate channel estimation. 
    
    The array errors are attributed not only to static factors, such as manufacturing errors, but also to time-varying factors, such as dynamic motion of the antenna arrays, aging of components, and heat dissipation of electronic devices~\cite{2022Liu_thermal_deform, 2023Xu_thermal_coupling, 2023Feng_thermal}.
    Due to heat dissipation in the device, the electrical characteristics of devices change, resulting in non-uniform gain and phase characteristics in each antenna element.
    Moreover, these effects cause a deformation of the array structure with offset from the ideal antenna positions, leading to changes in mutual coupling.
    Even a slight change in the antenna position, on the scale of a fraction of the wavelength, can significantly degrade the performance of array signal processing, such as channel estimation~\cite{2020Xie_CoDL, 2022Maity_CoDL_THz, 2023Xie_DADL, 2023Maity_CoDL_RIS} and beamforming~\cite{2023Cao_beam_stabilization}.
    Thus, the impact of array deformation becomes more pronounced at higher frequencies (\textit{i.e.}, with shorter wavelengths).
    Furthermore, in various communication scenarios, such as high-speed aircraft flight with dynamic wind loads~\cite{2020Wang_radome_coupling, 2023Cao_beam_stabilization}, satellite phased array~\cite{2009Takahshi_satellite_calibration}, and wearable array antennas~\cite{2018Song_wearable, 2021Fikes_wearable_coupling}, the array structure is deformed with time due to the motion of the array and fluctuations in surrounding environmental loads including temperature and pressure.
    Another scenario involving array deformation is \ac{MA} systems~\cite{2024Zhu_MA_Mag, 2023Ma_MA_CS}, including  \ac{FA} systems~\cite{2021Wong_fluid_antenna, 2022Wong_fluid_antenna_multiple_access}, where the antenna position is flexibly controlled to exploit spatial channel variation.
    However, due to limited position control accuracy, antenna spacing errors remain. 
    Moreover, mutual coupling changes with each adjustment of the antenna position, requiring frequent calibration after every adjustment.
    For the above reasons, frequent and fast calibration to mitigate time-varying array errors is needed~\cite{2023Zeng_fast_compensation_coupling, 2023Cao_beam_stabilization}.

    To compensate for array errors, various classic calibration methods have been proposed in the field of array signal processing~\cite{1996Boon_off_calib_ML, 2009Ng_calib_music, 1994Fuhrmann_calib_gain, 2015Pan_calib_bayes}. 
    In these methods, array errors, such as mutual coupling, gain/phase errors, and antenna spacing errors, are jointly estimated using calibration sources that are located at known positions and transmit reference signals for  calibration through offline operation.
    Despite achieving high estimation performance, these approaches are time-consuming and costly due to the requirement of calibration sources. 
    Additionally, the offline calibration is unable to address the time-varying array errors. 
    To address these limitations, blind calibration techniques without calibration sources have been proposed~\cite{1991Friedlander_Direction_finding, 2014Dai_calib_RARE, 2015Pan_calib_bayes}. 
    These methods can jointly estimate both array errors and \acp{AoA} from unknown sources. However, these methods assume narrowband fully-digital array architectures, focusing only on the receiver side for \ac{AoA} estimation. 
    Consequently, it is challenging to directly extend these methods to channel estimation algorithms in the considered communication systems operating at high carrier frequencies with wide bandwidth and hybrid array architectures that include both the transmitter and receiver.
    
    To calibrate array errors in wideband hybrid \ac{MIMO} systems without calibration sources, channel estimation methods based on dictionary learning techniques have been proposed~\cite{2020Xie_CoDL, 2022Maity_CoDL_THz, 2023Maity_CoDL_RIS}. 
    The authors of \cite{2020Xie_CoDL} proposed the \ac{CoDL} algorithm, where a combined dictionary matrix, including array errors at both the transmitter and receiver sides, is updated to calibrate these array errors using training pilots.
    While these dictionary learning methods do not require calibration sources, they require a large number of pilots to achieve sufficient channel estimation performance, since the dictionary matrix contains numerous unknown parameters.
    Therefore, these works~\cite{2020Xie_CoDL, 2022Maity_CoDL_THz, 2023Maity_CoDL_RIS} assume that
    the dictionary learning is performed offline with a large number of training pilots, after which channel estimation is performed online with a small number of pilots using the updated dictionary matrix.
    However, the channel estimation performance using the dictionary matrix trained offline degrades since array errors change with time~\cite{2022Liu_thermal_deform, 2023Xu_thermal_coupling,2023Cao_beam_stabilization, 2020Wang_radome_coupling, 2009Takahshi_satellite_calibration, 2018Song_wearable, 2021Fikes_wearable_coupling, 2024Zhu_MA_Mag, 2023Ma_MA_CS, 2021Wong_fluid_antenna, 2022Wong_fluid_antenna_multiple_access, 2023Zeng_fast_compensation_coupling}.
    \Kabuto{
    If the dictionary is updated online (\textit{i.e.}, dynamically updated whenever the array errors change), a significant number of additional pilots is required in addition to the pilots normally used for channel estimation~\cite{2020Xie_CoDL, 2022Maity_CoDL_THz}, which defeats the original purpose of \ac{CS}-based channel estimation in reducing pilot overhead.
    }

    Additionally, due to the significant increase in bandwidth and antenna aperture, the beam squint effect~\cite{2011Garakoui_squint}, also known as the spatial wideband effect~\cite{2018Wang_squint_mmWave}, becomes non-negligible at high-frequency bands, where the array responses depend on frequencies.
    \Kabuto{
    Although many studies~\cite{2024Xu_squint_CE_Bayesian_multi, 2020Wang_squint_CE_block, 2024Garg_squint_CE_JCDE, 2025Jang_squint_NN} have proposed channel estimation methods based on CS algorithms, these methods typically address only beam squint effects and neglect array errors.
    }
    To overcome this limitation, the authors in \cite{2023Xie_DADL} proposed the \ac{DLHWBS} algorithm, which considers both array errors and beam squint effects by accounting for the frequency-dependent array response.
    However, this dictionary learning method also relies on offline calibration, requiring a large number of training pilots, similar to conventional methods~\cite{2020Xie_CoDL, 2022Maity_CoDL_THz, 2023Maity_CoDL_RIS}.
    Consequently, the pilot overhead problem remains unresolved in scenarios involving both time-varying array errors and beam squint effects.

    In light of the above issues, we propose a channel estimation method considering time-varying array errors and beam squint effects by online calibration with small pilot overhead.
    The proposed method can be applied to wideband MIMO systems with hybrid architectures, where array errors and beam squint effects occur on both the transmitter and receiver sides.
    Our major contributions are summarized as follows\footnote{
    \Kabuto{
    An earlier version of this paper was presented in~\cite{2023Arai_vtc}.
    \KabutoSecond{The major differences from the earlier version are as follows.}
    1) Consider array errors not only at the BS side but also at the UE side in the case of both ULA and UPA cases.
    2) Consider beam squint effects in wideband MIMO-OFDM systems.
    3) Estimate the channel in the delay-angle domain unlike frequency-angle domain.
    4) Introduce the switching mechanism for estimating angles, delays, and path gains.
    5) Introduce the approximate mutual coupling model including ULA and UPA cases.
    6) Provide extensive numerical simulation results.
    }}:

    \begin{itemize}
        \item 
        \Kabuto{        
        The proposed channel estimation is operated online with small pilot overhead for array error calibration.
        This method uses only the current and a few past frames containing a small number of pilots normally used for channel estimation, not requiring additional pilot transmission for array calibration, unlike the conventional approaches~\cite{2020Xie_CoDL, 2022Maity_CoDL_THz, 2023Xie_DADL, 2023Maity_CoDL_RIS}.
        By decomposing the dictionary matrix containing many unknown parameters into a smaller number of physical parameters, which include \acp{AoA}, \acp{AoD}, \acp{ZoA}, \acp{ZoD}, path gains, delays, and array errors, these parameters can be efficiently estimated using an alternating optimization technique~\cite{2004Boyd_ConvexOpt} based on a \ac{ML} criteria.
        }
        \Kabuto{
        Furthermore, by exploiting angle-delay sparsity, channel estimation can be performed using a small number of subcarriers for pilot transmission, thereby reducing both pilot overhead and computational complexity.
        }

        \item
        To enhance convergence performance in the alternating optimization, we introduce a switching mechanism from an on-grid algorithm, based on a \ac{CS} method, to an off-grid algorithm, based on a gradient descent method, for estimating \Kabuto{angles, delays, and path gains.} 
        During early iterations when the estimation accuracy for array errors is limited, an on-grid algorithm is applied to avoid convergence to a local optimum. 
        In subsequent later iterations with the higher estimation accuracy for array errors, an off-grid algorithm is employed to \Kabuto{finely estimate angles and delays without angle and delay grids.}

        \item
        To enhance channel estimation performance and reduce computational complexity, an approximate mutual coupling model is introduced. 
        \Kabuto{
        The mutual coupling matrix can be modeled as a symmetric Toeplitz matrix in the case of a \ac{ULA}~\cite{1991Friedlander_Direction_finding}, and as a symmetric block Toeplitz matrix in the case of a \ac{UPA}~\cite{2008Ye_2D_Copuling}, respectively.
        However, antenna spacing errors distort this Toeplitz structure in practical systems.}
        To account for this distortion, the proposed method approximately decomposes the mutual coupling matrix into a non-Toeplitz part and a Toeplitz part. 
        Among the elements in the mutual coupling matrix, only a few elements, which have significant impact on estimation performance, are exactly modeled with a non-Toeplitz structure, while the remaining elements are approximately modeled with a Toeplitz structure.
        This approximate decomposition reduces the number of unknown parameters in the mutual coupling matrix compared to an exact model without a Toeplitz structure.
        Consequently, reducing the number of parameters not only lowers computational complexity but also helps avoid overfitting to noisy observations.
    \end{itemize}

    \textit{Notation}:
    The operators $(\cdot)^*$, $(\cdot)^\mathrm{T}$ and $(\cdot)^\mathrm{H}$ indicate conjugate, transpose and conjugate transpose, respectively.
    $\mathbf{I}_N$ and $\mathbf{0}_{N \times M}$ indicate the $N \times N$ identity matrix and $N \times M$ zero matrix, respectively. 
    \Kabuto{
    A diagonal matrix from a vector $\mathbf{a}$ and a block diagonal matrix from matrices $\mathbf{A}_1, \ldots, \mathbf{A}_N$ are represented as $\mathrm{diag}(\mathbf{a})$ and $\mathrm{blkdiag}(\mathbf{A}_1, \ldots \mathbf{A}_N)$, respectively.
    }
    The notation $[\mathbf{A}]_{i,j}$ and $[\mathbf{A}]_{:,j}$ indicate the $(i,j)$ element and the $j$-th column vector of the matrix $\mathbf{A}$, respectively.
    \Kabuto{
    The vector consisting of the $i$-th element to the $j$-th element of $\mathbf{a}$ ($i<j$) is denoted by $[\mathbf{a}]_{i:j}$.
    }
    \Kabuto{
    Let $\mathbf{e}_{N,i} \in \mathbb{R}^{N \times1}$ denote the $i$-th standard basis vector whose $i$-th entry is 1 and all others are $0$.}
    $\odot$, $\otimes$ and $\circ$ represent Hadamard, Kronecker, and Khatri-Rao product, respectively.
    $\| \mathbf{a} \|_p$ and $\|\mathbf{A}\|_\mathrm{F}$ indicate $p$-norm and Frobenius norm, respectively.
    $\mathfrak{R}[\mathbf{A}]$, $\mathbf{A}^\dagger$ and $\mathrm{vec}(\mathbf{A})$ denote the real part, pseudo-inverse matrix and vectorization of the matrix $\mathbf{A}$.
    $\mathcal{CN}(\bm{\mu}, \mathbf{C})$ denote a circularly symmetric complex Gaussian distribution with mean $\bm{\mu}$ and covariance $\mathbf{C}$.
    \Kabuto{$\emptyset$ denotes the empty set.}
    \Kabuto{The Cartesian product of two sets $A$ and $B$ is denoted by $A \times B$.}
    %

\vspace{-0.3cm}
\section{System Model}
    \label{sec:system_model}
    
    We consider an uplink \ac{MIMO} \ac{OFDM} system, 
    where a \ac{BS} and a \ac{UE} employ hybrid \ac{MIMO} architectures with \Kabuto{\acp{UPA}} and $N_\mathrm{r,RF}$ and $N_\mathrm{t,RF}$ \ac{RF} chains, respectively.
    \Kabuto{
    The UPAs are implemented in the $x$--$y$ plane at both the BS and UE sides, equipped with $N_\mathrm{r} = N_\mathrm{r}^x \times N_\mathrm{r}^y$ and $N_\mathrm{t} = N_\mathrm{t}^x \times N_\mathrm{t}^y$ antennas, respectively.
    }
    \Kabuto{
    When $N_\mathrm{r}^y=1$ and $N_\mathrm{t}^y=1$,  the antenna arrays correspond to the \ac{ULA}. This paper considers both \ac{ULA} and \ac{UPA} cases.
    }
    \Kabuto{
    The carrier frequency, the system bandwidth, the number of subcarriers, and the \ac{CP} length are denoted by $f_\mathrm{c}$, $B$, $K$, and $N_\mathrm{cp}$, respectively.}
    The carrier wavelength is denoted by \Kabuto{$\lambda_\mathrm{c} = c/f_\mathrm{c}$} with the speed of light $c$.
    The $k$-th subcarrier $f_k$ is given by $f_k = f_\mathrm{c} +\Delta f_k$ with $\Delta f_k  = -B/2 + (k-1)B/K,\ k \in \{1, 2, \ldots, K \}$.

    \vspace{-0.3cm}
    \subsection{Channel Model in MIMO-OFDM System}
    \label{subsec:channel_model}
    
    We consider a channel model with the frequency-wideband effect and the spatial wideband effect, also known as the dual-wideband effect \cite{2024Garg_squint_CE_JCDE,2020Wang_squint_CE_block,2024Xu_squint_CE_Bayesian_multi}.
    \Kabuto{
    The channel is composed of $L$ resolvable paths.
    For the $l$-th path, $\alpha_l,\ \tau_l,\ \phi_l^\mathrm{a},\ \phi_l^\mathrm{z},\ \theta_l^\mathrm{a},\ \text{and }\theta_l^\mathrm{z}$ are the path gain, delay, \ac{AoA}, \ac{ZoA}, \ac{AoD}, and \ac{ZoD}, respectively.
    }
    \Kabuto{
    For these angle parameters, let 
    $v_{\mathrm{r},l}^x \! \triangleq \! \sin(\phi_l^\mathrm{a}) \cos(\phi_l^\mathrm{z})$, 
    $v_{\mathrm{r},l}^y \! \triangleq \! \sin(\phi_l^\mathrm{a}) \sin(\phi_l^\mathrm{z})$, 
    $v_{\mathrm{t},l}^x \! \triangleq \! \sin(\theta_l^\mathrm{a}) \cos(\theta_l^\mathrm{z})$, and
    $v_{\mathrm{t},l}^y \! \triangleq \! \sin(\theta_l^\mathrm{a}) \sin(\theta_l^\mathrm{z})$
    denote spatial frequencies (also referred to as normalized angles) of the $l$-th path~\cite{2017Shafin_2DMIMO}.
    Using the angles $\mathbf{v}_{\mathrm{r},l} \! \triangleq \! \{ v^x_{\mathrm{r},l}, v^y_{\mathrm{r},l} \}$ and 
    $\mathbf{v}_{\mathrm{t},l} \! \triangleq \! \{ v^x_{\mathrm{t},l}, v^y_{\mathrm{t},l} \}$, the channel matrix between the \ac{BS} and \ac{UE} at the $k$-th subcarrier is expressed 
    as\footnote{
    Although the authors in \cite{2023Xie_DADL} consider the effect of pulse shaping for band-limitation, we assume ideal band-limitation in the OFDM system, as its impact is negligible for channel estimation performance, as in \cite{2024Garg_squint_CE_JCDE,2020Wang_squint_CE_block,2024Xu_squint_CE_Bayesian_multi}.
    }}
    \Kabuto{
    \begin{align}
        \label{eq:H_k}
        \mathbf{H}_k = \sum_{l=1}^L \alpha_l e^{-j2 \pi \Delta f_k \tau_l} \check{\mathbf{a}}_{\mathrm{r},k}(\mathbf{v}_{\mathrm{r},l})
        \check{\mathbf{a}}_{\mathrm{t},k}(\mathbf{v}_{\mathrm{t},l})^\mathrm{H},
    \end{align}
    where 
    $\check{\mathbf{a}}_{\mathrm{r},k}(\mathbf{v}_{\mathrm{r},l}) \in \mathbb{C}^{N_\mathrm{r} \times 1}$ and 
    $\check{\mathbf{a}}_{\mathrm{t},k}(\mathbf{v}_{\mathrm{t},l}) \in \mathbb{C}^{N_\mathrm{t} \times 1}$ are the array response vectors including array errors such as mutual coupling, gain/phase errors and antenna spacing errors at the \ac{BS} and \ac{UE} sides.
    }
    \Kabuto{
    These array response vectors are given by 
    \begin{subequations}
    \begin{align}
        \label{eq:a_bs_ck}
        \check{\mathbf{a}}_{\mathrm{r},k}(\mathbf{v}_{\mathrm{r},l}) &= \mathbf{C}_\mathrm{r} \mathbf{\Gamma}_\mathrm{r}  \mathbf{a}_{\mathrm{r},k}(\mathbf{v}_{\mathrm{r},l}, \bm{\varepsilon}_\mathrm{r}), \\
        \label{eq:a_ue_ck}
        \check{\mathbf{a}}_{\mathrm{t},k}(\mathbf{v}_{\mathrm{t},l}) &= \mathbf{C}_\mathrm{t} \mathbf{\Gamma}_\mathrm{t} \mathbf{a}_{\mathrm{t},k}(\mathbf{v}_{\mathrm{t},l}, \bm{\varepsilon}_\mathrm{t}), 
    \end{align}
    \end{subequations}
    where
    $\mathbf{a}_{\mathrm{r},k}(\mathbf{v}_{\mathrm{r},l}, \bm{\varepsilon}_\mathrm{r}) \in \mathbb{C}^{N_\mathrm{r} \times 1}$ and
    $\mathbf{a}_{\mathrm{t},k}(\mathbf{v}_{\mathrm{t},l}, \bm{\varepsilon}_\mathrm{t}) \in \mathbb{C}^{N_\mathrm{t} \times 1}$
    are the array response vectors including antenna spacing errors $\bm{\varepsilon}_\mathrm{r} \triangleq \{ \bm{\varepsilon}_\mathrm{r}^{x}, \bm{\varepsilon}_\mathrm{r}^{y} \}$ and 
    $\bm{\varepsilon}_\mathrm{t} \triangleq \{ \bm{\varepsilon}_\mathrm{t}^{x}, \bm{\varepsilon}_\mathrm{t}^{y} \}$.
    }
    \Kabuto{
    These array response vectors $\mathbf{a}_{\mathrm{r},k}(\mathbf{v}_{\mathrm{r},l}, \bm{\varepsilon}_\mathrm{r})$ and
    $\mathbf{a}_{\mathrm{t},k}(\mathbf{v}_{\mathrm{t},l}, \bm{\varepsilon}_\mathrm{t})$
    can be expressed as
    \begin{subequations}
    \begin{align}
        \label{eq:ar}
        \mathbf{a}_{\mathrm{r},k}(\mathbf{v}_{\mathrm{r},l}, \bm{\varepsilon}_\mathrm{r}) &= \mathbf{a}^y_{\mathrm{r},k}(v^y_{\mathrm{r},l}, \bm{\varepsilon}_\mathrm{r}^y) \otimes
         \mathbf{a}^x_{\mathrm{r},k}(v^x_{\mathrm{r},l}, \bm{\varepsilon}_\mathrm{r}^x), \\
        \label{eq:at}
        \mathbf{a}_{\mathrm{t},k}(\mathbf{v}_{\mathrm{t},l}, \bm{\varepsilon}_\mathrm{t}) &= \mathbf{a}^y_{\mathrm{t},k}(v^y_{\mathrm{t},l}, \bm{\varepsilon}_\mathrm{t}^y) \otimes
        \mathbf{a}^x_{\mathrm{t},k}(v^x_{\mathrm{t},l}, \bm{\varepsilon}_\mathrm{t}^x),
    \end{align}
    \end{subequations}
    where
    $\mathbf{a}^x_{\mathrm{r},k}(v^x_{\mathrm{r},l}, \bm{\varepsilon}_\mathrm{r}^x) \in \mathbb{C}^{N_\mathrm{r}^x \times 1}$ and 
    $\mathbf{a}^y_{\mathrm{r},k}(v^y_{\mathrm{r},l}, \bm{\varepsilon}_\mathrm{r}^y) \in \mathbb{C}^{N_\mathrm{r}^y \times 1}$ 
    are the array response vectors at the BS side for the $x$-axis and $y$-axis, respectively, and 
    $\mathbf{a}^x_{\mathrm{t},k}(v^x_{\mathrm{t},l}, \bm{\varepsilon}_\mathrm{t}^x) \in \mathbb{C}^{N_\mathrm{t}^x \times 1}$ and 
    $\mathbf{a}^y_{\mathrm{t},k}(v^y_{\mathrm{t},l}, \bm{\varepsilon}_\mathrm{t}^y) \in \mathbb{C}^{N_\mathrm{t}^y \times 1}$ 
    are the array response vectors at the UE side for the $x$-axis and $y$-axis, respectively.
    }
    \Kabuto{
    The $i$-th element of these array response vectors are given by
    \begin{subequations}
    \label{eq:ar_xy}
    \begin{align}
        \left [\mathbf{a}^x_{\mathrm{r},k}(v^x_{\mathrm{r},l}, \bm{\varepsilon}_\mathrm{r}^x) \right ]_i &= e^{j \frac{2 \pi}{\lambda_\mathrm{c}} \left( 1+\frac{\Delta f_k}{f_\mathrm{c}}\right ) (i-1) (d_\mathrm{r}^x + \varepsilon^x_{\mathrm{r},i}) v^x_{\mathrm{r},l} },  \\
        \left [\mathbf{a}^y_{\mathrm{r},k}(v^y_{\mathrm{r},l}, \bm{\varepsilon}_\mathrm{r}^y) \right ]_i &= e^{j \frac{2 \pi}{\lambda_\mathrm{c}} \left( 1+\frac{\Delta f_k}{f_\mathrm{c}}\right ) (i-1) (d_\mathrm{r}^y + \varepsilon^y_{\mathrm{r},i}) v^y_{\mathrm{r},l} },  \\
        \left [\mathbf{a}^x_{\mathrm{t},k}(v^x_{\mathrm{t},l}, \bm{\varepsilon}_\mathrm{t}^x) \right ]_i &= e^{j \frac{2 \pi}{\lambda_\mathrm{c}} \left( 1+\frac{\Delta f_k}{f_\mathrm{c}}\right ) (i-1) (d_\mathrm{t}^x + \varepsilon^x_{\mathrm{t},i}) v^x_{\mathrm{t},l} },  \\
        \left [\mathbf{a}^y_{\mathrm{t},k}(v^y_{\mathrm{t},l}, \bm{\varepsilon}_\mathrm{t}^y) \right ]_i &= e^{j \frac{2 \pi}{\lambda_\mathrm{c}} \left( 1+\frac{\Delta f_k}{f_\mathrm{c}}\right ) (i-1) (d_\mathrm{t}^y + \varepsilon^x_{\mathrm{t},i}) v^y_{\mathrm{t},l} }.
    \end{align}
    \end{subequations}
    where 
    $d_\mathrm{r}^x$, $d_\mathrm{r}^y$, $d_\mathrm{t}^x$ and $d_\mathrm{t}^y$ are the ideal antenna spacing at the BS and UE sides, respectively.
    With the angles
    \Kabuto{
    $\mathbf{v}_\mathrm{r}^x \triangleq \{v^x_{\mathrm{r},l} \}_{l=1}^L$, $\mathbf{v}_\mathrm{r}^y \triangleq \{v^y_{\mathrm{r},l} \}_{l=1}^L$, $\mathbf{v}_\mathrm{t}^x \triangleq \{v^x_{\mathrm{t},l} \}_{l=1}^L$, and 
    $\mathbf{v}_\mathrm{t}^y \triangleq \{v^y_{\mathrm{t},l} \}_{l=1}^L$, 
    }
    the array response matrices at the BS for the $x$-axis and the $y$-axis, and at the UE for $x$-axis and $y$-axis are defined as 
    $\mathbf{A}_\mathrm{r}^x (\mathbf{v}^x_\mathrm{r}, \bm{\varepsilon}^x_\mathrm{r}) \triangleq 
    \begin{bmatrix}
        \mathbf{a}_\mathrm{r}^x (v^x_{\mathrm{r},1}, \bm{\varepsilon}^x_\mathrm{r}), \ldots, \mathbf{a}_\mathrm{r}^x (v^x_{\mathrm{r},L}, \bm{\varepsilon}^x_\mathrm{r})
    \end{bmatrix} \in \mathbb{C}^{N^x_\mathrm{r} \times L}$, 
    $\mathbf{A}_\mathrm{r}^y (\mathbf{v}^y_\mathrm{r}, \bm{\varepsilon}^y_\mathrm{r}) \triangleq 
    \begin{bmatrix}
        \mathbf{a}_\mathrm{r}^y (v^y_{\mathrm{r},1}, \bm{\varepsilon}^y_\mathrm{r}), \ldots, \mathbf{a}_\mathrm{r}^y (v^y_{\mathrm{r},L}, \bm{\varepsilon}^y_\mathrm{r})
    \end{bmatrix} \in \mathbb{C}^{N^y_\mathrm{r} \times L}$, 
    $\mathbf{A}_\mathrm{t}^x (\mathbf{v}^x_\mathrm{t}, \bm{\varepsilon}^x_\mathrm{t}) \triangleq 
    \begin{bmatrix}
        \mathbf{a}_\mathrm{t}^x (v^x_{\mathrm{t},1}, \bm{\varepsilon}^x_\mathrm{t}), \ldots, \mathbf{a}_\mathrm{t}^x (v^x_{\mathrm{t},L}, \bm{\varepsilon}^x_\mathrm{t})
    \end{bmatrix} \in \mathbb{C}^{N^x_\mathrm{t} \times L}$, and
    $\mathbf{A}_\mathrm{t}^y (\mathbf{v}^y_\mathrm{t}, \bm{\varepsilon}^y_\mathrm{t}) \triangleq 
    \begin{bmatrix}
        \mathbf{a}_\mathrm{t}^y (v^y_{\mathrm{t},1}, \bm{\varepsilon}^y_\mathrm{t}), \ldots, \mathbf{a}_\mathrm{t}^y (v^y_{\mathrm{t},L}, \bm{\varepsilon}^y_\mathrm{t})
    \end{bmatrix} \in \mathbb{C}^{N^y_\mathrm{t} \times L}$, respectively.
    }
    %
    Due to the beam squint effect~\cite{2024Garg_squint_CE_JCDE,2020Wang_squint_CE_block,2024Xu_squint_CE_Bayesian_multi}, the array response vectors in \eqref{eq:ar_xy} depend on the subcarrier index $k$ because the approximation $ \Delta f_k / f_\mathrm{c} \simeq 0$ does not hold unlike the conventional MIMO systems~\cite{2018Rodriguez_SOMP,2020Xie_CoDL}.
    
    In \eqref{eq:a_bs_ck} and \eqref{eq:a_ue_ck}, 
    $\mathbf{\Gamma}_\mathrm{r} \triangleq \mathrm{diag}( \bm{\gamma}_\mathrm{r}) \in \mathbb{C}^{N_\mathrm{r} \times N_\mathrm{r}}$, and
    $\mathbf{\Gamma}_\mathrm{t} \triangleq \mathrm{diag}(\bm{\gamma}_\mathrm{t}) \in \mathbb{C}^{N_\mathrm{t} \times N_\mathrm{t}}$
    are the antenna gain/phase error matrices with 
    $\bm{\gamma}_\mathrm{r} \triangleq [
        g_{\mathrm{r},1} e^{j \nu_{\mathrm{r},1}},\ \ldots,\ g_{\mathrm{r},N_\mathrm{r}} e^{j \nu_{\mathrm{r},N_\mathrm{r}}}
    ]^\mathrm{T} \in \mathbb{C}^{N_\mathrm{r} \times 1}$, and
    $\bm{\gamma}_\mathrm{t} \triangleq [
        g_{\mathrm{t},1} e^{j \nu_{\mathrm{t},1}},\ \ldots,\ g_{\mathrm{t},N_\mathrm{t}} e^{j \nu_{\mathrm{t},N_\mathrm{t}}}
    ]^\mathrm{T} \in \mathbb{C}^{N_\mathrm{t} \times 1}$, 
    where
    $\{g_{\mathrm{r},n_\mathrm{r}}, \nu_{\mathrm{r}, n_\mathrm{r}} \}$ and $\{g_{\mathrm{t},n_\mathrm{t}}, \nu_{\mathrm{t}, n_\mathrm{t}} \}$ denote the gain and phase errors at the $n_\mathrm{r}$-th \ac{BS} antenna and the $n_\mathrm{t}$-th UE antenna, respectively.
    %
    $\mathbf{C}_\mathrm{r} \in \mathbb{C}^{N_\mathrm{r} \times N_\mathrm{r}}$ and
    $\mathbf{C}_\mathrm{t} \in \mathbb{C}^{N_\mathrm{t} \times N_\mathrm{t}}$
    are the mutual coupling matrices.
    \Kabuto{
    The mutual coupling matrix is ideally modeled by a symmetric Toeplitz matrix in the \ac{ULA} case~\cite{1991Friedlander_Direction_finding}, and a symmetric block Toeplitz matrix in the \ac{UPA} case~\cite{2008Ye_2D_Copuling, 2023Lan_2D_Coupling}.
    }
    However, the antenna spacing errors $\bm{\varepsilon}_{\mathrm{r}}$ and $\bm{\varepsilon}_{\mathrm{t}}$ distort the mutual coupling effects. 
    Therefore, calibration methods under the assumption of the symmetric Toeplitz structure leads to the performance deterioration due to the modeling errors.
    Thus, this paper considers the modeling errors as 
    will be described in Section~\ref{subsec:Est_c}.

    \Kabuto{
    By denoting the equivalent path gain at the $k$-th subcarrier as $\mathbf{z}_k \triangleq [z_{1,k}, \ldots, z_{L,k}]^\mathrm{T} \in \mathbb{C}^{L \times 1}$ with $z_{k,l} \triangleq \alpha_l e^{-j2 \pi \Delta f_k \tau_l}$,
    the channel matrix at the $k$-th subcarrier in \eqref{eq:H_k} is expressed as 
    \begin{align}
        \mathbf{H}_k = \check{\mathbf{A}}_{\mathrm{r},k}(\mathbf{v}_\mathrm{r}) \mathrm{diag}(\mathbf{z}_k) \check{\mathbf{A}}^\mathrm{H}_{\mathrm{t},k}(\mathbf{v}_\mathrm{t}),
    \end{align}
    where
    %
    $\check{\mathbf{A}}_{\mathrm{r},k}(\mathbf{v}_\mathrm{r}) \triangleq
    \begin{bmatrix}
        \check{\mathbf{a}}_{\mathrm{r},k} (\mathbf{v}_{\mathrm{r},1}) \ \cdots \ \check{\mathbf{a}}_{\mathrm{r},k} (\mathbf{v}_{\mathrm{r},L})
    \end{bmatrix} \in \mathbb{C}^{N_\mathrm{r} \times L}$ and
    $\check{\mathbf{A}}_{\mathrm{t},k}(\mathbf{v}_{\mathrm{t}}) \triangleq
    \begin{bmatrix}
        \check{\mathbf{a}}_{\mathrm{t},k} (\mathbf{v}_{\mathrm{t},1}) \ \cdots \ \check{\mathbf{a}}_{\mathrm{t},k} (\mathbf{v}_{\mathrm{t},L})
    \end{bmatrix} \in \mathbb{C}^{N_\mathrm{t} \times L}$ are the array response matrices including all array errors with 
    the angles at the BS $\mathbf{v}_{\mathrm{r}} \triangleq \{ \mathbf{v}_{\mathrm{r},l} \}_{l=1}^L$ and at the UE  $\mathbf{v}_{\mathrm{t}} \triangleq \{ \mathbf{v}_{\mathrm{t},l} \}_{l=1}^L$.    
    }
    For the sake of notation convenience, the array response matrices including antenna spacing errors at the BS and UE are defined as 
    \Kabuto{
    $\mathbf{A}_{\mathrm{r},k}(\mathbf{v}_{\mathrm{r}}, \bm{\varepsilon}_\mathrm{r}) \triangleq
    \begin{bmatrix}
        \mathbf{a}_{\mathrm{r},k} (\mathbf{v}_{\mathrm{r}, 1}, \bm{\varepsilon}_\mathrm{r}) \ \cdots \ \mathbf{a}_{\mathrm{r},k} (\mathbf{v}_{\mathrm{r}, L}, \bm{\varepsilon}_\mathrm{r})
    \end{bmatrix} \in \mathbb{C}^{N_\mathrm{r} \times L}$ and 
    %
    $\mathbf{A}_{\mathrm{t},k}(\mathbf{v}_{\mathrm{t}}, \bm{\varepsilon}_\mathrm{t}) \triangleq
    \begin{bmatrix}
        \mathbf{a}_{\mathrm{t},k} (\mathbf{v}_{\mathrm{t}, 1}, \bm{\varepsilon}_\mathrm{t}) \ \cdots \ \mathbf{a}_{\mathrm{t},k} (\mathbf{v}_{\mathrm{t}, L}, \bm{\varepsilon}_\mathrm{t})
    \end{bmatrix} \in \mathbb{C}^{N_\mathrm{t} \times L}$, 
    which satisfy  
    $\check{\mathbf{A}}_{\mathrm{r},k}(\mathbf{v}_{\mathrm{r}}) = 
    \mathbf{C}_\mathrm{r} \mathbf{\Gamma}_\mathrm{r}
    \mathbf{A}_{\mathrm{r},k}(\mathbf{v}_{\mathrm{r}}, \bm{\varepsilon}_\mathrm{r})$ and 
    $\check{\mathbf{A}}_{\mathrm{t},k}(\mathbf{v}_{\mathrm{t}}) = 
    \mathbf{C}_\mathrm{t} \mathbf{\Gamma}_\mathrm{t}
    \mathbf{A}_{\mathrm{t},k}(\mathbf{v}_{\mathrm{t}}, \bm{\varepsilon}_\mathrm{t})$.
    }

    Based on the property of vectorization, $\mathrm{vec}(\mathbf{A} \mathrm{diag}(\mathbf{b}) \mathbf{C}) = (\mathbf{C}^\mathrm{T} \circ \mathbf{A}) \mathbf{b}$, the vectorized channel at the $k$-th subcarrier $\mathbf{h}_k \triangleq \mathrm{vec} (\mathbf{H}_{k}) \in \mathbb{C}^{N_\mathrm{r} N_\mathrm{t} \times 1} $ is given by
    \Kabuto{
    \begin{align}
        \label{eq:H_k_vec_z}
        \mathbf{h}_k =
        \left \{ \check{\mathbf{A}}_{\mathrm{t},k}^\ast(\mathbf{v}_{\mathrm{t}}) \circ \check{\mathbf{A}}_{\mathrm{r},k}(\mathbf{v}_{\mathrm{r}}) \right \}
        \mathbf{z}_k.
    \end{align}
    }
    \Kabuto{
    For the $k$-th subcarrier, the delay response is defined as $\mathbf{b}_k (\bm{\tau}) \triangleq [e^{-j2 \pi \Delta f_k \tau_1}, \ldots, e^{-j2 \pi \Delta f_k \tau_L}]^\mathrm{T} \in \mathbb{C}^{L \times 1}$.
    Then, the equivalent path gain for the $k$-th subcarrier $\mathbf{z}_k$ is given by
    \begin{align}
        \label{eq:zk}
        \mathbf{z}_k =  \mathrm{diag} (\mathbf{b}_k( \bm{\tau})) \bm{\alpha}.
    \end{align}
    }
    \Kabuto{
    Substituting \eqref{eq:zk} into \eqref{eq:H_k_vec_z}, the channel vector at the $k$-th subcarrier $\mathbf{h}_k$ can be expressed as
    \begin{align}
        \label{eq:H_k_vec}
        \mathbf{h}_k 
        &= \underbrace{ \left \{ \mathbf{b}_k^\mathrm{T} (\bm{\tau} ) \circ \check{\mathbf{A}}_{\mathrm{t},k}^\ast(\mathbf{v}_\mathrm{t} ) \circ \check{\mathbf{A}}_{\mathrm{r},k}(\mathbf{v}_\mathrm{r} )
        \right \} }_{ \triangleq \mathbf{\Psi}_k^{\bm{\alpha}} \in \mathbb{C}^{N_\mathrm{r} N_\mathrm{t} \times L} }
        \bm{\alpha}.
    \end{align}
    }

    \Kabuto{
    Defining 
    $\mathbf{\Psi}^{\bm{\alpha}} \! \triangleq \! \begin{bmatrix} \mathbf{\Psi}^{\bm{\alpha} \mathrm{T}}_1, \ldots, \mathbf{\Psi}^{\bm{\alpha} \mathrm{T}}_K \end{bmatrix}^\mathrm{T} \!\! \in \! \mathbb{C}^{N_\mathrm{r} N_\mathrm{t} K \times L}$, 
    the channel vector $\mathbf{h} \triangleq \begin{bmatrix} \mathbf{h}^{\mathrm{T}}_1, \ldots, \mathbf{h}^{\mathrm{T}}_K \end{bmatrix}^\mathrm{T} \in \mathbb{C}^{N_\mathrm{r} N_\mathrm{t} K \times 1}$ is given by
    \begin{align}
        \label{eq:h_vec}
        \mathbf{h} = \mathbf{\Psi}^{\bm{\alpha}} \bm{\alpha}.
    \end{align}
    }

    \vspace{-0.3cm}
    \subsection{Received Signal Model in Hybrid MIMO-OFDM System}
    \Kabuto{
    To estimate the channel matrix, the UE transmits pilot symbols with the symbol length $N_\mathrm{p}$ to the BS in each training frame.
    Using the pilot symbols from the total $M$ training frames, including the past $M-1$ frames and the current frame, the BS calibrates the array errors and estimates channel for the current frame.
    Since the cause of array errors is attributed to physical deformation, their time scale is typically on the order of minutes~\cite{2020Wang_radome_coupling, 2022Liu_thermal_deform, 2023Xu_thermal_coupling, 2023Feng_thermal}, which is much larger than the time scale of channel variation at the high carrier frequencies. 
    Therefore, $M$ frames are selected such that the angles, delays, and path gains vary randomly, while the array errors remain constant over the $M$ frames.
    Let $m \in \{1,2,\ldots, M\}$ denote the frame index, where $m=1,2,\ldots, M-1$ correspond to the past frames, and $m=M$ corresponds to the current frame.
    For the $m$-th training frame, let $\mathbf{H}_k^{(m)}$, $\mathbf{v}_\mathrm{r}^{(m)}$, $\mathbf{v}_\mathrm{t}^{(m)}$, $\bm{\tau}^{(m)}$ and $\bm{\alpha}^{(m)}$ denote the channel matrix, the angles at the BS and UE, delay time, and path gains, respectively.
    Our objective is to estimate the channel matrix at the current frame $\mathbf{H}^{(M)}_k$ using the $M$ frames\footnote{
    \Kabuto{
    Although the channel fluctuates over time during data transmission, it is reasonable to assume that the channel remains constant during the pilot transmission, since the time duration of pilot transmission is much shorter than that of data transmission~\cite{2020Gonzalez_SIGW, 2024Uchimura_tracking}.}}.
    }

    \Kabuto{
    To consider pilot overhead and computational complexity in the wideband system, only $Q\ (\leq K)$ subcarriers are used for pilot transmission instead of $K$ subcarriers\footnote{\Kabuto{
    By allocating the remaining subcarriers to other UEs for pilot transmission, this method can be readily extended to multi-user scenarios.
    }}.
    Let $\mathcal{K} \triangleq \{1,2,\ldots, K\}$ denote the subcarrier index set, and 
    $\mathcal{Q} \triangleq \{k_1,k_2,\ldots, k_{Q}\} \subset \mathcal{K}$ denote the pilot subcarrier index set.
    }
    For the $p \! \in \! \{1, \ldots, N_\mathrm{p} \}$-th pilot symbol,
    $\mathbf{q}_{p} \! \in \! \mathbb{C}^{N_\mathrm{t, RF}}$ denotes the pilot generated by \ac{QPSK}, satisfying 
    $\mathbb{E}[\mathbf{q}_p \mathbf{q}_p^\mathrm{H} ] \!=\!\! P_\mathrm{s} \mathbf{I}_{N_\mathrm{r,RF}}$ 
    with the transmit power $P_\mathrm{s}$.
    \Kabuto{
    $\bar{\mathbf{W}}_{p} \! \in \! \mathbb{C}^{N_\mathrm{r,RF} \times N_\mathrm{r}}$ and 
    $\mathbf{F}_{p} \! \in \! \mathbb{C}^{N_\mathrm{t} \times N_\mathrm{t, RF}}$
    are the analog training combiner and precoder, designed by~ \cite{2018Rodriguez_SOMP}.
    }

    Then, the received pilot signal \Kabuto{(after CP removal)} at the $p$-th pilot symbol, the $k$-th subcarrier, and the $m$-th training frame, $\bar{\mathbf{y}}_{p,k}^{(m)} \in \mathbb{C}^{N_\mathrm{r, RF} \times 1}$ is expressed as
    \begin{align}
        \label{eq:y_pk}
        \bar{\mathbf{y}}_{p,k}^{(m)} = \bar{\mathbf{W}}_{p} \mathbf{H}_{k}^{(m)} \mathbf{F}_{p} \mathbf{q}_{p} + \bar{\mathbf{W}}_{p} \bar{\mathbf{n}}_{p,k}^{(m)},
    \end{align}
    where
    \Kabuto{
    $\bar{\mathbf{n}}_{p,k}^{(m)} \in \mathbb{C}^{N_\mathrm{r} \times 1}$} is an additive noise vector that follows \Kabuto{$\mathcal{CN} \left( \mathbf{0}_{N_{\mathrm{r}} \times 1}, \sigma^2 \mathbf{I}_{N_\mathrm{r}} \right)$} with noise variance $\sigma^2$.

    Since the combined noise vector 
    $\bar{\mathbf{W}}_{p} \bar{\mathbf{n}}_{p,k}^{(m)}$ in \eqref{eq:y_pk} follows 
    $\mathcal{CN}(\mathbf{0}_{N_\mathrm{r,RF} \times 1}, \mathbf{C}_p)$
    with the covariance matrix $\mathbf{C}_{p} = \sigma^2 \bar{\mathbf{W}}_{p} \bar{\mathbf{W}}_{p}^\mathrm{H}$,
    the received signal should be whitened by the pre-whitening matrix $\mathbf{D}_p = \sigma \mathbf{C}^{-\frac{1}{2}}_p$. 
    Then, the whitened received signal $\mathbf{y}^{(m)}_{p,k} \triangleq \mathbf{D}_p \bar{\mathbf{y}}^{(m)}_{p,k}\ \in \mathbb{C}^{N_\mathrm{r, RF} \times 1}$ is given by
    \begin{align}
        \label{eq:y_pk_white}
        \mathbf{y}_{p,k}^{(m)} 
        &= \mathbf{W}_{p} \mathbf{H}_{k}^{(m)} \mathbf{F}_{p} \mathbf{q}_{p} + \mathbf{n}_{p,k}^{(m)}, 
    \end{align}
    where $\mathbf{n}_{p,k}^{(m)} = \mathbf{D}_p \bar{\mathbf{W}}_{p} \bar{\mathbf{n}}_{p,k}^{(m)}$ is the whitened noise vector that follows $\mathcal{CN}(0, \sigma^2 \mathbf{I}_{N_\mathrm{r, RF}})$ and 
    $\mathbf{W}_p \triangleq \mathbf{D}_p \bar{\mathbf{W}}_p$ is the whitened analog combiner.

    Using the property of $\mathrm{vec}(\mathbf{AXC}) = (\mathbf{C}^\mathrm{T} \otimes \mathbf{A}) \mathrm{vec}(\mathbf{X})$, the received signal $\mathbf{y}_{p,k}^{(m)}$ in (\ref{eq:y_pk_white}) can be reformulated as
    \begin{align}
        \label{eq:y_pk_vec}
        \mathbf{y}_{p,k}^{(m)}
        & = \underbrace{\left ( \mathbf{q}_{{p}}^\mathrm{T} \mathbf{F}_{{p}}^\mathrm{T} \otimes \mathbf{W}_{p} \right )}_{\triangleq \mathbf{\Phi}_p \in \mathbb{C}^{N_\mathrm{r,RF} \times N_\mathrm{r} N_\mathrm{t}}} \mathbf{h}_k^{(m)}
        + \mathbf{n}_{p,k}^{(m)},
    \end{align}
    where 
    $\mathbf{h}_k^{(m)}$ is the vectorized channel defined in \eqref{eq:H_k_vec}, and 
    $\mathbf{\Phi}_p$ is the training beam matrix at the $p$-th pilot symbol.

    Stacking $\mathbf{y}_{p,k}^{(m)}$ over $N_p$ pilots, the received signal $\mathbf{y}_k^{(m)} \triangleq 
    [ \mathbf{y}_{1,k}^{(m) \mathrm{T}}, \ldots, \mathbf{y}_{N_\mathrm{p},k}^{(m)\mathrm{T}}
    ]^\mathrm{T} \in \mathbb{C}^{N_\mathrm{r,RF} N_\mathrm{p} \times 1}
    $ can be expressed as
    \begin{align}
        \label{eq:y_obs}
        \mathbf{y}_k^{(m)} = \mathbf{\Phi} \mathbf{h}_k^{(m)} + \mathbf{n}_k^{(m)},
    \end{align}
    where 
    $\mathbf{n}_k^{(m)} = [ 
    \mathbf{n}_{1,k}^{(m) \mathrm{T}}, \ldots, \mathbf{n}_{N_\mathrm{p},k}^{(m) \mathrm{T}} ]^\mathrm{T} \in \mathbb{C}^{N_\mathrm{r,RF} N_\mathrm{p} \times 1}$ and 
    $\mathbf{\Phi} = [
        \mathbf{\Phi}_{1}^\mathrm{T}, \ldots, \mathbf{\Phi}^\mathrm{T}_{N_\mathrm{p}}
    ]^\mathrm{T} \in \mathbb{C}^{N_\mathrm{r,RF} N_\mathrm{p} \times N_\mathrm{r} N_\mathrm{t}}$ are the noise vector and training beam matrix, including $N_\mathrm{p}$ pilot symbols.

    \Kabuto{
    Similarly, stacking $\mathbf{y}_k^{(m)} $ over $Q$ pilot subcarriers on $\mathcal{Q}  = \{k_1, \ldots k_{Q} \}$, 
    $\mathbf{y}^{(m)} \triangleq [ \mathbf{y}^{(m) \mathrm{T}}_{k_1}, \ldots, \mathbf{y}^{(m) \mathrm{T}}_{k_{Q}} ]^\mathrm{T} \in \mathbb{C}^{N_{\mathrm{r, RF}} N_\mathrm{p} Q \times 1}$ can be expressed as 
    \begin{align}
        \label{eq:y_obs_delay}
        \mathbf{y}^{(m)} = \check{\mathbf{\Phi}} \mathbf{h}_\mathcal{Q}^{(m)} + \mathbf{n}^{(m)},
    \end{align}
    where $\check{\mathbf{\Phi}} = \mathbf{I}_Q \otimes \mathbf{\Phi} \in \mathbb{C}^{N_\mathrm{r,RF} N_\mathrm{p} Q \times N_\mathrm{r} N_\mathrm{t} Q}$ is the training beam matrix, and 
    $\mathbf{h}_\mathcal{Q}^{(m)} =  [ \mathbf{h}_{k_1}^{(m)\mathrm{T}}, \ldots, \mathbf{h}_{k_Q}^{(m)\mathrm{T}} ]^\mathrm{T} \in \mathbb{C}^{QN_\mathrm{r} N_\mathrm{t} \times 1} $ is the channel vector, 
    $ \mathbf{n}^{(m)} = [\mathbf{n}_{k_1}^{(m)\mathrm{T}}, \ldots,  \mathbf{n}_{k_Q}^{(m)\mathrm{T}}]^\mathrm{T} \in \mathbb{C}^{N_\mathrm{r,RF} N_\mathrm{p} Q \times 1}$ is the noise vector.
    }

    \vspace{-0.3cm}
    \subsection{Channel Estimation Without Considering Array Errors}
    \label{subsec:Conv_CE}
    
    To accurately estimate the channel $\mathbf{h}^{(m)}_k $ in \eqref{eq:y_obs} using classical estimation methods, such as \ac{LS}, the number of subcarriers and pilots must satisfy $Q=K$ and $N_\mathrm{p} \geq \frac{N_\mathrm{t} N_\mathrm{t}}{N_\mathrm{r,RF}}$.
    This requirement results in an increase in pilot overhead as the number of subcarriers and antennas grows.
    
    \Kabuto{
    To reduce pilot overhead, channel estimation methods to exploit the channel sparsity in the angle-delay domain have been proposed in \cite{2016Heath_overview_sp, 2018Rodriguez_SOMP, 2024Garg_squint_CE_JCDE, 2020Wang_squint_CE_block, 2024Xu_squint_CE_Bayesian_multi}, based on a \ac{CS} framework.
    These methods utilize virtual channel representation with the angle and delay grids, generated by quantizing the angle-delay domain into $G_{\mathrm{r}}$, $G_{\mathrm{t}}$, and $G_\tau$ points.
    }
    \Kabuto{ 
    The angle grid set at the BS is designed with $G_\mathrm{r} = G^x_\mathrm{r} \times G^y_\mathrm{r}$ grid points as $\tilde{\mathbf{v}}_{\mathrm{r}} = \tilde{\mathbf{v}}_{\mathrm{r}}^x \times \tilde{\mathbf{v}}_{\mathrm{r}}^y$, where $\tilde{\mathbf{v}}_{\mathrm{r}}^x \triangleq \{ \tilde{v}_{\mathrm{r},g}^x \}_{g=1}^{G^x_\mathrm{r}}$ and $\tilde{\mathbf{v}}_{\mathrm{r}}^y \triangleq \{ \tilde{v}_{\mathrm{r},g}^y \}_{g=1}^{G^y_\mathrm{r}}$ are the angle grids for the $x$-axis and the $y$-axis, respectively.
    Similarly, the angle grid set at the UE is designed with $G_\mathrm{t} = G^x_\mathrm{t} \times G^y_\mathrm{t}$ grid points as $\tilde{\mathbf{v}}_{\mathrm{t}} = \tilde{\mathbf{v}}_{\mathrm{t}}^x \times \tilde{\mathbf{v}}_{\mathrm{t}}^y$, where $\tilde{\mathbf{v}}_{\mathrm{t}}^x \triangleq \{ \tilde{v}_{\mathrm{t},g}^x \}_{g=1}^{G^x_\mathrm{t}}$ and $\tilde{\mathbf{v}}_{\mathrm{t}}^y \triangleq \{ \tilde{v}_{\mathrm{t},g}^y \}_{g=1}^{G^y_\mathrm{t}}$ are the angle grids for the $x$-axis and the $y$-axis, respectively.
    These angle grids $\tilde{v}_{\mathrm{r},g}^x$, $\tilde{v}_{\mathrm{r},g}^y$, $\tilde{v}_{\mathrm{t},g}^x$, and $\tilde{v}_{\mathrm{t},g}^y$ are generated from the range $\in [-1, 1)$.
    The delay grid set is designed as $\tilde{\bm{\tau}} \triangleq \{ \tilde{\tau}_{g_\tau} \}_{g_\tau=1}^{G_\tau}$ with $\tilde{\tau}_{g_\tau} \in [0,\ \tau_\mathrm{max}]$.
    }
    \Kabuto{
    Using the virtual array responses
    $\check{\mathbf{A}}_{\mathrm{r},k} (\tilde{\mathbf{v}}_\mathrm{r}) \in \mathbb{C}^{N_\mathrm{r} \times G_\mathrm{t}}$ and $\check{\mathbf{A}}_{\mathrm{t},k} (\tilde{\mathbf{v}}_\mathrm{t}) \in \mathbb{C}^{N_\mathrm{t} \times G_\mathrm{t}}$ with the angle grids $\{ \tilde{\mathbf{v}}_\mathrm{r}, \tilde{\mathbf{v}}_\mathrm{t} \}$, and 
    the virtual delay response $\mathbf{b}_k (\tilde{\bm{\tau}}) \in \mathbb{C}^{G_\tau \times 1}$ with delay grids $\tilde{\bm{\tau}}$, 
    the channel vector in \eqref{eq:H_k_vec} can be approximated as
    }
    \Kabuto{
    \begin{align}
        \label{eq:hk_apx}
        \mathbf{h}_k^{(m)} \simeq 
        \underbrace{
        \left \{ \mathbf{b}_k^\mathrm{T} (\tilde{\bm{\tau}}) \otimes  \check{\mathbf{A}}_{\mathrm{t},k}^\ast(\tilde{\mathbf{v}}_\mathrm{t}) \otimes \check{\mathbf{A}}_{\mathrm{r},k}(\tilde{\mathbf{v}}_\mathrm{r}) \right \} 
        }_{\triangleq \tilde{\mathbf{\Psi}}_k^{\bm{\alpha}} \in \mathbb{C}^{N_\mathrm{r} N_\mathrm{t} \times G_\mathrm{r} G_\mathrm{t} G_\tau}}
        \tilde{\bm{\alpha}}^{(m)},
    \end{align}
    where 
    $\tilde{\mathbf{\Psi}}_k^{\bm{\alpha}}$ is the dictionary matrix, and
    $\tilde{\bm{\alpha}}^{(m)} \in \mathbb{C}^{G_\mathrm{r} G_\mathrm{t} G_\tau \times 1}$ is the sparse path gain vector corresponding to angle and delay grids.
    By designing the dictionary matrix to account for both the phase rotation from delay and the frequency-dependent array responses, beam squint effects can be considered.
    }

    \Kabuto{
    Stacking the channel vector $\mathbf{h}_k^{(m)}$ in \eqref{eq:hk_apx} over $Q$ subcarriers on $\mathcal{Q} = \{k_1, \ldots, k_Q \}$, the channel vector 
    $\mathbf{h}^{(m)}_\mathcal{Q}$
    is give by
    \begin{align}
        \label{eq:h_apx}
        \mathbf{h}^{(m)}_\mathcal{Q} \simeq \tilde{\mathbf{\Psi}}^{\bm{\alpha}}_\mathcal{Q} \tilde{\bm{\alpha}}^{(m)}, 
    \end{align}
    where $\tilde{\mathbf{\Psi}}^{\bm{\alpha}}_\mathcal{Q} \triangleq \begin{bmatrix}
    \tilde{\mathbf{\Psi}}^{\bm{\alpha} \mathrm{T}}_{k_1}, \ldots, \tilde{\mathbf{\Psi}}^{\bm{\alpha} \mathrm{T}}_{k_Q}
    \end{bmatrix}^\mathrm{T} \in \mathbb{C}^{N_\mathrm{t} N_\mathrm{r} Q \times G_\mathrm{r} G_\mathrm{t} G_\tau} $ is the dictionary matrix including $Q$ subcarriers.
    }

    \Kabuto{
    Substituting \eqref{eq:h_apx} into \eqref{eq:y_obs_delay}, the received signal can be approximated as
    \begin{align}
        \label{eq:y_apx}
        \mathbf{y}^{(m)} \simeq \check{\mathbf{\Phi}} \tilde{\mathbf{\Psi}}^{\bm{\alpha}}_\mathcal{Q} \tilde{\bm{\alpha}}^{(m)} + \mathbf{n}^{(m)}.
    \end{align}
    }
    \Kabuto{
    Based on the measurement equation \eqref{eq:y_apx}, the path gain vector, angles, and delays can be estimated with small pilot overhead by CS algorithms.
    Once the estimates of the path gain vector $\hat{\bm{\alpha}}$, the angles $\hat{\mathbf{v}}_\mathrm{r}^{(m)}$, $\hat{\mathbf{v}}_\mathrm{t}^{(m)}$, and delays $\hat{\bm{\tau}}^{(m)}$ are obtained, the channel vectors for all $K$ subcarriers, instead of $Q$ subcarriers, can be reconstructed as
    \begin{align}
        \label{eq:h_est}
        \hat{\mathbf{h}}_k^{(m)} = \hat{\mathbf{\Psi}}_k^{\bm{\alpha} (m)} \hat{\bm{\alpha}}^{(m)}, \quad \forall k \in \mathcal{K},
    \end{align}
    with
    $\hat{\mathbf{\Psi}}_k^{\bm{\alpha} (m)} = \left \{ \mathbf{b}_k^\mathrm{T} (\hat{\bm{\tau}}^{(m)} ) \circ \check{\mathbf{A}}_{\mathrm{t},k}^\ast(\hat{\mathbf{v}}_\mathrm{t}^{(m)} ) \circ \check{\mathbf{A}}_{\mathrm{r},k}(\hat{\mathbf{v}}_\mathrm{r}^{(m)} )
    \right \}$.
    }

    However, since the array responses $\check{\mathbf{A}}_{\mathrm{r},k}$ and $\check{\mathbf{A}}_{\mathrm{t},k}$ include array errors $\{ \mathbf{C}_\mathrm{r}, \mathbf{C}_\mathrm{t}, \mathbf{\Gamma}_\mathrm{r}, \mathbf{\Gamma}_\mathrm{t}, \bm{\varepsilon}_\mathrm{r}, \bm{\varepsilon}_\mathrm{t} \}$, the dictionary matrix $\tilde{\mathbf{\Psi}}^{\bm{\alpha}}_k$ cannot be designed. 
    Therefore, many works design the dictionary matrix under the assumption that array errors do not exist (\textit{i.e.}, $\mathbf{C}_\mathrm{r} = \mathbf{\Gamma}_\mathrm{r} = \mathbf{I}_{N_\mathrm{r}}$, $\mathbf{C}_\mathrm{t} = \mathbf{\Gamma}_\mathrm{t} = \mathbf{I}_{N_\mathrm{t}}$, $\bm{\varepsilon}_\mathrm{r} = \mathbf{0}$, $\bm{\varepsilon}_\mathrm{t} = \mathbf{0}$).
    This assumption causes severe performance deterioration due to model mismatch when using a \ac{CS} algorithm.
    Therefore, it is necessary to compensate for array errors while estimating angles, delays, and path gains to improve channel estimation performance.

\vspace{-0.3cm}
\section{Proposed channel estimation algorithm}
    \label{sec:prop}

    This section describes the proposed channel estimation algorithm considering array errors and beam squint effects.
    As described in Section \ref{subsec:Conv_CE}, channel estimation performance based on a CS framework is limited due to array errors. 
    
    To address this issue, the authors in \cite{2020Xie_CoDL,2022Maity_CoDL_THz,2023Xie_DADL} proposed dictionary learning methods, where the path gain vector and the dictionary matrix, which includes array errors, are alternately updated in an iterative manner.
    However, these approaches cause substantial pilot overhead for accurately updating the dictionary matrix, primarily due to the large dimension of the dictionary. 
    Therefore, the proposed method explicitly decomposes the dictionary matrix into a smaller number of physical parameters, and estimates these parameters instead of directly estimating the dictionary.
    Thus, the required pilot overhead can be effectively reduced.

    \Kabuto{
    From \eqref{eq:H_k_vec}, it can be seen that the channel vector consists of the 
    unknown parameters $\mathbf{\Omega} \triangleq  \big \{ \mathbf{C}_\mathrm{r}, \mathbf{C}_\mathrm{t}, \mathbf{\Gamma}_\mathrm{r}, \mathbf{\Gamma}_\mathrm{t}, \bm{\varepsilon}_\mathrm{r}, \bm{\varepsilon}_\mathrm{t},  \{\mathbf{v}_\mathrm{r}^{(m)}, \mathbf{v}_\mathrm{t}^{(m)}, \bm{\tau}^{(m)}, \bm{\alpha}^{(m)} \}_{m=1}^M  \big \}$.
    }
    In the proposed methods, these unknown parameters are alternately estimated based on a \ac{ML} criterion.
    From \eqref{eq:H_k_vec} and \eqref{eq:y_pk_vec}, the conditional \ac{PDF} given the unknown parameter set $\mathbf{\Omega}$ is written by    \footnote{
    \Kabuto{
    Since the conditional PDF $p \left( \mathbf{y}_{p,k}^{(m)}| \mathbf{\Omega} \right )$ is constructed over an arbitrary continuous angle-delay domain, the equality in \eqref{eq:py} holds exactly.
    }}
    \Kabuto{
    \begin{align}  
        \label{eq:py}
        p \left( \mathbf{y}_{p,k}^{(m)}| \mathbf{\Omega} \right ) = \mathcal{CN} \left ( \mathbf{\Phi}_p \mathbf{\Psi}_k^{\bm{\alpha}(m)} \bm{\alpha}_k^{(m)},\ \sigma^2        \mathbf{I}_{N_\mathrm{r,RF}} \right ).
    \end{align}
    }
    Since the log likelihood function $\mathcal{L} (\mathbf{\Omega})$ can be expressed as
    \Kabuto{
    \begin{align} 
        \mathcal{L} (\mathbf{\Omega}) &= \sum_{m=1}^{M} \sum_{p=1}^{N_\mathrm{p}} \sum_{k \in \mathcal{Q}}
        \ln p \left( \mathbf{y}_{p,k}^{(m)}| \mathbf{\Omega} \right ),
    \end{align}
    the \ac{ML} problem can be formulated as the minimization problem \eqref{eq:min} at the top of the next page.
    }
    \begin{figure*}
        \normalsize
        \begin{align}
            \label{eq:min}
            \underset{\mathbf{\Omega}}{\mathrm{minimize}} \ 
            \mathcal{F} (\mathbf{\Omega}) \! =  \!
            \sum_{m=1}^{M} \sum_{p=1}^{N_\mathrm{p}} \sum_{k \in \mathcal{Q}}
            & \Big \| \mathbf{y}_{p,k}^{(m)} \! - \! \mathbf{W}_p \mathbf{C}_\mathrm{r} \mathbf{\Gamma}_\mathrm{r} \mathbf{A}_{\mathrm{r},k} (\mathbf{v}_\mathrm{r}^{(m)}, \bm{\varepsilon}_\mathrm{r}) \mathrm{diag} \left ( \! \mathbf{b}_k( \bm{\tau}^{(m)}) \! \odot \! \bm{\alpha}^{(m)} \! \right) 
            \mathbf{A}_{\mathrm{t},k}^{\mathrm{H}} (\mathbf{v}_\mathrm{t}^{(m)}, \bm{\varepsilon}_\mathrm{t}) \mathbf{\Gamma}_\mathrm{t}^\mathrm{H} \mathbf{C}^{\mathrm{H}}_\mathrm{t} \mathbf{F}_p \mathbf{q}_p  \Big \|_2^2 .
        \end{align}
        \hrulefill
        \vspace*{-12pt}
    \end{figure*}

    Since the objective function $\mathcal{F} (\mathbf{\Omega})$ is not jointly convex for the unknown parameter set $\mathbf{\Omega}$,
    we employ an alternating optimization technique, where each parameter is iteratively updated with the other parameters fixed as tentative estimates.
    As described in the following subsections, the estimates for \Kabuto{$\hat{\bm{\alpha}}^{(m)}$, $\hat{\mathbf{C}}_\mathrm{r}$, $\hat{\mathbf{C}}_\mathrm{t}$, $\hat{\mathbf{\Gamma}}_\mathrm{r}$, and $\hat{\mathbf{\Gamma}}_\mathrm{t}$} are obtained in closed-form owing to the convexity of the objective function for each parameter in the alternating minimization process.
    In contrast, since the objective function is not convex with respect to \Kabuto{$\mathbf{v}_\mathrm{r}^{(m)}$, $\mathbf{v}_\mathrm{t}^{(m)}$, $ \bm{\tau}^{(m)}$, $\bm{\varepsilon}_\mathrm{r}$, and $\bm{\varepsilon}_\mathrm{t}$}, these parameters are updated via a gradient descent approach or a CS-based method.

    The following subsections describe the estimation method for each parameter, given the other tentative estimates.
    As for the initial tentative estimates in the algorithmic iteration, these parameters are set to 
    $\hat{\mathbf{C}}_\mathrm{r} = \hat{\mathbf{\Gamma}}_\mathrm{r} = \mathbf{I}_{N_\mathrm{r}}$, 
    $\hat{\mathbf{C}}_\mathrm{t} = \hat{\mathbf{\Gamma}}_\mathrm{t} = \mathbf{I}_{N_\mathrm{t}}$, 
    $\hat{\bm{\varepsilon}}_\mathrm{r} = \mathbf{0}$, 
    $\hat{\bm{\varepsilon}}_\mathrm{t} = \mathbf{0}$
    under the assumption of no array errors.

    \vspace{-0.3cm}
    \subsection{\Kabuto{Estimation for Angles, Delays and Path Gains}}
    In this estimation stage, angles, delays, and path gains $\{ \mathbf{v}_\mathrm{r}^{(m)}, \mathbf{v}_\mathrm{t}^{(m)}, \bm{\tau}^{(m)}, \bm{\alpha}^{(m)} \}_{m=1}^M$ are estimated, given the tentative estimates $ \{ \hat{\mathbf{C}}_\mathrm{r}, \hat{\mathbf{C}}_\mathrm{t}, \hat{\mathbf{\Gamma}}_\mathrm{r}, \hat{\mathbf{\Gamma}}_\mathrm{t}, \hat{\bm{\varepsilon}}_\mathrm{r}, \hat{\bm{\varepsilon}}_\mathrm{t} \}$, thorough on-grid and off-grid algorithms.
    Although the on-grid algorithm suffers from quantization errors due to the grid mismatch between the actual angles and the quantized grids, this method enables the coarse angle search from the grids over a wide range of angle candidates.
    In contrast, the off-grid algorithm treats angles as continuous variables, enabling fine estimation without the quantization errors. 
    In the proposed off-grid algorithm, the angles are finely updated using a gradient descent algorithm, with initial estimates obtained from the on-grid algorithm.
    Since the angle estimates might converge to a local optimum near the initial values, this method is incapable of exploring a wide range of the angle domain.
    Particularly, if the off-grid method is applied in the early phase of iterations without precise estimates of the array errors, the objective function might not sufficiently decrease, and the angle estimates might converge to a local optimum.
    Therefore, we employ the on-grid method in the early phase of iterations to search for angle estimates over a wide range of angle candidates.
    In the later phase with accurate estimates of the array errors, the off-grid algorithm is introduced to finely estimate angles with the initial values obtained from the on-grid method.

    The switch from the on-grid algorithm to the off-grid algorithm can be performed automatically when the rate of change in the objective function between the $t$-th and $(t+1)$-th iterations $\Delta \mathcal{F}$ falls below a threshold value $\mathrm{Th}_1$ as $\Delta \mathcal{F} \leq \mathrm{Th}_1$ with
    $\Delta \mathcal{F} \triangleq \left | \mathcal{F}(\mathbf{\Omega}_{t+1}) - \mathcal{F}(\mathbf{\Omega}_t) \right | / \mathcal{F}(\mathbf{\Omega}_t)$, where $\mathbf{\Omega}_{t}$ denotes the estimated parameter set at the $t$-th iteration.

    \Kabuto{
    Focusing on the angles, delays, and path gains, the objective function in \eqref{eq:min} can be separated for each training frame index $m \in \{1,\ldots,M\}$ as $\mathcal{F} = \sum_{m=1}^M \mathcal{F}^{(m)} (\mathbf{v}_\mathrm{r}^{(m)}, \mathbf{v}_\mathrm{t}^{(m)}, \bm{\tau}^{(m)}, \bm{\alpha}^{(m)})$.}
    Hence, these parameters are estimated in parallel for each training frame $m$. 
    In what follows, the on-grid algorithm and the off-grid algorithm are detailed in Section \ref{subsec:on_grid} and \ref{subsec:off_grid}, respectively.

    \subsubsection{Coarse Estimation Based on On-Grid Algorithm}
    \label{subsec:on_grid}

    \Kabuto{
    Using the virtual channel representation in \eqref{eq:y_apx} with the dictionary matrix $\tilde{\mathbf{\Psi}}_k^{\bm{\alpha}}$ constructed by the angle and delay grids and the tentative estimates for array errors, 
    the objective function for the $m$-th training frame in \eqref{eq:min} can be approximated as
    $ \mathcal{F}^{(m)} (\tilde{\mathbf{v}}_\mathrm{r}^{(m)}, \tilde{\mathbf{v}}_\mathrm{t}^{(m)}, \tilde{\bm{\tau}}^{(m)}, \tilde{\bm{\alpha}}^{(m)}) \simeq \left \|  \mathbf{y}^{(m)} - \check{\mathbf{\Phi}} \tilde{\mathbf{\Psi}}_\mathcal{Q}^{\bm{\alpha}} \tilde{\bm{\alpha}}^{(m)} \right \|_2^2$.
    }

    \Kabuto{
    By exploiting this sparsity of $\tilde{\bm{\alpha}}^{(m)}$, the minimization problem can be reformulated as
    \begin{subequations}
    \label{eq:min_z_CS}
    \begin{align}
        \label{eq:min_z_CS_obj}
        & \underset{\tilde{\bm{\alpha}}^{(m)}}{\mathrm{minimize}} \ \ 
        \left \|  \mathbf{y}^{(m)} - \check{\mathbf{\Phi}} \tilde{\mathbf{\Psi}}_\mathcal{Q}^{\bm{\alpha}} \tilde{\bm{\alpha}}^{(m)} \right \|_2^2  \\
        \label{eq:min_z_CS_constraint}
        & \mathrm{subject \ to} \ \ \left \| \tilde{\bm{\alpha}}^{(m)} \right \|_0 = \hat{L}.
    \end{align}
    \end{subequations}
    }
    \Kabuto{
    In practice, the path gain vector $\tilde{\bm{\alpha}}^{(m)}$ is not an exact $L$-sparse vector due to energy leakage caused by the quantization errors in the grids.
    Thus, the number of paths used for the constraint in \eqref{eq:min_z_CS} is set to an overestimated value $(\hat{L} > L)$\footnote{\Kabuto{
    Note that the number of actual paths $L$ can be obtained from long-term statistics or site-specific measurements because the temporal variation in the number of paths is significantly slower than that of the channel~\cite{2022Cui_PSOMP, 2024Tang_three_layer_BP}.}}. 
    }
    \Kabuto{
    To solve the optimization problem in \eqref{eq:min_z_CS}, we employ the OMP algorithm~\cite{1993Pati_OMP}, which can account for the sparsity of $\tilde{\bm{\alpha}}^{(m)}$.
    By solving the problem \eqref{eq:min_z_CS}, the path gain, angle, and delay estimates, $\hat{\bm{\alpha}}^{(m)} \in \mathbb{C}^{\hat{L} \times 1}$, $\hat{\mathbf{v}}_\mathrm{r}^{(m)}$, $\hat{\mathbf{v}}_\mathrm{t}^{(m)}$, and $\hat{\bm{\tau}}^{(m)} \in \mathbb{R}^{\hat{L} \times 1}$, associated with the non-zero entries of $\tilde{\bm{\alpha}}^{(m)}$, are obtained.
    }

    \subsubsection{Fine Estimation Based on Off-Grid Algorithm}
    \label{subsec:off_grid}

    \Kabuto{
    To further refine the estimation accuracy and mitigate the errors resulting from grid mismatch, we introduce an off-grid algorithm utilizing a gradient descent method, similar to the conventional algorithms~\cite{2020Gonzalez_SIGW, 2022Cui_PSOMP, 2024Lei_Hybrid_CE}. 
    The proposed off-grid algorithm, unlike these conventional algorithms, considers both the beam squint effects and array errors.
    }

    \Kabuto{
    Since the objective function in \eqref{eq:min}, given any $\mathbf{v}_\mathrm{r}^{(m)}$, $\mathbf{v}_\mathrm{t}^{(m)}$, and $\bm{\tau}^{(m)}$, is convex with respect to $\bm{\alpha}^{(m)}$, the path gain estimates can be derived in a closed-form expression as a function of $\mathbf{v}_\mathrm{r}^{(m)}$, $\mathbf{v}_\mathrm{t}^{(m)}$, and $\bm{\tau}^{(m)}$.
    Using the closed-form solution of $\bm{\alpha}^{(m)}$, the objective function is reformulated as a function of $\mathbf{v}_\mathrm{r}^{(m)}$, $\mathbf{v}_\mathrm{t}^{(m)}$, and $\bm{\tau}^{(m)}$.
    Then, $\mathbf{v}_\mathrm{r}^{(m)}$, $\mathbf{v}_\mathrm{t}^{(m)}$, and $\bm{\tau}^{(m)}$ are updated by a gradient descent manner.
    }

    \Kabuto{
    Let
    $\mathbf{t}^{z (m)}_{p,k} \triangleq \mathbf{A}_{\mathrm{t},k}^{\mathrm{H}} (\hat{\mathbf{v}}_\mathrm{t}^{(m)}, \hat{\bm{\varepsilon}}_\mathrm{t}) \hat{\mathbf{\Gamma}}_\mathrm{t}^\mathrm{H} \hat{\mathbf{C}}_\mathrm{t}^\mathrm{H} \mathbf{F}_p \mathbf{q}_p \in \mathbb{C}^{\hat{L} \times 1}$, 
    $\mathbf{T}_{p,k}^{z (m)} \triangleq \mathbf{W}_p \hat{\mathbf{C}}_\mathrm{r} \hat{\mathbf{\Gamma}}_\mathrm{r} \mathbf{A}_{\mathrm{r},k} (\hat{\mathbf{v}}_\mathrm{r}^{(m)}, \hat{\bm{\varepsilon}}_\mathrm{r}) \mathrm{diag} (\mathbf{t}_{p,k}^{z (m)}) \in \mathbb{C}^{N_\mathrm{r,RF} \times \hat{L}}$, and 
    $\mathbf{T}_{p,k}^{\alpha (m)} \triangleq \mathbf{T}_{p,k}^{z (m)} \mathrm{diag}(\mathbf{b}_k(\hat{\bm{\tau}}^{(m)}) ) \in \mathbb{C}^{N_\mathrm{r,RF} \times \hat{L}} $ be defined.
    Then, the minimization problem for $\bm{\alpha}^{(m)}$ is formulated as
    \begin{align}
        \label{eq:obj_alpha}
        \underset{\bm{\alpha}^{(m)}}{\mathrm{minimize}} \ \  \sum_{p=1}^{N_\mathrm{p}} \sum_{k \in \mathcal{Q}} \left \| \mathbf{y}_{p,k}^{(m)} - \mathbf{T}_{p,k}^{\alpha (m)} \bm{\alpha}^{(m)}
        \right \|_2^2.
    \end{align}
    }
    \Kabuto{
    The solution for $\bm{\alpha}^{(m)}$ is derived in a closed-form as
    \begin{align}
        \label{eq:alpha_opt}
        \hat{\bm{\alpha}}^{(m)} = \left( \mathbf{T}^{\alpha (m)} \right )^\dagger \mathbf{y}^{(m)},
    \end{align}
    where 
    $\mathbf{T}^{\alpha (m)} \triangleq 
    \begin{bmatrix} 
        \mathbf{T}_{k_1}^{\alpha (m) \mathrm{T}}, \ldots, \mathbf{T}_{k_Q}^{\alpha (m) \mathrm{T}} 
    \end{bmatrix}^\mathrm{T} \in \mathbb{C}^{N_\mathrm{r,RF} N_\mathrm{p} Q \times \hat{L}} $ 
    with
    $\mathbf{T}_{k}^{\alpha (m)} = \mathbf{T}_k^{z(m)} \mathrm{diag}(\mathbf{b}_k (\hat{\bm{\tau}})) \in \mathbb{C}^{N_\mathrm{r,RF} N_\mathrm{p} \times \hat{L}}$
    and 
    $\mathbf{T}_{k}^{z (m)} \triangleq 
    \begin{bmatrix} 
        \mathbf{T}_{1,k}^{z (m) \mathrm{T}}, \ldots, \mathbf{T}_{N_\mathrm{p},k}^{z (m) \mathrm{T}} 
    \end{bmatrix}^\mathrm{T} \in \mathbb{C}^{N_\mathrm{r,RF} N_\mathrm{p} \times \hat{L}},\ k \in \mathcal{Q}$.
    }

    \Kabuto{
    From \eqref{eq:zk}, the equivalent path gain estimate in the frequency-domain $\hat{\mathbf{z}}_k^{(m)} \in \mathbb{C}^{\hat{L} \times 1}$ can be expressed as 
    \begin{align}
        \label{eq:z_opt}
        \hat{\mathbf{z}}_k^{(m)} =  \mathrm{diag} (\mathbf{b}_k( \hat{\bm{\tau}}^{(m)})) \hat{\bm{\alpha}}^{(m)}.
    \end{align}
    }
    \Kabuto{
    Substituting \eqref{eq:alpha_opt} into \eqref{eq:obj_alpha}, the objective function for $\mathbf{v}_\mathrm{r}^{(m)}$, $\mathbf{v}_\mathrm{t}^{(m)}$, and $\bm{\tau}^{(m)}$ can be reformulated as
    \begin{align*}
        \mathcal{F}^{(m)}(\mathbf{v}_\mathrm{r}^{(m)},\mathbf{v}_\mathrm{t}^{(m)}, \bm{\tau}^{(m)}) = \left \|  
        \mathbf{y}^{(m)} - \mathbf{T}^{\alpha (m)} \left ( \mathbf{T}^{\alpha (m)} \right )^\dagger \mathbf{y}^{(m)}
        \right \|_2^2.
    \end{align*}
    }
    %
    \Kabuto{
    Using the gradients
    $\frac{\partial \mathcal{F}^{(m)} }{\partial \mathbf{v}_\mathrm{r}^{(m)}}$, $\frac{\partial \mathcal{F}^{(m)} }{\partial \mathbf{v}_\mathrm{t}^{(m)}}$, 
    $\frac{\partial \mathcal{F}^{(m)} }{\partial \bm{\tau}^{(m)}}$, 
    these parameters are updated by the gradient descent approach, where learning rates, determined by a backtracking line search to ensure that the objective function is non-increasing at every step~\cite{2004Boyd_ConvexOpt}.
    }
    %
    \Kabuto{
    The gradients of the angles at the BS side are calculated as 
    \begin{subequations}
    \label{eq:grad_v_r}
    \begin{align}
        \label{eq:grad_v_rx}
        \frac{\partial \mathcal{F}^{(m)}}{\partial v_{\mathrm{r},l}^{x(m)}}
        & \!\! = \! 2 \!\!\sum_{k \in \mathcal{Q}} \mathfrak{R} \left [ 
        \mathbf{d}_{\mathrm{r},k,l}^{x (m) \mathrm{H}} \mathbf{T}_k^{z(m)} \hat{\mathbf{z}}_k^{(m)} \! - \! \mathbf{y}_k^{(m) \mathrm{H}} \mathbf{d}_{\mathrm{r},k,l}^{x (m)}
        \right ], \\
        \label{eq:grad_v_ry}
        \frac{\partial \mathcal{F}^{(m)}}{\partial v_{\mathrm{r},l}^{y(m)}}
        & \!\! = \! 2 \!\! \sum_{k \in \mathcal{Q}} \mathfrak{R} \left [ 
        \mathbf{d}_{\mathrm{r},k,l}^{y (m) \mathrm{H}} \mathbf{T}_k^{z(m)} \hat{\mathbf{z}}_k^{(m)} \!-\! \mathbf{y}_k^{(m) \mathrm{H}} \mathbf{d}_{\mathrm{r},k,l}^{y (m)}
        \right ], 
    \end{align}
    \end{subequations}
    where $\mathbf{d}_{\mathrm{r},l,k}^{x (m)} \in \mathbb{C}^{N_\mathrm{r,RF} N_\mathrm{p}\times 1}$ and $\mathbf{d}_{\mathrm{r},l,k}^{y (m)} \in \mathbb{C}^{N_\mathrm{r,RF} N_\mathrm{p}\times 1}$ are calculated as \eqref{eq:d_rx} and \eqref{eq:d_ry} at the top of the next page, with
    \begin{align*}
        \mathbf{f}_{\mathrm{r},l,k}^{x (m)} &= j \frac{2 \pi}{\lambda_\mathrm{c}} \left( 1 + \frac{\Delta f_k}{f_\mathrm{c}} \right) \left \{ 
            (\mathbf{p}_\mathrm{r}^x + \hat{\bm{\varepsilon}}^x_\mathrm{r}) \odot \mathbf{a}^x_{\mathrm{r},k}(\hat{v}_{\mathrm{r},l}^{x(m)}, \hat{\bm{\varepsilon}}^x_\mathrm{r})
        \right \}, \\
        \mathbf{f}_{\mathrm{r},l,k}^{y (m)} &= j \frac{2 \pi}{\lambda_\mathrm{c}} \left( 1 + \frac{\Delta f_k}{f_\mathrm{c}} \right) \left \{ 
            (\mathbf{p}_\mathrm{r}^y + \hat{\bm{\varepsilon}}^y_\mathrm{r}) \odot \mathbf{a}^y_{\mathrm{r},k}(\hat{v}_{\mathrm{r},l}^{y(m)}, \hat{\bm{\varepsilon}}^y_\mathrm{r})
        \right \}, 
    \end{align*}
    where
    $\mathbf{p}^x_\mathrm{r} \in \mathbb{R}^{N_\mathrm{r}^x \times 1}$ and 
    $\mathbf{p}^y_\mathrm{r} \in \mathbb{R}^{N_\mathrm{r}^y \times 1}$
    are the ideal antenna position vectors defined as 
    $\mathbf{p}^x_\mathrm{r} = [0, d_\mathrm{r}^x/2, \ldots, (N_\mathrm{r}^x-1)d^x_\mathrm{r}/2]^\mathrm{T}$ and
    $\mathbf{p}^y_\mathrm{r} = [0, d_\mathrm{r}^y/2, \ldots, (N_\mathrm{r}^y-1)d^y_\mathrm{r}/2]^\mathrm{T}$.
    }
    %
    \begin{figure*}
        \begin{subequations}
        \begin{align}
            \label{eq:d_rx}
            \Kabuto{
            \mathbf{d}_{\mathrm{r},l,k}^{x (m)}} & \Kabuto{= 
            \begin{bmatrix}
                \left[ \mathbf{t}_{1,k}^{z (m)} \right]_{l} \mathbf{W}_1^\mathrm{T}, \ldots, 
                \left[ \mathbf{t}_{N_\mathrm{p},k}^{z (m)} \right]_{l} \mathbf{W}_{N_\mathrm{p}}^\mathrm{T}
            \end{bmatrix}^\mathrm{T} 
            \hat{\mathbf{C}}_\mathrm{r} \hat{\mathbf{\Gamma}}_\mathrm{r}
            \left( \mathbf{a}_{\mathrm{r},k}^y (\hat{v}_{\mathrm{r},l}^{y (m)}, \hat{\bm{\varepsilon}}_\mathrm{r}^y )
            \otimes \mathbf{f}_{\mathrm{r},l,k}^{x (m)}  \right) \hat{z}_{l,k}^{(m)} 
            \in \mathbb{C}^{N_{\mathrm{r,RF}} N_\mathrm{p} \times 1},}
            \\
            \label{eq:d_ry}
            \Kabuto{
            \mathbf{d}_{\mathrm{r},l,k}^{y (m)} } & \Kabuto{= 
            \begin{bmatrix}
                \left[ \mathbf{t}_{1,k}^{z (m)} \right]_{l} \mathbf{W}_1^\mathrm{T}, \ldots, 
                \left[ \mathbf{t}_{N_\mathrm{p},k}^{z (m)} \right]_{l} \mathbf{W}_{N_\mathrm{p}}^\mathrm{T}
            \end{bmatrix}^\mathrm{T} 
            \hat{\mathbf{C}}_\mathrm{r} \hat{\mathbf{\Gamma}}_\mathrm{r}
            \left(  
            \mathbf{f}_{\mathrm{r},l,k}^{y (m)}
            \otimes \mathbf{a}_{\mathrm{r},k}^x (\hat{v}_{\mathrm{r},l}^{x (m)}, \hat{\bm{\varepsilon}}_\mathrm{r}^x)   \right) \hat{z}_{l,k}^{(m)} 
            \in \mathbb{C}^{N_{\mathrm{r,RF}} N_\mathrm{p} \times 1},}
        \end{align}
        \end{subequations}
        %
        \vspace*{-30pt}
    \end{figure*}

    \Kabuto{
    Similarly, the gradients at the UE side are calculated as 
    \begin{subequations}
    \label{eq:grad_v_t}
    \begin{align}
        \label{eq:grad_v_tx}
        \frac{\partial \mathcal{F}^{(m)}}{\partial v_{\mathrm{t},l}^{x(m)}}
        & \!\! =\! 2 \!\! \sum_{k \in \mathcal{Q}} \mathfrak{R} \left [ 
        \mathbf{d}_{\mathrm{t},k,l}^{x (m) \mathrm{H}} \mathbf{T}_k^{z(m)} \hat{\mathbf{z}}_k^{(m)} \!- \! \mathbf{y}_k^{(m) \mathrm{H}} \mathbf{d}_{\mathrm{t},k,l}^{x (m)}
        \right ], \\
        \label{eq:grad_v_ty}
        \frac{\partial \mathcal{F}^{(m)}}{\partial v_{\mathrm{t},l}^{y(m)}}
        & \!\! =\! 2 \!\! \sum_{k \in \mathcal{Q}} \mathfrak{R} \left [ 
        \mathbf{d}_{\mathrm{t},k,l}^{y (m) \mathrm{H}} \mathbf{T}_k^{z(m)} \hat{\mathbf{z}}_k^{(m)} \! - \! \mathbf{y}_k^{(m) \mathrm{H}} \mathbf{d}_{\mathrm{t},k,l}^{y (m)}
        \right ], 
    \end{align}
    \end{subequations}
    where $\mathbf{d}_{\mathrm{t},l,k}^{x (m)} \in \mathbb{C}^{N_\mathrm{r,RF} N_\mathrm{p}\times 1}$ and $\mathbf{d}_{\mathrm{t},l,k}^{y (m)} \in \mathbb{C}^{N_\mathrm{r,RF} N_\mathrm{p}\times 1}$ are calculated as \eqref{eq:d_tx} and \eqref{eq:d_ty} at the top of the next page, with
    $c_{p,k,l}^{x(m)} = \left ( \mathbf{a}_{\mathrm{t},k}^y (\hat{\mathbf{v}}_{\mathrm{t},l}^{y (m)}, \hat{\bm{\varepsilon}}^y_\mathrm{t}) \otimes \mathbf{f}_{\mathrm{t},l,k}^{x(m)} \right )^\mathrm{H} \hat{\mathbf{\Gamma}}_\mathrm{t} \mathbf{F}_p \mathbf{q}_p$ and 
    $c_{p,k,l}^{y(m)} = \left ( \mathbf{f}_{\mathrm{t},l,k}^{y(m)} \otimes \mathbf{a}_{\mathrm{t},k}^x (\hat{\mathbf{v}}_{\mathrm{t},l}^{x(m)}, \hat{\bm{\varepsilon}}^x_\mathrm{t})  \right )^\mathrm{H} \hat{\mathbf{\Gamma}}_\mathrm{t} \mathbf{F}_p \mathbf{q}_p$, where $\mathbf{f}_{\mathrm{t},l,k}^{x (m)} \in \mathbb{C}^{N^x_\mathrm{t} \times 1}$ and $\mathbf{f}_{\mathrm{t},l,k}^{y (m)}  \in \mathbb{C}^{N^y_\mathrm{t} \times 1}$ are given by
    %
    \begin{align*}
        \mathbf{f}_{\mathrm{t},l,k}^{x (m)} &= j \frac{2 \pi}{\lambda_\mathrm{c}} \left( 1 + \frac{\Delta f_k}{f_\mathrm{c}} \right) \left \{ 
            (\mathbf{p}_\mathrm{t}^x + \hat{\bm{\varepsilon}}^x_\mathrm{t}) \odot \mathbf{a}^x_{\mathrm{t},k}(\hat{v}_{\mathrm{t},l}^{x(m)}, \hat{\bm{\varepsilon}}^x_\mathrm{t})
        \right \}, \\
        \mathbf{f}_{\mathrm{t},l,k}^{y (m)} &= j \frac{2 \pi}{\lambda_\mathrm{c}} \left( 1 + \frac{\Delta f_k}{f_\mathrm{c}} \right) \left \{ 
            (\mathbf{p}_\mathrm{t}^y + \hat{\bm{\varepsilon}}^y_\mathrm{t}) \odot \mathbf{a}^y_{\mathrm{r},k}(\hat{v}_{\mathrm{t},l}^{y(m)}, \hat{\bm{\varepsilon}}^y_\mathrm{t})
        \right \}, 
    \end{align*}
    %
    and
    $\mathbf{p}^x_\mathrm{t} \in \mathbb{R}^{N_\mathrm{t}^x \times 1}$ and 
    $\mathbf{p}^y_\mathrm{t} \in \mathbb{R}^{N_\mathrm{t}^y \times 1}$
    are the ideal antenna position vectors defined as 
    $\mathbf{p}^x_\mathrm{t} = [0, d_\mathrm{t}^x/2, \ldots, (N_\mathrm{t}^x-1)d^x_\mathrm{t}/2]^\mathrm{T}$ and
    $\mathbf{p}^y_\mathrm{t} = [0, d_\mathrm{t}^y/2, \ldots, (N_\mathrm{t}^y-1)d^y_\mathrm{t}/2]^\mathrm{T}$.
    }

    \begin{figure*}
        \begin{subequations}
        \begin{align}
            \label{eq:d_tx}
            \Kabuto{
            \mathbf{d}_{\mathrm{t},l,k}^{x (m)}} & \Kabuto{= 
            \begin{bmatrix}
                c_{1,k,l}^{x(m)} \mathbf{W}_{1}^\mathrm{T}
                , \ldots, 
                c_{N_\mathrm{p},k,l}^{x(m)} \mathbf{W}_{N_\mathrm{p}}^\mathrm{T}
            \end{bmatrix}^\mathrm{T}
            \hat{\mathbf{C}}_\mathrm{r} \hat{\mathbf{\Gamma}}_\mathrm{r} \mathbf{a}_{\mathrm{r},k} (\hat{\mathbf{v}}_{\mathrm{r},l}^{(m)}, \hat{\bm{\varepsilon}}_\mathrm{r}) \hat{z}_{k,l}^{(m)} \in \mathbb{C}^{N_{\mathrm{r,RF}} N_\mathrm{p} \times 1},
            } \\
            \label{eq:d_ty}
            \Kabuto{
            \mathbf{d}_{\mathrm{t},l,k}^{y (m)} } & \Kabuto{ = 
            \begin{bmatrix}
                c_{1,k,l}^{y(m)} \mathbf{W}_{1}^\mathrm{T}
                , \ldots, 
                c_{N_\mathrm{p},k,l}^{y(m)} \mathbf{W}_{N_\mathrm{p}}^\mathrm{T}
            \end{bmatrix}^\mathrm{T}
            \hat{\mathbf{C}}_\mathrm{r} \hat{\mathbf{\Gamma}}_\mathrm{r} \mathbf{a}_{\mathrm{r},k} (\hat{\mathbf{v}}_{\mathrm{r},l}^{(m)}, \hat{\bm{\varepsilon}}_\mathrm{r}) \hat{z}_{k,l}^{(m)} \in \mathbb{C}^{N_{\mathrm{r,RF}} N_\mathrm{p} \times 1},
            }
        \end{align}
        \end{subequations}
        \vspace*{-30pt}
    \end{figure*}

    \Kabuto{
    The gradient of the delay can be calculated as
    \begin{align}
        \label{eq:grad_tau}
        \frac{\partial \mathcal{F}^{(m)} }{\partial \tau_l^{(m)}} = 
        2\mathfrak{R} \left [ 
            \mathbf{d}_l^{\alpha(m) \mathrm{H}} \mathbf{T}^{\alpha (m)} \hat{\bm{\alpha}}^{(m)} - \mathbf{y}^{(m) \mathrm{H}} \mathbf{d}_l^{\alpha(m)}
        \right],
    \end{align}
    where $\mathbf{d}_l^{\alpha(m)}$ is given by \eqref{eq:dl_alpha} at the top of the next page.
    %
    \begin{figure*}
        \begin{align}
            \label{eq:dl_alpha}
            \Kabuto{
            \mathbf{d}_l^{\alpha(m)} = -j 2\pi \hat{\alpha}_l^{(m)} 
            \begin{bmatrix}
                \Delta f_{k_1} e^{-j 2 \pi \Delta f_{k_1} \hat{\tau}_{l}^{(m)}} \left[ \mathbf{T}_{k_1}^{z (m)} \right]_{:, l}^\mathrm{T} ,\  \ldots,\ 
                \Delta f_{k_Q} e^{-j 2 \pi \Delta f_{k_Q} \hat{\tau}_{l}^{(m)}} \left[ \mathbf{T}_{k_Q}^{z (m)} \right]_{:, l}^\mathrm{T}
            \end{bmatrix}^\mathrm{T}  \in \mathbb{C}^{N_\mathrm{r,RF} N_\mathrm{p} Q\times 1}.
            }
        \end{align}
        \hrulefill
        \vspace*{-12pt}
    \end{figure*}
    }

    \vspace{-0.3cm}
    \subsection{\Kabuto{Approximation of Mutual Coupling Matrix}}
    \label{subsec:Approx_c}

    \Kabuto{
    To effectively estimate the mutual coupling matrices $\mathbf{C}_\mathrm{r}$ and $\mathbf{C}_\mathrm{t}$, we introduce an approximate coupling model that reduces the number of parameters to be estimated.
    This subsection focuses only on the approximation model for $\mathbf{C}_\mathrm{r}$, since the procedures for $\mathbf{C}_\mathrm{r}$ and $\mathbf{C}_\mathrm{t}$ are the same.
    }
    
    \Kabuto{
    In the case of UPA, 
    $\mathbf{C}_\mathrm{r} \in \mathbb{C}^{N_\mathrm{r} \times N_\mathrm{r}}$ is modeled by 
    \begin{align}
        \label{eq:Cr}
        \mathbf{C}_\mathrm{r} =
        \begin{bmatrix}
            \mathbf{C}_{\mathrm{r}, 0,0} & \cdots & \mathbf{C}_{\mathrm{r}, 0,N_\mathrm{r}^y-1} \\
            \vdots & \ddots & \vdots \\
            \mathbf{C}_{\mathrm{r}, N_\mathrm{r}^y-1,0} & \cdots & \mathbf{C}_{\mathrm{r}, N_\mathrm{r}^y-1,N_\mathrm{r}^y-1} \\
        \end{bmatrix},
    \end{align}
    where the $(j,j^\prime)$-th block $\mathbf{C}_{\mathrm{r},j,j^\prime} \in \mathbb{C}^{N^x_\mathrm{r} \times N^x_\mathrm{r}}$ is modeled by 
    \begin{align}
        \label{eq:Cr_jj}
        \mathbf{C}_{\mathrm{r},j,j^\prime} =
        \begin{bmatrix}
            c_{\mathrm{r}, 0,0}^{(j,j^\prime)} & \cdots & c_{\mathrm{r}, 0,N^x_\mathrm{r}-1}^{(j,j^\prime)} \\
            \vdots & \ddots & \vdots \\
            c_{\mathrm{r}, N^x_\mathrm{r}-1, 0}^{(j,j^\prime)} & \cdots & c_{\mathrm{r}, N^x_\mathrm{r}-1, N^x_\mathrm{r}-1}^{(j,j^\prime)} \\
        \end{bmatrix}.
    \end{align}
    }
    \Kabuto{
    The $(i,i^\prime)$-th element of the $(j,j^\prime)$-th block, $c_{\mathrm{r}, i,i^\prime}^{(j,j^\prime)}$, represents the coefficient of mutual coupling between the $(i,j)$-th antenna and the $(i^\prime,j^\prime)$ antenna in the $x$-$y$ plane, where 
    $0 \leq i,i^\prime \leq N_\mathrm{r}^x-1$ and 
    $0 \leq j,j^\prime \leq N_\mathrm{r}^y-1$.
    When $j=j^\prime$ and $i=i^\prime=0$, the coefficient $c_{0,0}^{(j,j)}$ is 1, indicating that the diagonal elements of $\mathbf{C}_\mathrm{r}$ in \eqref{eq:Cr} are equal to 1.
    }

    As described in Section~\ref{subsec:channel_model}, in the case of UPA, the mutual coupling matrix $\mathbf{C}_\mathrm{r}$ has \Kabuto{a symmetric block Toeplitz structure~\cite{2008Ye_2D_Copuling}.}
    This property reduces the number of parameters to be estimated to \Kabuto{$N_\mathrm{r}-N^x_\mathrm{r}$}, leading to more efficient estimation for $\mathbf{C}_\mathrm{r}$.
    However, antenna spacing errors $\bm{\varepsilon}_\mathrm{r}$ distort the symmetric Toeplitz structure, resulting in a deterioration of estimation performance when utilizing the Toeplitz structure.
    When $\mathbf{C}_\mathrm{r}$ is exactly modeled without assuming the Toeplitz structure as written in \eqref{eq:Cr}, the number of parameters to be estimated rises to $N_\mathrm{r} (N_\mathrm{r} - 1) / 2$ because $\mathbf{C}_\mathrm{r}$ is a $N_\mathrm{r} \times N_\mathrm{r}$ symmetric matrix.
    The increase in parameters leads to overfitting to noisy observations and higher computational complexity.

    \Kabuto{
    Therefore, to balance the accuracy of modeling and the number of parameters, the proposed method approximately decomposes the mutual coupling matrix $\mathbf{C}_\mathrm{r}$ into a Toeplitz part $\mathbf{C}_\mathrm{r}^\mathrm{TP} \in \mathbb{C}^{N_\mathrm{r} \times N_\mathrm{r}}$ and a non-Toeplitz part $\mathbf{C}_\mathrm{r}^\mathrm{NTP} \in \mathbb{C}^{N_\mathrm{r} \times N_\mathrm{r}}$.
    Since the effects of mutual coupling monotonically decrease with distance~\cite{2021Fikes_wearable_coupling, 2016Liu_couoling_model}, the coupling coefficients of \Kabuto{the first $q^x_\mathrm{r}$-antenna and $q^y_\mathrm{r}$-antenna elements in the $x \text{-} y$ plane from the reference antenna}, which have significant impact on performance, are modeled using an exact coupling model with a non-Toeplitz structure, while the other coupling coefficients are modeled using a Toeplitz structure.
    }
    Then, the mutual coupling matrix $\mathbf{C}_\mathrm{r}$ is approximately decomposed as 
    \begin{align}
        \label{eq:C_TP_NTP}
        \mathbf{C}_\mathrm{r} \simeq \mathbf{C}_\mathrm{r}^{\mathrm{TP}} + \mathbf{C}_\mathrm{r}^{\mathrm{NTP}},
    \end{align}
    \Kabuto{
    whose $(j,j^\prime)$-th block $\mathbf{C}_{\mathrm{r},j,j^\prime}$ in \eqref{eq:Cr_jj} is approximated by 
    \begin{align}
        \label{eq:C_TP_NTP_jj}
        \mathbf{C}_{\mathrm{r},j,j^\prime} \simeq \mathbf{C}^\mathrm{TP}_{\mathrm{r}, |j - j^\prime|} + \mathbf{C}^\mathrm{NTP}_{\mathrm{r}, j, j^\prime},
    \end{align}
    where the Toeplitz part at the $(j,j^\prime)$-th block, $\mathbf{C}^\mathrm{TP}_{\mathrm{r}, |j - j^\prime|}$, depends only on the absolute index difference $|j-j^\prime|$, which reduces the number of parameters.
    }
    \Kabuto{
    The Toeplitz part $\mathbf{C}^\mathrm{TP}_{\mathrm{r}, |j - j^\prime|}$ and the non-Toeplitz part $\mathbf{C}^\mathrm{NTP}_{\mathrm{r}, j, j^\prime}$ in \eqref{eq:C_TP_NTP_jj} 
     are given by the following equations in \eqref{eq:C_TP_1}-\eqref{eq:C_TP_2} and 
     \eqref{eq:C_NTP_1}-\eqref{eq:C_NTP_3}, respectively.
    Note that $\mathbf{C}^\mathrm{TP}_{\mathrm{r}, |j - j^\prime|}$ and $\mathbf{C}^\mathrm{NTP}_{\mathrm{r}, j, j^\prime}$ are symmetric matrices.
    }

    \Kabuto{
    \textit{{Toeplitz part $\mathbf{C}^\mathrm{TP}_{\mathrm{r}, |j - j^\prime|}$}}:
    \begin{itemize}
        \item 
        For $|j-j^\prime| \leq q_\mathrm{r}^y$, the $(i,i^\prime)$-th element of $\mathbf{C}^\mathrm{TP}_{\mathrm{r}, |j - j^\prime|}$ is
        \begin{align}
            \label{eq:C_TP_1}
            &\left [ \mathbf{C}_{\mathrm{r}, |j-j^\prime|}^\mathrm{TP} \right ]_{i,i^\prime} =
            \begin{cases}
            0 & (i < i^\prime \leq i+q^x_\mathrm{r}) \\
            c_{\mathrm{r},|i-i^\prime|}^{|j-j^\prime|} & (i+q^x_\mathrm{r} < i^\prime \leq N_\mathrm{r}^x-1)
            \end{cases}.
        \end{align}
        \item 
        For $|j-j^\prime| > q_\mathrm{r}^y$, the $(i,i^\prime)$-th element of $\mathbf{C}^\mathrm{TP}_{\mathrm{r}, |j - j^\prime|}$ is 
        \begin{align}
            \label{eq:C_TP_2}
            &\left [ \mathbf{C}_{\mathrm{r}, |j-j^\prime|}^\mathrm{TP} \right ]_{i,i^\prime} =
            c_{\mathrm{r},|i-i^\prime|}^{|j-j^\prime|} & (0 < i,i^\prime \leq N_\mathrm{r}^x-1).
        \end{align}
    \end{itemize}
    }

    \Kabuto{
    From \eqref{eq:C_TP_1} and \eqref{eq:C_TP_2}, the Toeplitz part at the 
    $(j,j^\prime)$-th block $\mathbf{C}^\mathrm{TP}_{\mathrm{r}, |j - j^\prime|}$ includes 
    unknown parameters $\mathbf{u}_{|j-j^\prime|}^\mathrm{TP}$, given by
    \begin{align}
        \label{eq:ur_TP}
        \hspace{-0.2cm} \mathbf{u}^\mathrm{TP}_{\mathrm{r}, |j-j^\prime|} \! = \! 
        \begin{cases}
            \left[ c_{\mathrm{r}, q_\mathrm{r}^x+1}^{|j-j^\prime|}, \ldots, c_{\mathrm{r}, N^x_\mathrm{r}-1}^{|j-j^\prime|} \right ]^\mathrm{T} 
            & (|j-j^\prime| \leq q^y_\mathrm{r}) \\
            \left[ c_{\mathrm{r}, 0}^{|j-j^\prime|}, \ldots, c_{\mathrm{r}, N^x_\mathrm{r}-1}^{|j-j^\prime|} \right ]^\mathrm{T} & (|j-j^\prime| > q^y_\mathrm{r})
        \end{cases}.
    \end{align}
    }

    \Kabuto{
    Stacking the vectors $\mathbf{u}^\mathrm{TP}_{\mathrm{r}, |j-j^\prime|}$ in \eqref{eq:ur_TP}, the unknown parameters in the Toeplitz part $\mathbf{C}_\mathrm{r}^{\mathrm{TP}}$ in \eqref{eq:C_TP_NTP} is obtained as 
    \begin{align}
        \label{eq:u_TP}
        \mathbf{u}^\mathrm{TP}_\mathrm{r} = \begin{bmatrix}
            (\mathbf{u}^\mathrm{TP}_{\mathrm{r}, 0})^\mathrm{T}, \ldots, (\mathbf{u}^\mathrm{TP}_{\mathrm{r},N^y_{\mathrm{r}}-1})^\mathrm{T}
        \end{bmatrix}^\mathrm{T} \in \mathbb{C}^{Q_\mathrm{r}^\mathrm{TP} \times 1},
    \end{align}
    where $Q_\mathrm{r}^\mathrm{TP} = N^x_\mathrm{r} (N^y_\mathrm{r} - q^y_\mathrm{r} - 1) + (q^y_\mathrm{r} + 1) (N^x_\mathrm{r} - q^x_\mathrm{r} - 1)$ is the number of unknown parameters in the Toeplitz part.
    }

    \Kabuto{
    \textit{{Non-Toeplitz part $\mathbf{C}^\mathrm{NTP}_{\mathrm{r}, j, j^\prime}$}}:
    \begin{itemize}
        \item 
        For $|j-j^\prime| = 0$, 
        the $(i,i^\prime)$-th element of $\mathbf{C}^\mathrm{NTP}_{\mathrm{r}, j,j^\prime}$ is
        \begin{align}
        \label{eq:C_NTP_1}
        &\left [ \mathbf{C}_{\mathrm{r}, j,j^\prime}^\mathrm{NTP} \right ]_{i,i^\prime} \! =
        \begin{cases}
        1 & (i = i^\prime) \\
        c_{i,i^\prime}^{(j,j^\prime)} & (i < i^\prime \leq i+q_\mathrm{r}^x) \\
        0 & (i+q^x_\mathrm{r} < i^\prime \leq N_\mathrm{r}^x-1)
        \end{cases}.
        \end{align}
        \item 
        For $1 \leq |j-j^\prime| \leq q^y_\mathrm{r}$, 
        the $(i,i^\prime)$-th element of $\mathbf{C}^\mathrm{NTP}_{\mathrm{r}, j,j^\prime}$ is
        \begin{align}
        \label{eq:C_NTP_2}
        &\left [ \mathbf{C}_{\mathrm{r}, j,j^\prime}^\mathrm{NTP} \right ]_{i,i^\prime} \! =
        \begin{cases}
        c_{i,i^\prime}^{(j,j^\prime)} & (i < i^\prime \leq i+q_\mathrm{r}^x) \\
        0 & (i+q^x_\mathrm{r} < i^\prime \leq N_\mathrm{r}^x-1)
        \end{cases}.
        \end{align}
        \item 
        For $|j-j^\prime| > q^y_\mathrm{r}$, 
        $\mathbf{C}^\mathrm{NTP}_{\mathrm{r}, j,j^\prime}$ is expressed as
        \begin{align}
        \label{eq:C_NTP_3}
        \mathbf{C}_{\mathrm{r}, j,j^\prime}^\mathrm{NTP} = \mathbf{0}_{N^x_\mathrm{r} \times N^x_\mathrm{r}}.
        \end{align}
    \end{itemize}
    }

    \Kabuto{
    From \eqref{eq:C_NTP_1}-\eqref{eq:C_NTP_3}, the non-Toeplitz part at the $(j,j^\prime)$-th block $\mathbf{C}^\mathrm{NTP}_{\mathrm{r}, j,j^\prime}$ includes 
    unknown parameters, when $|j-j^\prime| \leq q^y_\mathrm{r}$, as 
    \begin{align}
        \label{eq:u_NTP_jj}
        \mathbf{u}_{\mathrm{r}, j,j^\prime}^{\mathrm{NTP}} = \begin{bmatrix}
            \left( \mathbf{u}_{\mathrm{r},0}^{(j,j^\prime)} \right )^\mathrm{T}, \ldots, \left( \mathbf{u}_{\mathrm{r},N^x_\mathrm{r}-1}^{(j,j^\prime)} \right )^\mathrm{T}
        \end{bmatrix}^\mathrm{T} \in \mathbb{C}^{ (Q_\mathrm{r}^x - N^x_\mathrm{r} \delta_{j,j^\prime}) \times 1},
    \end{align}
    where $Q_\mathrm{r}^x = N_\mathrm{r}^x + q^x_\mathrm{r} (N^x_\mathrm{r} -1) - q^x_\mathrm{r}(q^x_\mathrm{r} - 1)/2$ is the number of unknown parameters, $\delta_{j,j^\prime}$ is the Kronecker delta, and $\mathbf{u}_{\mathrm{r},i}^{(j,j^\prime)},\ i \in \{0, 1,\ldots,N^x_\mathrm{r} -1 \}$ is given by 
    \begin{align}
        \label{eq:ur}
        \mathbf{u}_{\mathrm{r}, i}^{(j,j^\prime)} \! = \!
        \begin{cases}
            [\emptyset] & (i = 0 \text{ and } j = j) \\
            \begin{bmatrix}
                c^{(j,j^\prime)}_{0,0}, \ldots, c^{(j,j^\prime)}_{N^x_\mathrm{r}-1, N^x_\mathrm{r}-1}
            \end{bmatrix}^\mathrm{T} & (i = 0 \text{ and } j \neq j) \\
            \begin{bmatrix}
                c^{(j,j^\prime)}_{i-1,i}, \ldots, c^{(j,j^\prime)}_{i-1, i + q^x_\mathrm{r}-1}
            \end{bmatrix}^\mathrm{T} &
            (1 \le i \leq N^x_\mathrm{r} - q^x_\mathrm{r}) \\
            \begin{bmatrix}
                c^{(j,j^\prime)}_{i-1,i}, \ldots, c^{(j,j^\prime)}_{i-1, N^x_\mathrm{r}-1}
            \end{bmatrix}^\mathrm{T} &  \hspace{-5ex} (N^x_\mathrm{r} - q^x_\mathrm{r} < i < N^x_\mathrm{r} - 1)
        \end{cases}.
    \end{align}
    }

    \Kabuto{
    To collect the unknown parameters of the non-Toeplitz part in \eqref{eq:u_NTP_jj}, we define the vector $\mathbf{u}_{\mathrm{r}, j}^\mathrm{NTP}$, expressed as \eqref{eq:u_NTP_j} at the top of the next page.
    }
    %
    \begin{figure*}
    \Kabuto{
    \begin{align}
        \label{eq:u_NTP_j}
        \mathbf{u}_{\mathrm{r}, j}^\mathrm{NTP} \triangleq 
        \begin{cases}
            \begin{bmatrix}
                \left( \mathbf{u}_{\mathrm{r},0,0}^\mathrm{NTP} \right )^\mathrm{T}, \ldots, 
                \left( \mathbf{u}_{\mathrm{r},N^y_\mathrm{r}-1, N^y_\mathrm{r}-1}^\mathrm{NTP} \right )^\mathrm{T} 
            \end{bmatrix}^\mathrm{T} \in \mathbb{C}^{N^y_\mathrm{r} (Q_\mathrm{r}^x - N^x_\mathrm{r}) \times 1}
            & (j = 0) \\
            \begin{bmatrix}
                \left( \mathbf{u}_{\mathrm{r}, j-1,j}^\mathrm{NTP} \right )^\mathrm{T}, \ldots, 
                \left( \mathbf{u}_{\mathrm{r}, j-1,j+q^y_\mathrm{r}-1}^\mathrm{NTP} \right )^\mathrm{T}
            \end{bmatrix}^\mathrm{T} 
            \in \mathbb{C}^{ Q_\mathrm{r}^x q^y_\mathrm{r} \times 1}
            & (1 \leq j \leq N^y_\mathrm{r} - q^y_\mathrm{r}) \\
            \begin{bmatrix}
                \left( \mathbf{u}_{\mathrm{r}, j-1,j}^\mathrm{NTP} \right )^\mathrm{T}, \ldots, 
                \left( \mathbf{u}_{\mathrm{r}, j-1, N^y_\mathrm{r}-1}^\mathrm{NTP} \right )^\mathrm{T}
            \end{bmatrix}^\mathrm{T} 
            \in \mathbb{C}^{ Q_\mathrm{r}^x (N^y_\mathrm{r} - j) \times 1}
            & (N^y_\mathrm{r} - q^y_\mathrm{r} < j \leq N_\mathrm{r}^y-1) \\
        \end{cases}.
    \end{align}
    }
    \hrulefill
    \vspace*{-12pt}
    \end{figure*}
    %
    \Kabuto{
    Stacking the vectors $\mathbf{u}^\mathrm{NTP}_{\mathrm{r}, j}$ in \eqref{eq:u_NTP_j}, the unknown parameters of $\mathbf{C}_\mathrm{r}^{\mathrm{NTP}}$ in \eqref{eq:C_TP_NTP} is obtained as 
    \begin{align}
        \label{eq:u_NTP}
        \mathbf{u}_\mathrm{r}^\mathrm{NTP} = \begin{bmatrix}
            (\mathbf{u}^\mathrm{NTP}_{\mathrm{r}, 0})^\mathrm{T}, \ldots, (\mathbf{u}^\mathrm{NTP}_{\mathrm{r},N^y_{\mathrm{r}}-1})^\mathrm{T}
        \end{bmatrix}^\mathrm{T} \in \mathbb{C}^{Q_\mathrm{r}^\mathrm{NTP} \times 1},
    \end{align}
    where $Q_\mathrm{r}^\mathrm{NTP} = N^y_\mathrm{r}(Q^x_\mathrm{r} - N^x_\mathrm{r}) + Q^x_\mathrm{r} q^y_\mathrm{r} (N^y_\mathrm{r} - q^y_\mathrm{r}) + Q^x_\mathrm{r} q^y_\mathrm{r}(q^y_\mathrm{r}-1)/2 $ is the number of unknown parameters in the non-Toeplitz part.
    Let $Q_\mathrm{r}$ be defined as $Q_\mathrm{r} \triangleq Q_\mathrm{r}^\mathrm{TP} + Q_\mathrm{r}^\mathrm{NTP}$, denoting the total number of unknown parameters, including both the Toeplitz and non-Toeplitz parts.
    }


    \Kabuto{
    Based on the unknown parameters $\mathbf{u}_\mathrm{r}^\mathrm{TP}$ in \eqref{eq:u_TP}, $\mathbf{u}_\mathrm{r}^\mathrm{NTP}$ in \eqref{eq:u_NTP}, and the mutual coupling matrices $\mathbf{C}_\mathrm{r}^\mathrm{TP}$, $\mathbf{C}_\mathrm{r}^\mathrm{NTP}$ in \eqref{eq:C_TP_NTP}, the following property holds for any vector $\mathbf{b} \in \mathbb{C}^{N_\mathrm{r}\times 1}$ as
    \begin{align}
        \label{eq:Cb}
        \mathbf{C}_\mathrm{r} \mathbf{b} 
        &\simeq 
        \underbrace{\mathbf{C}_\mathrm{r}^\mathrm{TP} \mathbf{b}}_{\overset{(a)}{=} \mathbf{Q}_\mathrm{r}^\mathrm{TP} (\mathbf{b}) \mathbf{u}_\mathrm{r}^\mathrm{TP}} 
        + \underbrace{\mathbf{C}_\mathrm{r}^\mathrm{NTP} \mathbf{b}}_{\overset{(b)}{=} \mathbf{Q}_\mathrm{r}^\mathrm{NTP} (\mathbf{b}) \mathbf{u}_\mathrm{r}^\mathrm{NTP}} \nonumber \\
        &= \underbrace{\begin{bmatrix}
            \mathbf{Q}_\mathrm{r}^\mathrm{TP} (\mathbf{b}),\ \mathbf{Q}_\mathrm{r}^\mathrm{NTP} (\mathbf{b}) 
        \end{bmatrix}}_{\triangleq \mathbf{Q}_\mathrm{r} (\mathbf{b}) \in \mathbb{C}^{N_\mathrm{r} \times Q_\mathrm{r}}} 
        \underbrace{
        \begin{bmatrix}
                \mathbf{u}_\mathrm{r}^\mathrm{TP} \\
                \mathbf{u}_\mathrm{r}^\mathrm{NTP} 
        \end{bmatrix}}_{\triangleq \mathbf{u}_\mathrm{r} \in \mathbb{C}^{Q_\mathrm{r} \times1}},
    \end{align}
    where $\mathbf{Q}_\mathrm{r}^\mathrm{TP} (\mathbf{b})$ and $\mathbf{Q}_\mathrm{r}^\mathrm{NTP} (\mathbf{b})$ are given by \eqref{eq:Q_TP} and \eqref{eq:Q_NTP}, which are described in the following subsections.
    }

    \Kabuto{
    Exploiting the property in \eqref{eq:Cb}, the unknown parameters $\mathbf{u}_\mathrm{r}$ can be explicitly extracted from the mutual coupling matrix $\mathbf{C}_\mathrm{r}$. 
    Consequently, the estimation of $\mathbf{u}_\mathrm{r}$ can be efficiently performed in a closed-form, as will be described in Section~\ref{subsec:Est_c}.
    In the following subsections, the property $(a)$ in \eqref{eq:Cb}, which corresponds to the Toeplitz part, is described in Section~\ref{subsec:C_TP}, and the property $(b)$ in \eqref{eq:Cb}, which corresponds to the non-Toeplitz part, is described in Section~\ref{subsec:C_NTP}.
    }

    
    \subsubsection{\Kabuto{Property of Toeplitz Part}}
    \label{subsec:C_TP}
    \Kabuto{
    For any vector $\mathbf{a} = [a_0, \ldots, a_{N_\mathrm{r}^x-1}]^\mathrm{T} \in \mathbb{C}^{N^x_\mathrm{r} \times 1} $, the following property with respect to $\mathbf{u}^\mathrm{TP}_{\mathrm{r}, d}$ in \eqref{eq:ur_TP} and $\mathbf{C}^\mathrm{TP}_{\mathrm{r},d}$ in \eqref{eq:C_TP_NTP_jj}, $\forall d \triangleq |j-j^\prime| \in \{0,1,\ldots, N^y_\mathrm{r} -1\}$, holds as 
    \begin{align}
        \label{eq:Ca_TP_j}
        \mathbf{C}^\mathrm{TP}_{\mathrm{r}, d} \mathbf{a} = 
        \begin{cases}
             \tilde{\mathbf{Q}}^\mathrm{TP}_\mathrm{r} (\mathbf{a}) \mathbf{u}^\mathrm{TP}_{\mathrm{r},d}, & \hspace{-1cm} (q^y_\mathrm{r} < d < N^y_\mathrm{r} - 1)  \\
             \left [\tilde{\mathbf{Q}}^\mathrm{TP}_\mathrm{r} (\mathbf{a}) \right]_{:,(q^x_\mathrm{r}+1):(N^x_\mathrm{r}-1)} \mathbf{u}^\mathrm{TP}_{\mathrm{r},d}, & (0 \leq d \leq q^y_\mathrm{r} )
        \end{cases}
    \end{align}
    where $\tilde{\mathbf{Q}}^\mathrm{TP}_{\mathrm{r}}(\mathbf{a}) \in \mathbb{C}^{N^x_\mathrm{r} \times (N^x_\mathrm{r} - 1)}$ is defined in \eqref{eq:Q_tilde_TP} at the top of the next page.
    }
    \begin{figure*}
        \begin{align}
            \label{eq:Q_tilde_TP}
            \Kabuto{
            \tilde{\mathbf{Q}}^\mathrm{TP}_{\mathrm{r}} (\mathbf{a}) = 
            \begin{bmatrix} 
                a_{1} & a_{2} & \cdots & a_{N^x_\mathrm{r}-2} & a_{N^x_\mathrm{r}-1} \\
                a_{q_\mathrm{r}+2} & a_{3} & \cdots & a_{N^x_\mathrm{r}-1} & 0 \\
                \vdots & \vdots & \ddots & \vdots & \vdots \\
                a_{N^x_\mathrm{r}-1} & 0 & \cdots & 0 & 0 \\
                & & \mathbf{0}_{1 \times (N^x_\mathrm{r}-1)} & & 
            \end{bmatrix}
            +
            \begin{bmatrix} 
                & & \mathbf{0}_{1 \times (N^x_\mathrm{r}-1)} & & \\
                a_0 & 0 & \cdots & 0 & 0 \\
                a_1 & a_0 & \cdots & 0 & 0 \\
                \vdots & \vdots & \ddots & \vdots & \vdots \\
                a_{N^x_\mathrm{r}-2} & a_{N^x_\mathrm{r}-3} & \cdots & a_1 & a_0 \\
            \end{bmatrix}.
            }
        \end{align}
        \hrulefill
        \vspace*{-12pt}
    \end{figure*}

    \Kabuto{
    For any vector $\mathbf{b} = \begin{bmatrix}
        \mathbf{b}_0^\mathrm{T}, \ldots, \mathbf{b}_{N^y_\mathrm{r} - 1}^\mathrm{T}
    \end{bmatrix}^\mathrm{T} \in \mathbb{C}^{N^x_\mathrm{r} N^y_\mathrm{r} \times 1}$ with $\mathbf{b}_{d} \in \mathbb{C}^{N_\mathrm{r}^x \times 1}, \ d \in \{0,1, \ldots, N^y_\mathrm{r} -1 \}$, 
    the following property with respect to 
    $\mathbf{u}_\mathrm{r}^\mathrm{TP}$ in \eqref{eq:u_TP} and $\mathbf{C}_\mathrm{r}^\mathrm{TP}$ in \eqref{eq:C_TP_NTP} holds as
    \begin{align}
        \mathbf{C}_\mathrm{r}^\mathrm{TP} \mathbf{b} = \mathbf{Q}_\mathrm{r}^\mathrm{TP} (\mathbf{b}) \mathbf{u}_\mathrm{r}^\mathrm{TP},
    \end{align}
    with 
    \begin{align}
        \label{eq:Q_TP}
        \mathbf{Q}_\mathrm{r}^\mathrm{TP} (\mathbf{b})  = \begin{bmatrix}
        \hat{\mathbf{Q}}_{\mathrm{r},0}^\mathrm{TP} (\mathbf{b}), \ldots, 
        \hat{\mathbf{Q}}_{\mathrm{r},N^y_\mathrm{r} - 1}^\mathrm{TP} (\mathbf{b})
        \end{bmatrix} \in \mathbb{C}^{N^x_\mathrm{r} N^y_\mathrm{r} \times Q_\mathrm{r}^\mathrm{TP}},
    \end{align}
    where $\hat{\mathbf{Q}}^\mathrm{TP}_{\mathrm{r}, d}(\mathbf{b}),\ d\in \{0,1, \ldots, N_\mathrm{r}^y - 1\}$ are calculated as the following equations in \eqref{eq:Q_hat_TP_0}, \eqref{eq:Q_hat_TP_1}, and \eqref{eq:Q_hat_TP_2}.
    }

    \Kabuto{
    \textit{{Calculation of $\hat{\mathbf{Q}}_{\mathrm{r},d}^\mathrm{TP} (\mathbf{b})$}}:
    \begin{itemize}
        \item 
        For $d=0$, $\hat{\mathbf{Q}}_{\mathrm{r},d}^\mathrm{TP} (\mathbf{b}) \in \mathbb{C}^{N^x_\mathrm{r} N^x_\mathrm{r} \times (N^x_\mathrm{r}-q^x_\mathrm{r}-1)}$ is given by 
        \begin{align}
            \label{eq:Q_hat_TP_0}
            \hat{\mathbf{Q}}_{\mathrm{r},d}^\mathrm{TP} (\mathbf{b}) = 
            \Big [&
                \left( \tilde{\mathbf{Q}}_{\mathrm{r}}^\mathrm{TP} (\mathbf{b}_{0}) \right)^\mathrm{T}, \ldots, \nonumber \\ 
                &\left (\tilde{\mathbf{Q}}_{\mathrm{r}}^\mathrm{TP} (\mathbf{b}_{N^y_\mathrm{r} -1}) \right)^\mathrm{T}
            \Big ]^\mathrm{T}_{:, (q^x_\mathrm{r}+1):(N^x_\mathrm{r}-1)},
        \end{align}
        where $\tilde{\mathbf{Q}}_{\mathrm{r}}^\mathrm{TP} (\mathbf{b}_{d})$ is provided by \eqref{eq:Q_tilde_TP}.
        %
        \item 
        For $1 \leq d \leq q^y_\mathrm{r}$, $\hat{\mathbf{Q}}_{\mathrm{r},d}^\mathrm{TP} (\mathbf{b}) \in \mathbb{C}^{N^x_\mathrm{r} N^x_\mathrm{r} \times (N^x_\mathrm{r}-q^x_\mathrm{r}-1)}$ is
        \begin{align}
            \label{eq:Q_hat_TP_1}
            \hspace{-0.1cm}
            \hat{\mathbf{Q}}_{\mathrm{r},d}^\mathrm{TP} (\mathbf{b}) \!\! = \!\! \left[ \! \hat{\mathbf{Q}}_{\mathrm{r},d}^\mathrm{TP(A)} \!(\mathbf{b}) \!+\! 
            \hat{\mathbf{Q}}_{\mathrm{r},d}^\mathrm{TP(B)} \!(\mathbf{b})
            \! \right ]_{:, (q^x_\mathrm{r}+1):(N^x_\mathrm{r}-1)}
        \end{align}
        with 
        \begin{align}
            \label{eq:Q_hat_TP_A}
            \hat{\mathbf{Q}}_{\mathrm{r},d}^\mathrm{TP(A)} = 
            \Big [
                &\left( \tilde{\mathbf{Q}}^\mathrm{TP}_{\mathrm{r}} (\mathbf{b}_d) \right )^\mathrm{T} , \ \ldots,\ \left(\tilde{\mathbf{Q}}^\mathrm{TP}_{\mathrm{r}} (\mathbf{b}_{N^y_\mathrm{r}-1}) \right )^\mathrm{T} , \nonumber \\  
                &\mathbf{0}^\mathrm{T}_{d N^x_\mathrm{r} \times (N^x_\mathrm{r} \ -1)}
            \Big ]^\mathrm{T} \in \mathbb{C}^{N^x_\mathrm{r} N^y_\mathrm{r} \times (N^x_\mathrm{r}-1)} \\
            \label{eq:Q_hat_TP_B}
            \hat{\mathbf{Q}}_{\mathrm{r},d}^\mathrm{TP(B)} = 
            \Big [
                &\mathbf{0}^\mathrm{T}_{d N^x_\mathrm{r} \times (N^x_\mathrm{r} -1)},\ 
                \left( \tilde{\mathbf{Q}}^\mathrm{TP}_{\mathrm{r}} (\mathbf{b}_0) \right )^\mathrm{T}, \ldots, \nonumber \\ 
                & \hspace{-0.3cm} \left(\tilde{\mathbf{Q}}^\mathrm{TP}_{\mathrm{r}} (\mathbf{b}_{N^y_\mathrm{r}-d}) \right )^\mathrm{T}
            \Big ]^\mathrm{T} \! \! \! \! \in \mathbb{C}^{N^x_\mathrm{r} N^y_\mathrm{r} \times (N^x_\mathrm{r}-1)},
        \end{align}
        where $\tilde{\mathbf{Q}}_{\mathrm{r}}^\mathrm{TP} (\mathbf{b}_{d})$ is provided by \eqref{eq:Q_tilde_TP}.
        %
        \item
        For $q^y_\mathrm{r} < d \leq N^y_\mathrm{r} - 1$, $\hat{\mathbf{Q}}_{\mathrm{r},d}^\mathrm{TP} (\mathbf{b}) \in \mathbb{C}^{N^x_\mathrm{r} N^x_\mathrm{r} \times N^x_\mathrm{r}}$ is 
        \begin{align}
            \label{eq:Q_hat_TP_2}
            \hat{\mathbf{Q}}_{\mathrm{r},d}^\mathrm{TP} (\mathbf{b}) = \hat{\mathbf{Q}}_{\mathrm{r},d}^\mathrm{TP(A)} (\mathbf{b}) + 
            \hat{\mathbf{Q}}_{\mathrm{r},d}^\mathrm{TP(B)} (\mathbf{b}),
        \end{align}
        where $\hat{\mathbf{Q}}_{\mathrm{r},d}^\mathrm{TP(A)} (\mathbf{b})$, $\hat{\mathbf{Q}}_{\mathrm{r},d}^\mathrm{TP(A)} (\mathbf{b})$ are given by \eqref{eq:Q_hat_TP_A}, \eqref{eq:Q_hat_TP_B}. 
    \end{itemize}
    }

    \subsubsection{\Kabuto{Property of Non-Toeplitz Part}}
    \label{subsec:C_NTP}
    \Kabuto{
    For any vector $\mathbf{a} = [a_0, \ldots, a_{N_\mathrm{r}^x-1}]^\mathrm{T} \in \mathbb{C}^{N^x_\mathrm{r} \times 1} $, the following property with respect to $\mathbf{u}^\mathrm{NTP}_{\mathrm{r},j,j^\prime}$ in \eqref{eq:u_NTP_jj} and $\mathbf{C}_{\mathrm{r},j,j^\prime}^\mathrm{NTP}$ in \eqref{eq:C_TP_NTP_jj} holds as 
    \begin{align}
        \label{eq:Ca_NTP}
        \mathbf{C}_{\mathrm{r}, j, j^\prime}^\mathrm{NTP} \mathbf{a} = 
        \begin{cases}
            \tilde{\mathbf{Q}}^\mathrm{NTP}_\mathrm{r} (\mathbf{a}) \mathbf{u}_{\mathrm{r}, j,j^\prime}^\mathrm{NTP}, & (j=j^\prime \text{ and } |j-j^\prime| \leq q^y_\mathrm{r}) \\
            \check{\mathbf{Q}}^\mathrm{NTP}_\mathrm{r} (\mathbf{a}) \mathbf{u}_{\mathrm{r}, j,j^\prime}^\mathrm{NTP}, & (j \neq j^\prime \text{ and } |j-j^\prime| \leq q^y_\mathrm{r}) \\
            \mathbf{0}, & (|j-j^\prime| > q^y_\mathrm{r})
        \end{cases},
    \end{align}
    with
    \begin{align}
        \label{eq:Q_check}
        & \hspace{-0.1cm}
        \check{\mathbf{Q}}^\mathrm{NTP}_\mathrm{r}(\mathbf{a}) 
        = \begin{bmatrix} \mathrm{diag}(\mathbf{a}), \ \tilde{\mathbf{Q}}^\mathrm{NTP}_\mathrm{r} (\mathbf{a}) \end{bmatrix} \in \mathbb{C}^{N^x_\mathrm{r} \times Q^x_\mathrm{r}},
        \\
        \label{eq:Q_tilde}
        & \hspace{-0.1cm}
        \tilde{\mathbf{Q}}^\mathrm{NTP}_\mathrm{r} \!(\mathbf{a}) 
        \!=\!\! \begin{bmatrix} \!
            \tilde{\mathbf{Q}}^\mathrm{NTP}_{\mathrm{r}, 1} \! (\mathbf{a}), \! \ldots \!, 
            \tilde{\mathbf{Q}}^\mathrm{NTP}_{\mathrm{r}, N^x_\mathrm{r}-1}\! (\mathbf{a}) \!
        \end{bmatrix} \! \in \! \mathbb{C}^{N_\mathrm{r}^x \times (Q^x_\mathrm{r} - N_\mathrm{r}^x)}, \! \!
    \end{align}
    where $\tilde{\mathbf{Q}}^\mathrm{NTP}_{\mathrm{r}, i}(\mathbf{a}),\ i\in \{1,2, \ldots, N_\mathrm{r}^x -1\}$ are given by the following equations in \eqref{eq:Q_i_1} and \eqref{eq:Q_i_2}.
    }

    \Kabuto{
    \textit{{Calculation of $\tilde{\mathbf{Q}}^\mathrm{NTP}_{\mathrm{r}, i}(\mathbf{a})$}}:
    \begin{itemize}
        \item 
        For $1 \leq i \leq N^x_\mathrm{r} - q^x_\mathrm{r}$,  $\tilde{\mathbf{Q}}^\mathrm{NTP}_{\mathrm{r}, i}(\mathbf{a}) \in \mathbb{C}^{N^x_\mathrm{r} \times q^x_\mathrm{r}}$ is given by
        \begin{align}
            \label{eq:Q_i_1}
            & \tilde{\mathbf{Q}}^\mathrm{NTP}_{\mathrm{r},i} (\mathbf{a}) = 
            \begin{bmatrix} 
                & & \mathbf{0}_{(i-1) \times q^x_\mathrm{r}} & & \\
                a_{i} & a_{i+1} & \cdots & a_{i+q^x_\mathrm{r}-1} \!\!\!\!\!\!\! \\
                a_{i-1} & 0 & \cdots & 0 \\
                0 & a_{i-1} & \cdots & 0 \\
                \vdots & \vdots & \ddots & \vdots \\
                0 & 0 & \cdots & a_{i-1} \\
                & & \mathbf{0}_{(N^x_\mathrm{r}-i-q^x_\mathrm{r}) \times q^x_\mathrm{r}} & & 
            \end{bmatrix}.
        \end{align} 
        %
        \item 
        For $N^x_\mathrm{r}-q^x_\mathrm{r} < i \leq N^x_\mathrm{r}-1$, $\tilde{\mathbf{Q}}^\mathrm{NTP}_{\mathrm{r}, i}(\mathbf{a}) \in \mathbb{C}^{N^x_\mathrm{r} \times (N^x_\mathrm{r} - i)}$ is
        \begin{align}
            \label{eq:Q_i_2}
            & \tilde{\mathbf{Q}}^\mathrm{NTP}_{\mathrm{r}, i} (\mathbf{a}) = 
            \begin{bmatrix} 
                & & \mathbf{0}_{(i-1) \times (N^x_\mathrm{r}-i)} & & \\
                a_{i} & a_{i+1} & \cdots & a_{N^x_\mathrm{r}-1} \\
                a_{i-1} & 0 & \cdots & 0 \\
                0 & a_{i-1} & \cdots & 0 \\
                \vdots & \vdots & \ddots & \vdots \\
                0 & 0 & \cdots & a_{i-1}
            \end{bmatrix}.
        \end{align}
    \end{itemize}
    }

    \Kabuto{ 
    For any vector $\mathbf{b} = \begin{bmatrix}
        \mathbf{b}_0^\mathrm{T}, \ldots, \mathbf{b}_{N^y_\mathrm{r} - 1}^\mathrm{T}
    \end{bmatrix}^\mathrm{T} \in \mathbb{C}^{N^x_\mathrm{r} N^y_\mathrm{r} \times 1}$ with $\mathbf{b}_{j} \in \mathbb{C}^{N_\mathrm{r}^x \times 1}, \ j \in \{0,1, \ldots, N^y_\mathrm{r} -1 \}$, 
    the following property with respect to 
    $\mathbf{u}_\mathrm{r}^\mathrm{NTP}$ in \eqref{eq:u_NTP} and
    $\mathbf{C}_\mathrm{r}^\mathrm{NTP}$ in \eqref{eq:C_TP_NTP} holds as
    \begin{align}
        \mathbf{C}_\mathrm{r}^\mathrm{NTP} \mathbf{b} = \mathbf{Q}_\mathrm{r}^\mathrm{NTP} (\mathbf{b}) \mathbf{u}_\mathrm{r}^\mathrm{NTP} + \mathbf{b} ,
    \end{align}
    with 
    \begin{align}
        \label{eq:Q_NTP}
        \mathbf{Q}_\mathrm{r}^\mathrm{NTP} (\mathbf{b})  = \begin{bmatrix}
        \hat{\mathbf{Q}}_{\mathrm{r},0}^\mathrm{NTP} (\mathbf{b}), \ldots, 
        \hat{\mathbf{Q}}_{\mathrm{r},N^y_\mathrm{r} - 1}^\mathrm{NTP} (\mathbf{b})
        \end{bmatrix} \in \mathbb{C}^{N^x_\mathrm{r} N^y_\mathrm{r} \times Q_\mathrm{r}^\mathrm{NTP}},
    \end{align}
    where $\hat{\mathbf{Q}}^\mathrm{NTP}_{\mathrm{r}, j}(\mathbf{b}),\ j\in \{0,1, \ldots, N_\mathrm{r}^y - 1\}$ are calculated as the following equations in \eqref{eq:Q_hat_NTP_0}, \eqref{eq:Q_hat_NTP_1}, and \eqref{eq:Q_hat_NTP_2}.
    }

    \Kabuto{
    \textit{{Calculation of $\hat{\mathbf{Q}}^\mathrm{NTP}_{\mathrm{r}, j}(\mathbf{b})$}}:
    \begin{itemize}
        \item 
        For $j=0$, $\hat{\mathbf{Q}}_{\mathrm{r},0}^\mathrm{NTP} (\mathbf{b}) \! \in \! \mathbb{C}^{N_\mathrm{r}^x N_\mathrm{r}^y \times N^y_\mathrm{r}(Q^x_\mathrm{r} - N^x_\mathrm{r})}$ is 
        \begin{align}
            \label{eq:Q_hat_NTP_0}
            \hspace{-0.4cm}
            \hat{\mathbf{Q}}^\mathrm{NTP}_{\mathrm{r},0} \! (\mathbf{b}) \!=\! \mathrm{blkdiag} \! \left ( \!
                \tilde{\mathbf{Q}}^\mathrm{NTP}_{\mathrm{r}}\! (\mathbf{b}_0), \ldots, \tilde{\mathbf{Q}}^\mathrm{NTP}_{\mathrm{r}} \!(\mathbf{b}_{N^y_\mathrm{r}-1}) \!
            \right ) \!, \!\!\!\!
        \end{align}
        where $\tilde{\mathbf{Q}}^\mathrm{NTP}_{\mathrm{r}} \! (\mathbf{b}_s),\ s \! \in \! \{0,1,\ldots, N^y_\mathrm{r} \!-\! 1 \}$ is given by \eqref{eq:Q_tilde}.
        %
        \item 
        For $1 \leq j \leq N^y_\mathrm{r} - q^y_\mathrm{r}$, $\hat{\mathbf{Q}}_{\mathrm{r},j}^\mathrm{NTP} (\mathbf{b}) \! \in \! \mathbb{C}^{N_\mathrm{r}^x N_\mathrm{r}^y \times Q^x_\mathrm{r} q^y_\mathrm{r}}$ is 
        \begin{align}
            \label{eq:Q_hat_NTP_1}
            & 
            \hat{\mathbf{Q}}_{\mathrm{r},j}^\mathrm{NTP} (\mathbf{b}) \!= \! \begin{bmatrix}
                \hat{\mathbf{Q}}_{\mathrm{r},j-1,j}^\mathrm{NTP} (\mathbf{b}), \ldots , \hat{\mathbf{Q}}_{\mathrm{r},j-1, j+q^y_\mathrm{r} - 1}^\mathrm{NTP} (\mathbf{b})
            \end{bmatrix} \!\!, \!\!
        \end{align}
        where $\hat{\mathbf{Q}}_{\mathrm{r},j-1,j+s}^\mathrm{NTP} (\mathbf{b}) \in \mathbb{C}^{N^x_\mathrm{r} N^y_\mathrm{r} \times Q^x_\mathrm{r}},\ s \in \{0,1,\ldots, q^y_\mathrm{r}-1 \}$ is given by
        \begin{align}
            \label{eq:Q_hat_NTP_j}
            & 
            \hat{\mathbf{Q}}_{\mathrm{r},j-1,j+s}^\mathrm{NTP} (\mathbf{b}) \! = \! \Big [
            \mathbf{0}_{(j-1)N^x_\mathrm{r} \times Q^x_\mathrm{r}}^\mathrm{T}, \left ( \check{\mathbf{Q}}^\mathrm{NTP}_{\mathrm{r}} \!(\mathbf{b}_{j+s}) \right)^\mathrm{T}\!\!, \nonumber \\
            & \mathbf{0}_{s N^x_\mathrm{r} \times Q^x_\mathrm{r}}^\mathrm{T}, \left ( \check{\mathbf{Q}}^\mathrm{NTP}_{\mathrm{r}} \!(\mathbf{b}_{j-1}) \right)^\mathrm{T}\!\!, \mathbf{0}_{(N^y_\mathrm{r}-j-s-1)N^x_\mathrm{r} \times Q^x_\mathrm{r}}^\mathrm{T}
            \Big ]^\mathrm{T} \!\!\! , \!\!\!
        \end{align}
        where $\check{\mathbf{Q}}^\mathrm{NTP}_{\mathrm{r}} (\mathbf{b}_{j+s})$ is calculated by \eqref{eq:Q_check}.
        %
        \item 
        For $N^y_\mathrm{r} - q^y_\mathrm{r} < j \leq N^y_\mathrm{r} - 1$, 
        $\hat{\mathbf{Q}}_{\mathrm{r},j}^\mathrm{NTP} (\mathbf{b}) \in \mathbb{C}^{N_\mathrm{r}^x N_\mathrm{r}^y \times Q^x_\mathrm{r} (N^y_\mathrm{r} - j)}$ is expressed as
        \begin{align}
            \label{eq:Q_hat_NTP_2}
            & \hspace{-0.1cm}
            \hat{\mathbf{Q}}_{\mathrm{r},j}^\mathrm{NTP} (\mathbf{b}) = \begin{bmatrix}
                \hat{\mathbf{Q}}_{\mathrm{r},j-1,j}^\mathrm{NTP} (\mathbf{b}), \ldots , \hat{\mathbf{Q}}_{\mathrm{r},j-1, N^y_\mathrm{r} - 1}^\mathrm{NTP} (\mathbf{b})
            \end{bmatrix},
        \end{align}
        where $\hat{\mathbf{Q}}_{\mathrm{r},j-1,j+s}^\mathrm{NTP} (\mathbf{b}) \in \mathbb{C}^{N^x_\mathrm{r} N^y_\mathrm{r} \times Q^x_\mathrm{r}},\ s \in \{0,1,\ldots, N^y_\mathrm{r} - j - 1 \}$ is provided by \eqref{eq:Q_hat_NTP_j}.
    \end{itemize}
    }

    \subsection{Estimation for Mutual Coupling}     
    \label{subsec:Est_c}
    
    In this estimation stage, the mutual coupling matrices $\mathbf{C}_\mathrm{r}$ and $\mathbf{C}_\mathrm{t}$ are estimated given the tentative estimates \Kabuto{$ \{ \hat{\mathbf{\Gamma}}_\mathrm{r}, \hat{\mathbf{\Gamma}}_\mathrm{t}, \hat{\bm{\varepsilon}}_\mathrm{r}, \hat{\bm{\varepsilon}}_\mathrm{t}, \{\hat{\mathbf{v}}_\mathrm{r}^{(m)}, \hat{\mathbf{v}}_\mathrm{t}^{(m)}, \hat{\bm{\tau}}^{(m)},\hat{\bm{\alpha}}^{(m)} \}_{m=1}^M \}$}.

    Estimating the unknown parameters $\mathbf{u}_\mathrm{r}$, instead of directly estimating $\mathbf{C}_\mathrm{r}$, reduces computational complexity. 
    Additionally, to account for the property that the coefficients of mutual coupling at close distances have similar values, a regularizer for the estimation of $\mathbf{u}_\mathrm{r}$ is introduced as 
    \Kabuto{
    \begin{align} 
        \label{eq:S_reg} 
        r_\mathrm{c_\mathrm{r}} \! \triangleq \!
        &\sum_{j=0}^{N^y_\mathrm{r}-2} \left \| \mathbf{u}_{\mathrm{r}, j,j}^\mathrm{NTP} - \mathbf{u}_{j+1,j+1}^\mathrm{NTP}  \right \|_2^2  + \!\!\!
        \sum_{j=1}^{N^y_\mathrm{r} - q^y_\mathrm{r} -1} \!\!\!\!\! \left \| \mathbf{u}_{\mathrm{r},j}^\mathrm{NTP} - \mathbf{u}_{\mathrm{r},j+1}^\mathrm{NTP} \right \|_2^2 \nonumber \\
        & +  \sum_{j=N^y_\mathrm{r}-q^y_\mathrm{r}+1}^{N^y_\mathrm{r}-1} \sum_{d=1}^{N^y_\mathrm{r} - j - 2} 
        \left \| \mathbf{u}_{\mathrm{r}, j,j+d}^\mathrm{NTP} - \mathbf{u}_{j+1,j+d+1}^\mathrm{NTP}  \right \|_2^2 \nonumber \\
        =& \| \mathbf{S}_\mathrm{r} \mathbf{u}_\mathrm{r} \|_2^2,
    \end{align}
    where $\mathbf{S}_\mathrm{r} \in \mathbb{R}^{Q_\mathrm{r} \times Q_\mathrm{r}}$ is the regularization matrix whose entries are $0$, $1$, or $-1$, ensuring that the equation \eqref{eq:S_reg} is satisfied. \\
    }

    By defining \Kabuto{$\mathbf{t}_{p,k}^{\mathbf{c}_\mathrm{r} (m)}
    \triangleq \hat{\mathbf{\Gamma}}_\mathrm{r} 
    \mathbf{A}_{\mathrm{r},k} (\hat{\mathbf{v}_\mathrm{r}}^{(m)}, \hat{\bm{\varepsilon}}_\mathrm{r}) 
    \mathrm{diag} (\hat{\mathbf{z}}_k^{(m)})$ 
    $\mathbf{A}^\mathrm{H}_{\mathrm{t},k} (\hat{\mathbf{v}_\mathrm{t}}^{(m)}, \hat{\bm{\varepsilon}}_\mathrm{t}) 
    \hat{\mathbf{\Gamma}}_\mathrm{t}^\mathrm{H} \hat{\mathbf{C}}_\mathrm{t}^\mathrm{H} \mathbf{F}_p \mathbf{q}_p
    \in \mathbb{C}^{N_\mathrm{r} \times 1}$},
    the objective function for $\mathbf{u}_\mathrm{r}$ with the regularizer in \eqref{eq:S_reg} is formulated as
    \begin{align}
        \label{eq:loss_Cr}
        \mathcal{F} &(\mathbf{u}_\mathrm{r}) \! = \! 
        \sum_{m=1}^{M} \sum_{p=1}^{N_\mathrm{p}} \sum_{k=1}^{K}
        \left \| \mathbf{y}_{p,k}^{(m)} -  \mathbf{W}_p \mathbf{C}_\mathrm{r} \mathbf{t}_{p,k}^{\mathbf{c}_\mathrm{r} (m)} \right \|_2^2 
        + \lambda^{\mathbf{c}_\mathrm{r}} \left \| \mathbf{S}_\mathrm{r} \mathbf{u}_\mathrm{r} \right \|_2^2 \nonumber \\
        & \overset{(c)}{=}
        \sum_{m=1}^{M} \sum_{p=1}^{N_\mathrm{p}} \sum_{k=1}^{K}
        \left \| \mathbf{y}_{p,k}^{(m)} - \mathbf{W}_p \mathbf{t}_{p,k}^{\mathbf{c}_\mathrm{r} (m)} - \mathbf{W}_p \mathbf{Q}_\mathrm{r} (\mathbf{t}_{p,k}^{\mathbf{c}_\mathrm{r} (m)}) \mathbf{u}_\mathrm{r} \right \|_2^2  \nonumber \\
        & \qquad + \lambda^{\mathbf{c}_\mathrm{r}} \left \| \mathbf{S}_\mathrm{r} \mathbf{u}_\mathrm{r} \right \|_2^2,
    \end{align}
    where $\lambda^{\mathbf{c}_\mathrm{r}}$ is the penalty coefficient, and \Kabuto{the equality in $(c)$ holds using the property \eqref{eq:Cb}}.
    Since the regularization term in \eqref{eq:loss_Cr} prevents overfitting caused by the increase in parameters due to the introduction of the non-Toeplitz part, the penalty coefficient is set as $\lambda^{\mathbf{c}_\mathrm{r}} \propto \sigma^2$ to mitigate overfitting in the low \ac{SNR} region.
    
    The optimal solution of $\mathbf{u}_\mathrm{r}$ is calculated in closed-form as 
    \begin{align}
        \label{eq:ur_opt}
        \hat{\mathbf{u}}_\mathrm{r} = 
        & \left \{ 
            \sum_{m,p,k}
            \mathbf{Q}^\mathrm{H}_\mathrm{r} (\mathbf{t}_{p,k}^{\mathbf{c}_\mathrm{r} (m)})\mathbf{W}_p^\mathrm{H} 
            \mathbf{W}_p \mathbf{Q}_\mathrm{r} (\mathbf{t}_{p,k}^{\mathbf{c}_\mathrm{r} (m)})
            + \lambda^{\mathbf{c}_\mathrm{r}}
            \mathbf{S}_\mathrm{r}^\mathrm{H} \mathbf{S}_\mathrm{r}
        \right \}^{-1} \nonumber \\
        & \!\!\!\! \times \left \{ \sum_{m,p,k}
            \mathbf{Q}^\mathrm{H}_\mathrm{r} (\mathbf{t}_{p,k}^{\mathbf{c}_\mathrm{r} (m)})\mathbf{W}_p^\mathrm{H}  
            \left( \mathbf{y}_{p,k}^{(m)} - \mathbf{W}_p \mathbf{t}_{p,k}^{\mathbf{c}_\mathrm{r} (m)} \right )
        \right \}.
    \end{align}

    Following a similar procedure in the derivation of \eqref{eq:ur_opt}, the optimal solution for $\mathbf{u}_\mathrm{t}$ can be obtained as
    \begin{align}
        \label{eq:ut_opt}
        \hat{\mathbf{u}}_\mathrm{t} & \!=\!\!
        \left \{ \!
            \sum_{m,p,k} \!\!
            \mathbf{Q}^\mathrm{H}_\mathrm{t} ( \mathbf{F}_p \mathbf{q}_p ) \mathbf{R}_{p,k}^{ \mathbf{c}_\mathrm{t} (m) \mathrm{H}} \!
            \mathbf{R}_{p,k}^{ \mathbf{c}_\mathrm{t} (m)} \mathbf{Q}_\mathrm{t} ( \mathbf{F}_p \mathbf{q}_p )
            \! + \! \lambda^{\mathbf{c}_\mathrm{t}}
            \mathbf{S}_\mathrm{t}^\mathrm{H} \mathbf{S}_\mathrm{t} \! \!
        \right \}^{\! -1} \nonumber \\
        & \!\!\! \times \left \{ \! \sum_{m,p,k} \!\!
            \mathbf{Q}^\mathrm{H}_\mathrm{t} ( \mathbf{F}_p \mathbf{q}_p ) \mathbf{R}_{p,k}^{ \mathbf{c}_\mathrm{t} (m) \mathrm{H}} \!\!
            \left( \mathbf{y}_{p,k}^{(m)} - \mathbf{R}_{p,k}^{ \mathbf{c}_\mathrm{t} (m)} \mathbf{F}_p \mathbf{q}_p \! \right ) \!
        \right \} \!, \!\!
    \end{align}
    with 
    \Kabuto{
    $\mathbf{R}_{p,k}^{ \mathbf{c}_\mathrm{t} (m)} \triangleq \mathbf{W}_p \hat{\mathbf{C}}_\mathrm{r} \hat{\mathbf{\Gamma}}_\mathrm{r} \mathbf{A}_{\mathrm{r},k} (\hat{\mathbf{v}}_\mathrm{r}^{(m)}, \hat{\bm{\varepsilon}}_\mathrm{r}) \mathrm{diag}(\mathbf{z}^{(m)}_k) $
    $\mathbf{A}^{\mathrm{H}}_{\mathrm{t},k} (\hat{\mathbf{v}}_\mathrm{t}^{(m)}, \hat{\bm{\varepsilon}}_\mathrm{t} ) \hat{\mathbf{\Gamma}}_\mathrm{t}^\mathrm{H}
    \in \mathbb{C}^{N_{\mathrm{r,RF}} \times N_\mathrm{t}}$.}
    Using the estimates $\hat{\mathbf{u}}_\mathrm{r}$ and $\hat{\mathbf{u}}_\mathrm{t}$, the mutual coupling matrices $\hat{\mathbf{C}}_\mathrm{r}$ and $\hat{\mathbf{C}}_\mathrm{t}$ can be reconstructed based on the structure in \Kabuto{\eqref{eq:C_TP_1}-\eqref{eq:C_NTP_3}}.

    In a similar way to the switching mechanism for the on-grid and the off-grid algorithms described in Section~\ref{subsec:on_grid} and ~\ref{subsec:off_grid}, the approximate coupling model in \eqref{eq:C_TP_NTP} is introduced depending on the algorithmic iterations.
    Due to the low estimation accuracy of each parameter in the early algorithmic iterations, the use of the approximate model in \eqref{eq:C_TP_NTP} results in poor convergence performance because of an excessive number of parameters in the mutual coupling matrix, which might cause overfitting, particularly with large $q^x_\mathrm{r}$, $q^y_\mathrm{r}$, $q^x_\mathrm{t}$ and $q^y_\mathrm{t}$.
    Thus, in the early iterations, only the Toeplitz part (\textit{i.e.}, $q^x_\mathrm{r}=q^y_\mathrm{r}=q^x_\mathrm{t}=q^y_\mathrm{t}=0$), without the non-Toeplitz part, is utilized to avoid overfitting induced by inaccurate tentative estimates.
    When the rate of change in the objective function $\Delta \mathcal{F}$ falls below a threshold $\mathrm{Th}_2$ as $\Delta \mathcal{F} \leq \mathrm{Th}_2$, the approximate model in \eqref{eq:C_TP_NTP} is introduced to mitigate model mismatch in the mutual coupling.

    \subsection{Estimation for Gain/Phase Errors}
    
    In this stage, we estimate $\bm{\gamma}_\mathrm{r}$ and $\bm{\gamma}_\mathrm{t}$ (because of $\mathbf{\Gamma}_\mathrm{r} = \mathrm{diag} (\bm{\gamma}_\mathrm{r}), \mathbf{\Gamma}_\mathrm{t} = \mathrm{diag} (\bm{\gamma}_\mathrm{t})$) given the tentative estimates \Kabuto{$ \{ \hat{\mathbf{C}}_\mathrm{r}, \hat{\mathbf{C}}_\mathrm{t}, \hat{\bm{\varepsilon}}_\mathrm{r}, \hat{\bm{\varepsilon}}_\mathrm{t}, \{\hat{\mathbf{v}}_\mathrm{r}^{(m)}, \hat{\mathbf{v}}_\mathrm{t}^{(m)}, \hat{\bm{\tau}}^{(m)},\hat{\bm{\alpha}}^{(m)} \}_{m=1}^M \}$}.
    
    Defining
    \Kabuto{
    $\mathbf{t}_{p,k}^{\bm{\gamma}_\mathrm{r} (m)} \triangleq \mathbf{A}_{\mathrm{r},k}(\hat{\mathbf{v}}_\mathrm{r}^{(m)}, \hat{\bm{\varepsilon}}_\mathrm{r}) \mathrm{diag} (\hat{\mathbf{z}}_k^{(m)}) \mathbf{A}^\mathrm{H}_{\mathrm{t},k}(\hat{\mathbf{v}}_\mathrm{t}^{(m)}, \hat{\bm{\varepsilon}}_\mathrm{t}) $
    }
    $\hat{\mathbf{\Gamma}}_\mathrm{t}^\mathrm{H} \hat{\mathbf{C}}_\mathrm{t}^\mathrm{H} \mathbf{F}_p \mathbf{q}_p \in \mathbb{C}^{N_\mathrm{r} \times 1}$, 
    the objective function is given by 
    \begin{align}
        \label{eq:obj_gamma_r}
        \mathcal{F}(\bm{\gamma}_\mathrm{r}) 
        & = \sum_{m=1}^M \sum_{p=1}^{N_\mathrm{p}} \sum_{k \in \mathcal{Q}} \left \| 
        \mathbf{y}_{p,k}^{(m)} - \mathbf{T}_{p,k}^{\bm{\gamma}_\mathrm{r} (m)} \bm{\gamma}_\mathrm{r}
        \right  \|_2^2,
    \end{align}
    with
    $\mathbf{T}_{p,k}^{\bm{\gamma}_\mathrm{r} (m)} \triangleq \mathbf{W}_p \hat{\mathbf{C}}_\mathrm{r} \hat{\mathbf{\Gamma}}_\mathrm{r} \mathrm{diag} (\mathbf{t}_{p,k}^{\bm{\gamma}_\mathrm{r} (m)}) \in \mathbb{C}^{N_\mathrm{r,RF} \times N_\mathrm{r}}$.
    The minimization of \eqref{eq:obj_gamma_r} can be solved in closed form as 
    \begin{align}   
        \label{eq:gamma_r_opt}
        \hat{\bm{\gamma}}_\mathrm{r} = \left ( \!
        \sum_{m,p,k} 
        \mathbf{T}_{p,k}^{\bm{\gamma}_\mathrm{r} (m) \mathrm{H}}
        \mathbf{T}_{p,k}^{\bm{\gamma}_\mathrm{r} (m)} \!\! \right)^{\!-1} \!\!\!\!
        \left ( \! \sum_{m,p,k} \mathbf{T}_{p,k}^{\bm{\gamma}_\mathrm{r} (m) \mathrm{H}} \mathbf{y}_{p,k}^{(m)} \! \right).
    \end{align}

    Through the same procedure as the derivation of $\hat{\bm{\gamma}}_\mathrm{r}$ in \eqref{eq:gamma_r_opt}, the gain/phase errors at the UE side $\hat{\bm{\gamma}}_\mathrm{t}$ can be obtained as 
    \begin{align}
        \label{eq:gamma_t_opt}
        \hat{\bm{\gamma}}_\mathrm{t} \!=\! \left ( \!
        \sum_{m,p,k} \mathbf{T}_{p,k}^{\bm{\gamma}_\mathrm{t} (m) \mathrm{H}}
        \mathbf{T}_{p,k}^{\bm{\gamma}_\mathrm{t} (m)} \!\! \right)^{-1} \!\!\!\!
        \left (  \sum_{m,p,k}  \mathbf{T}_{p,k}^{\bm{\gamma}_\mathrm{t} (m) \mathrm{H}} \mathbf{y}_{p,k}^{(m)} \! \right),
    \end{align}
    where 
    \Kabuto{
    $\mathbf{T}_{p,k}^{\bm{\gamma}_\mathrm{t} (m)} \!\! \triangleq \! \mathbf{W}_p \hat{\mathbf{C}}_\mathrm{r} \hat{\mathbf{\Gamma}}_\mathrm{r} \mathbf{A}_{\mathrm{r},k} (\hat{\mathbf{v}}_\mathrm{r}^{(m)}, \hat{\bm{\varepsilon}}_\mathrm{r}) \mathrm{diag}(\hat{\mathbf{z}}^{(m)}_k) \mathbf{A}^{\mathrm{H}}_{\mathrm{t},k} (\hat{\mathbf{v}}_\mathrm{t}^{(m)}, $
    $ \hat{\bm{\varepsilon}}_\mathrm{t})  \mathrm{diag} ( \mathbf{t}_p^{\bm{\gamma}_\mathrm{t}} ) \in \mathbb{C}^{N_{\mathrm{r,RF}} \times N_\mathrm{r}}$,
    }
    with 
    $\mathbf{t}_p^{\bm{\gamma}_\mathrm{t}} = \hat{\mathbf{C}}_\mathrm{t} \mathbf{F}_p \mathbf{q}_p \in \mathbb{C}^{N_\mathrm{t} \times 1}$.

    \subsection{Estimation for Antenna Spacing Errors}
    
    In this stage, antenna spacing errors $\bm{\varepsilon}_\mathrm{r}$ and $\bm{\varepsilon}_\mathrm{t}$ are estimated by a gradient descent method.
    
    \Kabuto{
    Substituting the optimal solution $\hat{\mathbf{z}}_k^{(m)}$ in \eqref{eq:z_opt} into \eqref{eq:min},
    the objective function for $\bm{\varepsilon}_\mathrm{r},\bm{\varepsilon}_\mathrm{t}$ is given by
    \begin{align}
        \label{eq:obj_AoD}
        \mathcal{F}(\bm{\varepsilon}_\mathrm{r},\bm{\varepsilon}_\mathrm{t}) \! = \! \sum_{m=1}^M \sum_{k \in \mathcal{Q}} \left \| 
        \mathbf{y}_k^{(m)} - \mathbf{T}_k^{z (m)} \left ( \mathbf{T}_k^{z (m)} \right )^\dagger \mathbf{y}_k^{(m)}
        \! \right \|_2^2 \!\!. \!\!
    \end{align}
    }

    \Kabuto{
    The gradients of $\bm{\varepsilon}_\mathrm{r}^x$ and $\bm{\varepsilon}_\mathrm{r}^y$ at the BS are calculated as
    \begin{subequations}
    \label{eq:grad_e_r}
    \begin{align}
        \label{eq:grad_e_rx}
        \frac{\partial \mathcal{F}}{\partial \varepsilon_{\mathrm{r},n}^{x}}
        & \!=\! 2 \! \sum_{m,k} \mathfrak{R} \left [ 
        \mathbf{d}_{\mathrm{r},k,n}^{\varepsilon,x (m) \mathrm{H}} \mathbf{T}_k^{z(m)} \hat{\mathbf{z}}_k^{(m)} \!-\! \mathbf{y}_k^{(m)} \mathbf{d}_{\mathrm{r},k,n}^{\varepsilon, x (m)}
        \right ], \\
        \label{eq:grad_e_ry}
        \frac{\partial \mathcal{F}}{\partial \varepsilon_{\mathrm{r},n}^{y}}
        & \!=\! 2 \! \sum_{m,k} \mathfrak{R} \left [ 
        \mathbf{d}_{\mathrm{r},k,n}^{\varepsilon,y (m) \mathrm{H}} \mathbf{T}_k^{z(m)} \hat{\mathbf{z}}_k^{(m)} \!-\! \mathbf{y}_k^{(m)} \mathbf{d}_{\mathrm{r},k,n}^{\varepsilon, y (m)}
        \right ],
    \end{align}
    \end{subequations}
    where $\mathbf{d}_{\mathrm{r},k,n}^{\varepsilon,x (m)}$ and $\mathbf{d}_{\mathrm{r},k,n}^{\varepsilon,y (m)}$ are calculated as \eqref{eq:dr_eps_x} and \eqref{eq:dr_eps_y} at the top of the next page.
    }
    \begin{figure*}
        \begin{subequations}
        \begin{align}
            \label{eq:dr_eps_x}
            \Kabuto{
            \mathbf{d}_{\mathrm{r},k,n}^{\varepsilon,x (m)} = 
            \begin{bmatrix}
            \left \{
            \mathbf{W}_1 \hat{\mathbf{C}}_\mathrm{r} \hat{\mathbf{\Gamma}}_\mathrm{r}  \left( \mathbf{g}_{\mathrm{r},1,k,n}^{x(m)} \otimes \mathbf{e}_{N^x_\mathrm{r}, n}  \right ) \right \}^\mathrm{T}, \ldots, 
            \left \{ \mathbf{W}_{N_\mathrm{p}} \hat{\mathbf{C}}_\mathrm{r} \hat{\mathbf{\Gamma}}_\mathrm{r}  \left( \mathbf{g}_{\mathrm{r},N_\mathrm{p},k,n}^{x(m)} \otimes \mathbf{e}_{N^x_\mathrm{r}, n}  \right ) \right \}^\mathrm{T}
            \end{bmatrix}^\mathrm{T} \in \mathbb{C}^{N_{\mathrm{r,RF}} N_\mathrm{p} \times 1}} \\
            \label{eq:dr_eps_y}
            \Kabuto{
            \mathbf{d}_{\mathrm{r},k,n}^{\varepsilon,x (m)} = 
            \begin{bmatrix}
            \left \{ \mathbf{W}_1 \hat{\mathbf{C}}_\mathrm{r} \hat{\mathbf{\Gamma}}_\mathrm{r}  \left( \mathbf{e}_{N^y_\mathrm{r}, n} \otimes\mathbf{g}_{\mathrm{r},1,k,n}^{y(m)}  \right ) \right \}^\mathrm{T}, \ldots, 
            \left \{ \mathbf{W}_{N_\mathrm{p}} \hat{\mathbf{C}}^\mathrm{T}_\mathrm{r} \hat{\mathbf{\Gamma}}_\mathrm{r}  \left( \mathbf{e}_{N^y_\mathrm{r}, n} \otimes\mathbf{g}_{\mathrm{r},N_\mathrm{p},k,n}^{y(m)}  \right ) \right \}^\mathrm{T}
            \end{bmatrix}^\mathrm{T} \in \mathbb{C}^{N_{\mathrm{r,RF}} N_\mathrm{p} \times 1}}
        \end{align}
        \end{subequations}
        \vspace*{-30pt}
    \end{figure*}
    %
    \Kabuto{
    In \eqref{eq:dr_eps_x} and \eqref{eq:dr_eps_y}, $\mathbf{g}_{\mathrm{r},p,k,n}^{x(m)} \in \mathbb{C}^{N^y_\mathrm{r} \times 1}$ and $\mathbf{g}_{\mathrm{r},p,k,n}^{y(m)} \in \mathbb{C}^{N^x_\mathrm{r} \times 1}$ are expressed as
    \begin{subequations}
    \begin{align}
        \mathbf{g}_{\mathrm{r},p,k,n}^{x(m)} & \!\! = \!
        \mathbf{A}_{\mathrm{r},k}^y ( \hat{\mathbf{v}}_\mathrm{r}^{y(m)}, \hat{\bm{\varepsilon}}_\mathrm{r}^y)
        \left(  \mathbf{t}_{p,k}^{z(m)} \odot \hat{\mathbf{z}}_{k}^{(m)} \odot \mathbf{f}_{\mathrm{r},n,k}^{\varepsilon,x (m)} \right), \\
        \mathbf{g}_{\mathrm{r},p,k,n}^{y(m)} & \!\! = \! 
        \mathbf{A}_{\mathrm{r},k}^x ( \hat{\mathbf{v}}_\mathrm{r}^{x(m)}, \hat{\bm{\varepsilon}}_\mathrm{r}^x)
        \left(  \mathbf{t}_{p,k}^{z(m)} \odot \hat{\mathbf{z}}_{k}^{(m)} \odot \mathbf{f}_{\mathrm{r},n,k}^{\varepsilon,y (m)} \right),
    \end{align}
    \end{subequations}
    where 
    $\mathbf{f}_{\mathrm{r},n,k}^{\varepsilon,x (m)} \in \mathbb{C}^{\hat{L} \times 1}$ and $\mathbf{f}_{\mathrm{r},n,k}^{\varepsilon,y (m)} \in \mathbb{C}^{\hat{L} \times 1}$ are given by
    \begin{align*}
        \mathbf{f}_{\mathrm{r},n,k}^{\varepsilon,x (m)} \!\! &= \!\! \left [ j\frac{2 \pi}{\lambda_\mathrm{c}} \left \{  \left ( 1 + \frac{\Delta f_k}{f_\mathrm{c}} \right ) \hat{\mathbf{v}}_\mathrm{r}^{x(m)}) \right \} \right ] \!\! \odot \!\!
        \left [\mathbf{A}_{\mathrm{r},k}^{x \mathrm{T}} (\hat{\mathbf{v}}_\mathrm{r}^{x(m)}, \hat{\bm{\varepsilon}}_\mathrm{r}^x) \right ]_{:, n} \!\! , \\
        \mathbf{f}_{\mathrm{r},n,k}^{\varepsilon,y (m)} \!\! &= \!\! \left [ j\frac{2 \pi}{\lambda_\mathrm{c}} \left \{  \left ( 1 + \frac{\Delta f_k}{f_\mathrm{c}} \right ) \hat{\mathbf{v}}_\mathrm{r}^{y(m)}) \right \} \right ] \!\! \odot \!\!
        \left [\mathbf{A}_{\mathrm{r},k}^{y \mathrm{T}} (\hat{\mathbf{v}}_\mathrm{r}^{y(m)}, \hat{\bm{\varepsilon}}_\mathrm{r}^y) \right ]_{:, n} \!\!.
    \end{align*}
    }

    \Kabuto{
    Similarly, the gradients of $\bm{\varepsilon}_\mathrm{t}^x$, $\bm{\varepsilon}_\mathrm{t}^y$ at the UE are given by
    \begin{subequations}
    \label{eq:grad_e_t}
    \begin{align}
        \label{eq:grad_e_tx}
        \frac{\partial \mathcal{F}}{\partial \varepsilon_{\mathrm{t},n}^{x}}
        & \! = \! 2  \sum_{m,k} \mathfrak{R} \left [ 
        \mathbf{d}_{\mathrm{t},k,n}^{\varepsilon,x (m) \mathrm{H}} \mathbf{T}_k^{z(m)} \hat{\mathbf{z}}_k^{(m)} \!-\! \mathbf{y}_k^{(m)} \mathbf{d}_{\mathrm{t},k,n}^{\varepsilon, x (m)}
        \right ], \!\! \\
        \label{eq:grad_e_ty}
        \frac{\partial \mathcal{F}}{\partial \varepsilon_{\mathrm{t},n}^{y}}
        & \!=\! 2  \sum_{m,k} \mathfrak{R} \left [ 
        \mathbf{d}_{\mathrm{t},k,n}^{\varepsilon,y (m) \mathrm{H}} \mathbf{T}_k^{z(m)} \hat{\mathbf{z}}_k^{(m)} \!-\! \mathbf{y}_k^{(m)} \mathbf{d}_{\mathrm{t},k,n}^{\varepsilon, y (m)}
        \right ], \!\!
    \end{align}
    \end{subequations}
    where $\mathbf{d}_{\mathrm{t},k,n}^{\varepsilon,x (m)}$ and $\mathbf{d}_{\mathrm{t},k,n}^{\varepsilon,y (m)}$ are calculated as \eqref{eq:dt_eps_x} and \eqref{eq:dt_eps_y} at the top of the next page.
    }
    %
    \begin{figure*}
        \begin{subequations}
        \begin{align}
            \label{eq:dt_eps_x}
            \Kabuto{
            \mathbf{d}_{\mathrm{t},k,n}^{\varepsilon,x (m)} = 
            \begin{bmatrix}
                \left \{ \mathbf{W}^\mathrm{T}_1 \hat{\mathbf{C}}^\mathrm{T}_\mathrm{r} \hat{\mathbf{\Gamma}}^\mathrm{T}_\mathrm{r}   
                \mathbf{A}_{\mathrm{r},k}(\hat{\mathbf{v}}_{\mathrm{r}}^{(m)}, \hat{\bm{\varepsilon}}_\mathrm{r})
                \mathbf{g}_{\mathrm{t},1,k,n}^{x(m)} \right \}^\mathrm{T}, \ldots,  
                \left \{ \mathbf{W}^\mathrm{T}_{N_\mathrm{p}} \hat{\mathbf{C}}^\mathrm{T}_\mathrm{r} \hat{\mathbf{\Gamma}}^\mathrm{T}_\mathrm{r}   
                \mathbf{A}_{\mathrm{r},k}(\hat{\mathbf{v}}_{\mathrm{r}}^{(m)}, \hat{\bm{\varepsilon}}_\mathrm{r})
                \mathbf{g}_{\mathrm{t},N_\mathrm{p},k,n}^{x(m)} \right \}^\mathrm{T}
            \end{bmatrix}^\mathrm{T} \in \mathbb{C}^{N_{\mathrm{r,RF}} N_\mathrm{p} \times 1}} \\
            \label{eq:dt_eps_y}
            \Kabuto{
            \mathbf{d}_{\mathrm{t},k,n}^{\varepsilon,y (m)} = 
            \begin{bmatrix}
                \left \{ \mathbf{W}^\mathrm{T}_1 \hat{\mathbf{C}}^\mathrm{T}_\mathrm{r} \hat{\mathbf{\Gamma}}^\mathrm{T}_\mathrm{r}   
                \mathbf{A}_{\mathrm{r},k}(\hat{\mathbf{v}}_{\mathrm{r}}^{(m)}, \hat{\bm{\varepsilon}}_\mathrm{r})
                \mathbf{g}_{\mathrm{t},1,k,n}^{y(m)} \right \}^\mathrm{T}, \ldots,  
                \left \{ \mathbf{W}^\mathrm{T}_{N_\mathrm{p}} \hat{\mathbf{C}}^\mathrm{T}_\mathrm{r} \hat{\mathbf{\Gamma}}^\mathrm{T}_\mathrm{r}   
                \mathbf{A}_{\mathrm{r},k}(\hat{\mathbf{v}}_{\mathrm{r}}^{(m)}, \hat{\bm{\varepsilon}}_\mathrm{r})
                \mathbf{g}_{\mathrm{t},N_\mathrm{p},k,n}^{y(m)} \right \}^\mathrm{T}
            \end{bmatrix}^\mathrm{T} \in \mathbb{C}^{N_{\mathrm{r,RF}} N_\mathrm{p} \times 1}} 
        \end{align}
        \end{subequations}
        \hrulefill
        \vspace*{-12pt}
    \end{figure*}
    %
    \Kabuto{
    In \eqref{eq:dt_eps_x} and \eqref{eq:dt_eps_y}, $\mathbf{g}_{\mathrm{t},p,k,n}^{x(m)} \in \mathbb{C}^{\hat{L} \times 1}$ and $\mathbf{g}_{\mathrm{t},p,k,n}^{y(m)} \in \mathbb{C}^{\hat{L} \times 1}$ are expressed as
    \begin{subequations}
    \begin{align}
        \mathbf{g}_{\mathrm{t},p,k,n}^{x(m)} = 
        &\mathrm{diag}(\mathbf{z}_k^{(m)})
        \left \{ \mathbf{A}_{\mathrm{t},k}^{y} (\hat{\mathbf{v}}_{\mathrm{t}}^{y(m)},  \hat{\bm{\varepsilon}}_\mathrm{r}^y) \circ 
        \left( \mathbf{e}_{N_\mathrm{t}^x,n} \mathbf{f}_{\mathrm{t},n,k}^{\varepsilon, x (m) \mathrm{T}} \right)  
        \right \}^\mathrm{H} \nonumber \\
        & \times \hat{\mathbf{\Gamma}}^\mathrm{H}_\mathrm{t} \hat{\mathbf{C}}^\mathrm{H}_\mathrm{t} \mathbf{F} \mathbf{q}_p, \\
        \mathbf{g}_{\mathrm{t},p,k,n}^{y(m)} = 
        &\mathrm{diag}(\mathbf{z}_k^{(m)})
        \left \{ \left( \mathbf{e}_{N_\mathrm{t}^y,n} \mathbf{f}_{\mathrm{t},n,k}^{\varepsilon, y (m) \mathrm{T}} \right)   \circ \mathbf{A}_{\mathrm{t},k}^{x} (\hat{\mathbf{v}}_{\mathrm{t}}^{x(m)},  \hat{\bm{\varepsilon}}_\mathrm{r}^x)
        \right \}^\mathrm{H} \nonumber \\
        & \times \hat{\mathbf{\Gamma}}^\mathrm{H}_\mathrm{t} \hat{\mathbf{C}}^\mathrm{H}_\mathrm{t} \mathbf{F} \mathbf{q}_p,
    \end{align}
    \end{subequations}
    where 
    $\mathbf{f}_{\mathrm{t},n,k}^{\varepsilon,x (m)} \in \mathbb{C}^{\hat{L} \times 1}$ and $\mathbf{f}_{\mathrm{t},n,k}^{\varepsilon,y (m)} \in \mathbb{C}^{\hat{L} \times 1}$ are given by
    \begin{align*}
        \mathbf{f}_{\mathrm{t},n,k}^{\varepsilon,x (m)} \!\! &= \!\! \left [ j\frac{2 \pi}{\lambda_\mathrm{c}} \left \{  \left ( 1 + \frac{\Delta f_k}{f_\mathrm{c}} \right ) \hat{\mathbf{v}}_\mathrm{t}^{x(m)}) \right \} \right ] \!\! \odot \!\!
        \left [\mathbf{A}_{\mathrm{t},k}^{x \mathrm{T}} (\hat{\mathbf{v}}_\mathrm{t}^{x(m)}, \hat{\bm{\varepsilon}}_\mathrm{t}^x) \right ]_{:, n} \!\! , \\
        \mathbf{f}_{\mathrm{t},n,k}^{\varepsilon,y (m)} \!\! &= \!\! \left [ j\frac{2 \pi}{\lambda_\mathrm{c}} \left \{  \left ( 1 + \frac{\Delta f_k}{f_\mathrm{c}} \right ) \hat{\mathbf{v}}_\mathrm{t}^{y(m)}) \right \} \right ] \!\! \odot \!\!
        \left [\mathbf{A}_{\mathrm{t},k}^{y \mathrm{T}} (\hat{\mathbf{v}}_\mathrm{t}^{y(m)}, \hat{\bm{\varepsilon}}_\mathrm{t}^y) \right ]_{:, n} \!\!.
    \end{align*}
    }

    Finally, from the estimated parameters 
    $\hat{\mathbf{\Omega}} =  \left \{ \hat{\mathbf{C}}_\mathrm{r}, \hat{\mathbf{C}}_\mathrm{t}, \hat{\mathbf{\Gamma}}_\mathrm{r}, \hat{\mathbf{\Gamma}}_\mathrm{t}, \hat{\bm{\varepsilon}}_\mathrm{r}, \hat{\bm{\varepsilon}}_\mathrm{t},  \{\hat{\mathbf{v}}_\mathrm{r}^{(m)}, \hat{\mathbf{v}}_\mathrm{t}^{(m)}, \hat{\bm{\tau}}^{(m)} , \hat{\bm{\alpha}}^{(m)} \}_{m=1}^M  \right \}$,
    the channel for the current frame $m=M$ is reconstructed as
    \begin{align}
        \label{eq:H_est}
        \hat{\mathbf{H}}_k^{(M)} = 
        & \hat{\mathbf{C}}_\mathrm{r} \hat{\mathbf{\Gamma}}_\mathrm{r} 
        \mathbf{A}_{\mathrm{r},k}(\hat{\mathbf{v}}_\mathrm{r}^{(M)}, \hat{\bm{\varepsilon}}_\mathrm{r}) 
        \mathrm{diag} \left( \hat{\bm{\alpha}}^{(M)} \odot \mathbf{b}_k (\hat{\bm{\tau}}^{(M)}) \right) \nonumber \\
        & \times \mathbf{A}^\mathrm{H}_{\mathrm{t},k}(\hat{\mathbf{v}}_\mathrm{t}^{(M)}, \hat{\bm{\varepsilon}}_\mathrm{t})
        \hat{\mathbf{\Gamma}}_\mathrm{t}^\mathrm{H} \hat{\mathbf{C}}_\mathrm{t}^\mathrm{H}, \quad \forall k \in \mathcal{K}.
    \end{align}
    The flow of the proposed algorithm is provided in Algorithm~\ref{alg:prop}.

    \setlength{\textfloatsep}{5pt}
\begin{algorithm}[t]
    \caption[]{Proposed channel estimation algorithm}
    \label{alg:prop}
    \hrulefill
    \begin{algorithmic}[1]
        \vspace{-0.5ex}
        %
        \Statex \textbf{Input:}  \Kabuto{$T_\mathrm{iter}, \mathrm{Th}_1, \mathrm{Th}_2, q^x_\mathrm{r}, q^y_\mathrm{r}, q^x_\mathrm{t}, q^y_\mathrm{t}$},\ 
        \KabutoSecond{\textbf{Output:}} \KabutoSecond{$\hat{\mathbf{H}}^{(M)}$}
        \vspace{-1.5ex}
        \Statex \hspace{-3ex} \hrulefill

        \Statex \textbf{// Initialization} 
        \Statex 
            $\hat{\mathbf{C}}_\mathrm{r} = \hat{\mathbf{\Gamma}}_\mathrm{r} = \mathbf{I}_{N_\mathrm{r}}$, 
            $\hat{\mathbf{C}}_\mathrm{t} = \hat{\mathbf{\Gamma}}_\mathrm{t} = \mathbf{I}_{N_\mathrm{t}}$, 
            $\hat{\bm{\varepsilon}}_\mathrm{r} = \mathbf{0}$, 
            $\hat{\bm{\varepsilon}}_\mathrm{t} = \mathbf{0}$

        \Statex \textbf{// ML algorithm via alternating optimization}
        \For{$t=1, 2, \ldots, T_\mathrm{iter}$} 

            \If{$\Delta \mathcal{F} > \mathrm{Th}_1$}
                \State \Kabuto{Update $\hat{\bm{\alpha}}^{(m)}$, $\hat{\bm{\tau}}^{(m)}$, $\hat{\mathbf{v}}_\mathrm{r}^{(m)}$, $\hat{\mathbf{v}}_\mathrm{t}^{(m)}$ by solving \eqref{eq:min_z_CS} }
                \Statex \hspace{27pt} \Kabuto{via OMP~\cite{1993Pati_OMP}}
                \State Update $\hat{\mathbf{C}}_\mathrm{r}$ and $\hat{\mathbf{C}}_\mathrm{t}$ from \eqref{eq:ur_opt} and \eqref{eq:ut_opt}
                \Statex \hspace{27pt} \Kabuto{with $q^x_\mathrm{r} = q^y_\mathrm{r} = q^x_\mathrm{t} = q^y_\mathrm{t} = 0$}
                %
            \Else 
                \State \Kabuto{Update $\hat{\bm{\alpha}}^{(m)}$, $\hat{\bm{\tau}}^{(m)}$, $\hat{\mathbf{v}}_\mathrm{r}^{(m)}$, $\hat{\mathbf{v}}_\mathrm{t}^{(m)}$ by the gradient}
                \Statex \hspace{27pt} \Kabuto{descent from \eqref{eq:alpha_opt}, \eqref{eq:grad_v_r}, \eqref{eq:grad_v_t}, \eqref{eq:grad_tau}.}
                \If{$\Delta \mathcal{F} > \mathrm{Th}_2$}
                    \State Update $\hat{\mathbf{C}}_\mathrm{r}$ and $\hat{\mathbf{C}}_\mathrm{t}$ from \eqref{eq:ur_opt} and \eqref{eq:ut_opt} 
                    \Statex \hspace{42pt} \Kabuto{with $q^x_\mathrm{r} = q^y_\mathrm{r} = q^x_\mathrm{t} = q^y_\mathrm{t} = 0$}
                \Else
                    \State Update $\hat{\mathbf{C}}_\mathrm{r}$ and $\hat{\mathbf{C}}_\mathrm{t}$ from \eqref{eq:ur_opt} and \eqref{eq:ut_opt} 
                    \Statex \hspace{42pt} \Kabuto{with $q^x_\mathrm{r}, q^y_\mathrm{r}, q^x_\mathrm{t}, q^y_\mathrm{t}$}
                \EndIf
            \EndIf
            \State Update $\hat{\mathbf{\Gamma}}_\mathrm{r}$ and $\hat{\mathbf{\Gamma}}_\mathrm{t}$ from \eqref{eq:gamma_r_opt} and \eqref{eq:gamma_t_opt}
            \State Update $\hat{\bm{\varepsilon}}_\mathrm{r}$ and $\hat{\bm{\varepsilon}}_\mathrm{t}$ from \eqref{eq:grad_e_r} and \eqref{eq:grad_e_t}
        \EndFor
        \State \Kabuto{Calculate $\hat{\mathbf{H}}^{(M)}_k$ from \eqref{eq:H_est} }  \KabutoSecond{\textbf{// Channel estimation}} 
    \end{algorithmic}
\end{algorithm}


    \vspace{-0.3cm}
    \subsection{Complexity Analysis}
    
    Computational complexity of the proposed algorithm is evaluated in terms of the number of complex multiplications as \ac{FLOPs}. 
    The proposed method, as shown in Algorithm~\ref{alg:prop}, alternately estimates each parameter \Kabuto{$ \big \{ \mathbf{C}_\mathrm{r}, \mathbf{C}_\mathrm{t}, \mathbf{\Gamma}_\mathrm{r}, \mathbf{\Gamma}_\mathrm{t}, \bm{\varepsilon}_\mathrm{r}, \bm{\varepsilon}_\mathrm{t},  \{\mathbf{v}_\mathrm{r}^{(m)}, \mathbf{v}_\mathrm{t}^{(m)}, \bm{\tau}^{(m)}, \bm{\alpha}^{(m)} \}_{m=1}^M  \big \}$}. 
    The FLOPs for estimating each parameter are provided in Table~\ref{table:FLOPs}, 
    \Kabuto{
    where the complexity of estimating $\{\mathbf{v}_\mathrm{r}^{(m)}, \mathbf{v}_\mathrm{t}^{(m)}, \bm{\tau}^{(m)}, \bm{\alpha}^{(m)} \}$ using both the on-grid OMP~\cite{1993Pati_OMP} and the off-grid method from \eqref{eq:alpha_opt}, \eqref{eq:grad_v_r}, \eqref{eq:grad_v_t}, and \eqref{eq:grad_tau}} is presented since the estimation methods are switched from the on-grid to the off-grid method based on the rate of change in the objective function.
     
    In the case of using the exact mutual coupling model in \eqref{eq:Cr} (\textit{i.e.}, setting \Kabuto{$q_\mathrm{r}^x = N_\mathrm{r}^x -1,\ q_\mathrm{r}^y = N_\mathrm{r}^y -1,\ q_\mathrm{t}^x = N_\mathrm{t}^x -1$ and $q^y_\mathrm{t} = N^y_\mathrm{t} -1$} in \eqref{eq:ur}), the dominant complexity in the entire algorithm is \Kabuto{$ \mathcal{O} \left ( T_\mathrm{iter} \{ M  Q N_\mathrm{p} N_\mathrm{r,RF} (N_\mathrm{r}^4 + N_\mathrm{t}^4 ) + N_\mathrm{r}^6 + N_\mathrm{t}^6 \} \right )$} for estimating $\mathbf{C}_\mathrm{r}$ and $\mathbf{C}_\mathrm{t}$ 
    because equations \eqref{eq:ur_opt}, \eqref{eq:ut_opt} require matrix multiplications and inverse operations with size $Q_\mathrm{r} \times Q_\mathrm{r}$ and $Q_\mathrm{t} \times Q_\mathrm{t}$ matrices.
    Conversely, the proposed algorithm can reduce the complexity by selecting small values for \Kabuto{$q^x_\mathrm{r}$, $q^y_\mathrm{r}$, $q^x_\mathrm{t}$ and $q^y_\mathrm{t}$ (\textit{e.g.}, $q^x_\mathrm{r} = q^y_\mathrm{r}=1, q^x_\mathrm{t}=2, q^y_\mathrm{t}=0$)} in the simulation of Section~\ref{sec:sim}) by considering the property of mutual coupling.
    When \Kabuto{$q_\mathrm{r}^x=q_\mathrm{r}^y=q_\mathrm{t}^x=q_\mathrm{t}^y=0$}, indicating that the mutual coupling matrix is assumed to be modeled as a block Toeplitz matrix, the complexity can be minimized. 
    However the estimation performance significantly degrades due to the modeling errors.
    %

    \begin{table}[t!]
        \caption{\Kabuto{Computational complexity of the proposed algorithm}} 
        \label{table:FLOPs}
        \centering
        \begin{tabular*}{9cm}{c|l}
            \hline
            \multicolumn{1}{c}{Parameters} & FLOPs	\\
            \hline
            \Kabuto{
            \makecell[l]{
            $ \bm{\alpha}, \bm{\tau}, \mathbf{v}_\mathrm{r}, \mathbf{v}_\mathrm{t}$ \\ (Off-grid)}
            }
            & 
            \Kabuto{
            \makecell[l]{
            $ \!\!\! \mathcal{O} \Big( T_\mathrm{iter} T_\mathrm{bct} \big \{M Q N_\mathrm{p} N_\mathrm{r,RF} \hat{L} ( N_\mathrm{r} + \hat{L})$ \\ $\quad +N_\mathrm{r}^2 (N_\mathrm{p} N_\mathrm{r,RF} + M Q \hat{L}) \big \} \Big)$
            }}\\
            %
            \Kabuto{
            \makecell[l]{
            $\bm{\alpha}, \bm{\tau}, \mathbf{v}_\mathrm{r}, \mathbf{v}_\mathrm{t}$ \\ (On-grid) }}
            & 
            \Kabuto{
            \makecell[l]{
            $ \!\!\! \mathcal{O} \Big ( T_\mathrm{iter} \big \{ M\hat{L} (Q G_\mathrm{t} G_\mathrm{t} G_\tau + Q N_\mathrm{p} G_\mathrm{r} G_\mathrm{t}$ \\ $\quad + Q N_\mathrm{p} N_\mathrm{r,RF}G_\mathrm{\tau} ) + M\hat{L}^4 \big \} \Big )$ }} 	\\
            %
            \Kabuto{
            $\mathbf{C}_\mathrm{r}, \mathbf{C}_\mathrm{t}$ in \eqref{eq:ur_opt}, \eqref{eq:ut_opt}}
            & \Kabuto{$ \!\!\! \mathcal{O} \left ( T_\mathrm{iter} \big \{ M Q N_\mathrm{p} N_\mathrm{r,RF} (Q_\mathrm{r}^2 + Q_\mathrm{t}^2 ) + Q_\mathrm{r}^3 + Q_\mathrm{t}^3 \big \} \right )$} 	\\
            %
            \Kabuto{
            $\mathbf{\Gamma}_\mathrm{r}, \mathbf{\Gamma}_\mathrm{t}$ in \eqref{eq:gamma_r_opt}, \eqref{eq:gamma_t_opt}}
            & \Kabuto{$ \!\!\! \mathcal{O} \left ( T_\mathrm{iter} \big \{ M Q N_\mathrm{p} N_\mathrm{r,RF}(N_\mathrm{r}^2 + N_\mathrm{t}^2 ) + N_\mathrm{r}^3 + N_\mathrm{t}^3 \big \} \right )$} 	\\
            %
            \Kabuto{
            $\bm{\varepsilon}_\mathrm{r}, \bm{\varepsilon}_\mathrm{t}$ 
            in \eqref{eq:grad_e_r},\eqref{eq:grad_e_t}}
            & 
            \Kabuto{
            \makecell[l]{
            $ \!\!\! \mathcal{O} \Big ( T_\mathrm{iter} T_\mathrm{bct} \big \{ M Q N_\mathrm{p} \{N_\mathrm{r,RF} (\hat{L} + N_\mathrm{r})$ \\
            $ \quad \times (N_\mathrm{r}^x + N_\mathrm{r}^y) + N_\mathrm{t} \hat{L} (N_\mathrm{t}^x + N_\mathrm{t}^y) \} \big \} \Big )$ }
            }  \\
            \hline
        \end{tabular*}
        \vspace{1mm}
        \\{Note: $T_\mathrm{iter}$ and $T_\mathrm{bct}$ denote the number of algorithmic iterations and the maximum number of steps in the backtracking line search, respectively.}
    \end{table}

\vspace{-0.3cm}
\section{Simulation Results}
    \label{sec:sim}

    \subsection{Simulation Setup}
    This section evaluates the performance of the proposed channel estimation algorithm under the following simulation parameters.
    \Kabuto{
    The carrier frequency is $f_\mathrm{c}=50\ \mathrm{GHz}$, 
    the system bandwidth is $B=2.5\ \mathrm{GHz}$, 
    the number of subcarriers is $K=64$\footnote{ \KabutoSecond{
    In our simulations, to reduce the simulation time in Monte Carlo evaluations, a small number of subcarriers with a large subcarrier spacing is used as in \cite{2020Xie_CoDL, 2023Xie_DADL, 2024Garg_squint_CE_JCDE}.
    Note that even if channel variation occurs within each subcarrier, the frequency-domain signal model in (10) still holds as long as the delay time does not exceed the CP duration~\cite{2005David_Fundamental}.}}, 
    the number of RF chains is $N_\mathrm{r,RF}=N_\mathrm{t,RF}=2$, 
    the number of training frames is $M=5$, 
    the number of pilot symbols is $N_\mathrm{p}=50$, 
    the ideal antenna spacings are $d_\mathrm{r}^x=d_\mathrm{r}^y=d_\mathrm{t}^x=d_\mathrm{t}^y=\lambda_\mathrm{c}/2$, and 
    the number of pilot subcarriers is $Q=16$. 
    The subcarriers allocation for pilots is determined by the allocation method proposed in~\cite{2016He_allocation_MIMO_OFDM}.
    }

    \Kabuto{
    In this simulation, the following two scenarios are evaluated.
    (a) \textit{ULA case;} both the BS and UE are equipped with the ULAs, where $(N_\mathrm{r}^x, N_\mathrm{r}^y)=(32,1)$ and $(N_\mathrm{t}^x, N_\mathrm{t}^y)=(8,1)$. 
    (b) \textit{UPA case;} the BS is equipped with the UPA, where $(N_\mathrm{r}^x, N_\mathrm{r}^y)=(8,4)$, and the UE is equipped with ULA, where $(N_\mathrm{t}^x, N_\mathrm{t}^y)=(8,1)$}.
    \Kabuto{
    In sub-THz bands, ULAs are commonly used from a practical implementation perspective, considering the extremely short wavelength and the RF loss on printed circuit boards (PCBs)~[49]–[51].
    For these reasons, we evaluated not only the UPA case but also the ULA case.}
    
    \Kabuto{
    The parameters related to the proposed algorithm are set as follows. 
    The number of estimated paths is $\hat{L}=12$, 
    the number of iterations is $T_\mathrm{iter}=250$ in the ULA case, and $T_\mathrm{iter}=1000$ in the UPA case,
    the maximum number of steps in the backtracking line search is $T_\mathrm{bct}=6$,
    the penalty coefficients are $\lambda^{\mathbf{c}_\mathrm{r}}=\lambda^{\mathbf{c}_\mathrm{t}}=10^2/\mathrm{SNR}$, where $\mathrm{SNR} \triangleq \frac{P_\mathrm{s}}{\sigma^2}$.
    Since the thresholds $\mathrm{Th}_1$ and $\mathrm{Th}_2$ must be sufficiently small to detect convergence of the objective function, these thresholds are set to $\mathrm{Th}_1=0.001$ and $\mathrm{Th}_2=0.001$.
    }
    
    The channels defined in \eqref{eq:H_k} are generated by the composition of total $L=6$ paths, where
    the path gain $\alpha_l$ is generated as $\alpha_l \sim \mathcal{CN}(0, 1/L)$, 
    the angles $\phi_l^\mathrm{a}$, $\phi_l^\mathrm{z}$, $\theta_l^\mathrm{a}$, $\theta_l^\mathrm{z}$ are uniformly distributed in $[-\pi, \pi)$, and 
    delay $\tau_l$ is uniformly distributed in $\left [0, (N_\mathrm{cp}-1)/B \right]$, as in \cite{2020Xie_CoDL,2023Xie_DADL,2018Rodriguez_SOMP} with $N_\mathrm{cp}=16$.
    \Kabuto{
    The number of grids $(G^x_\mathrm{r}, G^y_\mathrm{r}, G^x_\mathrm{t}, G^y_\mathrm{t}, G_\tau)$ is set to $(2N_\mathrm{r}^x, 2N_\mathrm{r}^y, 2N_\mathrm{t}^x, 2N_\mathrm{t}^y, 2N_\mathrm{cp})$ in the UPA case and $(2N_\mathrm{r}^x, 1, 2N_\mathrm{t}^x, 1, 2N_\mathrm{cp})$ in the ULA case, as in~\cite{2018Rodriguez_SOMP, 2022Cui_PSOMP, 2024Xu_squint_CE_Bayesian_multi}.
    }
    
    The array errors are generated by the following models in \cite{2020Xie_CoDL,2022Maity_CoDL_THz,2023Xie_DADL,2023Maity_CoDL_RIS,2016Liu_couoling_model}.
    The gain and phase errors are generated as
    $g_{\mathrm{r}, n_\mathrm{r}}, g_{\mathrm{t}, n_\mathrm{t}} \sim \mathcal{N}(1, 0.05^2)$ and 
    $\nu_{\mathrm{r}, n_\mathrm{r}}, \nu_{\mathrm{t}, n_\mathrm{t}} \sim \mathcal{N}(0, (20 \pi / 180)^2)$, and
    the antenna spacing errors are generated as 
    $\varepsilon^x_{\mathrm{r}, n_\mathrm{r}^x}, \varepsilon^y_{\mathrm{r}, n^y_\mathrm{r}}, \varepsilon^x_{\mathrm{t}, n^x_\mathrm{t}}, \varepsilon^y_{\mathrm{t}, n^y_\mathrm{t}} \sim \mathcal{U}(-0.1 \lambda_\mathrm{c},\ 0.1\lambda_\mathrm{c})$.
    Since effects of the mutual coupling decrease with distance from a reference antenna~\cite{2021Fikes_wearable_coupling, 2016Liu_couoling_model}, the coefficient of the mutual coupling can be modeled as a function of distance as $c_{\mathrm{}} (\tilde{r}) = \frac{c_{\mathrm{} 0}}{\tilde{r}} e^{-j (\tilde{r}-1) \pi / 8}$, 
    where $\tilde{r}$ is the distance normalized by half wavelength $\lambda_\mathrm{c} / 2$, and
    $c_0$ is the coupling coefficient at $\tilde{r}=1$, set to $c_{0} = 0.2 e^{j \frac{\pi}{3}}$ in the simulation.

    \Kabuto{The channel estimation performance is evaluated by \ac{NMSE}, defined as
    $\mathrm{NMSE}(\bm{\mathbf{H}}) \triangleq \mathbb{E} \left[ \sum_{k \in \mathcal{K}} {\| \bm{\mathbf{H}}_k^{(M)} - \hat{\bm{\mathbf{H}}}_k^{(M)} \|_\mathrm{F}^2} \Big/ \sum_{k \in \mathcal{K}} {\| \bm{\mathbf{H}}_k^{(M)} \|_\mathrm{F}^2} \right]$,
    where $\mathbf{H}_k^{(M)}$ and $\hat{\mathbf{H}}_k^{(M)}$ are a true channel and estimated channel matrices at the current frame.}
    To assess the pilot overhead in channel estimation, the compression ratio $\nu \triangleq \frac{N_\mathrm{r,RF} N_\mathrm{p}}{N_\mathrm{r} N_\mathrm{t}}$ is defined as the ratio between the number of observations and the number of unknown channel coefficients. 
    If the compression ratio $\nu$ is less than one, the measurement equation in \eqref{eq:y_obs} becomes an underdetermined system.
    In the simulations, the number of pilots $N_\mathrm{p}$ ranges from $32$ to $160$, corresponding to compression ratios between $0.25$ and $1.25$.

    To assess the effectiveness of the proposed algorithm, the following methods are used as benchmarks.
    \begin{itemize}
        \item[1)] \textbf{SOMP w/ CoDL}~\cite{2020Xie_CoDL}: 
        a dictionary learning-based channel estimation method, where the dictionary matrix is directly updated with training pilots to compensate for the array errors. 
        \item[2)] \textbf{DA-OMP-BS w/ DLHWBS}~\cite{2023Xie_DADL}: 
        a dictionary learning-based channel estimation method, where the dictionary matrix is decomposed into two dictionaries at the transmitter and receiver sides, which are alternately updated with training pilots to compensate for the array errors. 
        \Kabuto{
        \item[3)] \textbf{Off-grid SBL}~\cite{2024Xu_squint_CE_Bayesian_multi}: 
        an off-grid based channel estimation method that accounts for beam squint effects but does not account for array errors.
        \item[4)] \textbf{Prop. (w/o switch)}: 
        the proposed method under the assumption that the mutual coupling has Toeplitz structure (\textit{i.e.}, $(q^x_\mathrm{r}, q^y_\mathrm{r}, q^x_\mathrm{t}, q^y_\mathrm{t})=(0,0,0,0)$). 
        This method does not utilize the switching from the on-grid to off-grid algorithms (\textit{i.e.}, $\mathrm{Th}_1 \rightarrow \infty$).
        The initial estimates $\{ \hat{\mathbf{v}}_\mathrm{r}^{(m)}, \hat{\mathbf{v}}_\mathrm{t}^{(m)}, \bm{\tau}^{(m)} \}$ are generated by OMP~\cite{1993Pati_OMP}.
        \item[5)] \textbf{Prop. (w/ switch)}: 
        the proposed method under Toeplitz assumption (\textit{i.e.}, $(q^x_\mathrm{r}, q^y_\mathrm{r}, q^x_\mathrm{t}, q^y_\mathrm{t})=(0,0,0,0)$) with the switching from the on-grid to off-grid algorithms.
        \item[6)] \textbf{Prop. (w/ switch and approx. model)}: 
        the proposed method with the switching mechanism and the approximate coupling model in \eqref{eq:C_TP_NTP}, where
        $(q^x_\mathrm{r}, q^y_\mathrm{r}, q^x_\mathrm{t}, q^y_\mathrm{t})=(2,0,2,0)$ in the ULA case, and 
        $(q^x_\mathrm{r}, q^y_\mathrm{r}, q^x_\mathrm{t}, q^y_\mathrm{t})=(1,1,2,0)$ in the UPA case.
        \item[8)] \textbf{Perfect calibration}: 
        the proposed method with perfect knowledge of all array errors.
        \item[9)] \textbf{Genie-aided LS}: 
        channel estimation based on the least square with perfect knowledge of all array errors and angle parameters (including AoA, ZoA, AoD, and ZoD).
        }
        \vspace{-0.3cm}
    \end{itemize}
    \Kabuto{
    Since the conventional methods, DA-OMP-BS w/ DLHWBS~\cite{2023Xie_DADL} and Off-grid SBL~\cite{2024Xu_squint_CE_Bayesian_multi} are applicable only to the ULA case,  their performance is evaluated in the ULA case.}

    \vspace{-0.3cm}
    \subsection{Channel Estimation Performance}

    \begin{figure}[t]
        \begin{minipage}{1.0\columnwidth}
            \centering
            \includegraphics[scale=0.5]{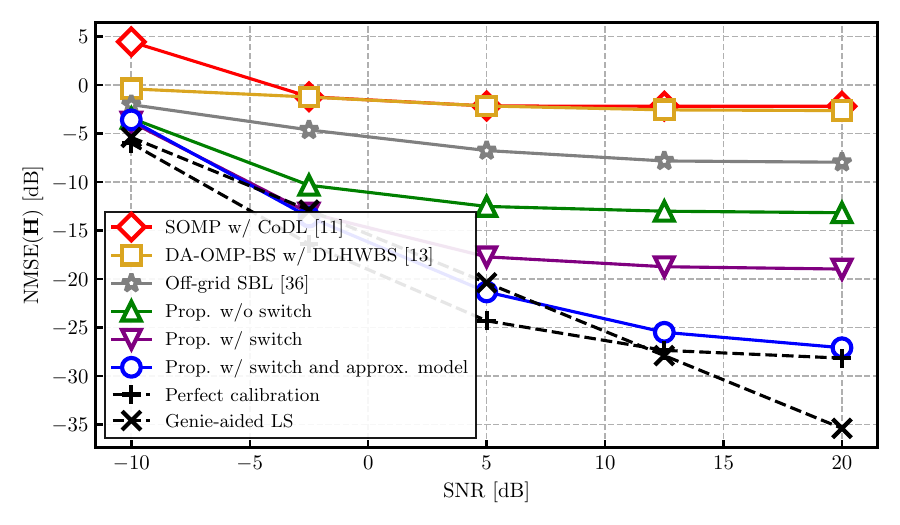}
            \subcaption{ULA case}
            \label{fig:NMSE_vs_SNR_ULA}
        \end{minipage}
        \\
        \begin{minipage}{1.0\columnwidth}
            \centering
            \includegraphics[scale=0.5]{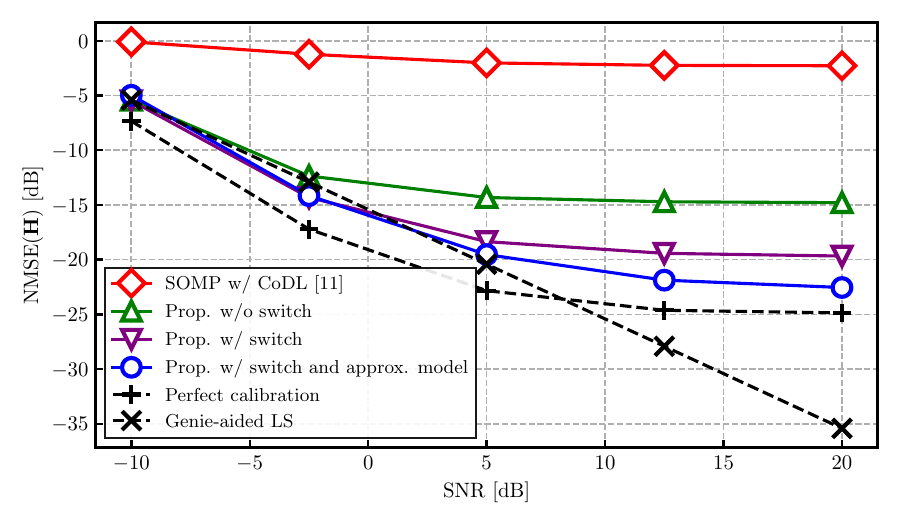}
            \subcaption{UPA case} 
            \label{fig:NMSE_vs_SNR_UPA}
        \end{minipage}
        \caption{\KabutoSecond{NMSE$(\mathbf{H})$ versus SNR with $N_\mathrm{p}=50$ (compression ratio $\nu=0.391$)}.} 
        \label{fig:NMSE_vs_SNR}
    \end{figure}

    \begin{figure}[t]
        \begin{minipage}{1.0\columnwidth}
            \centering
            \includegraphics[scale=0.5]{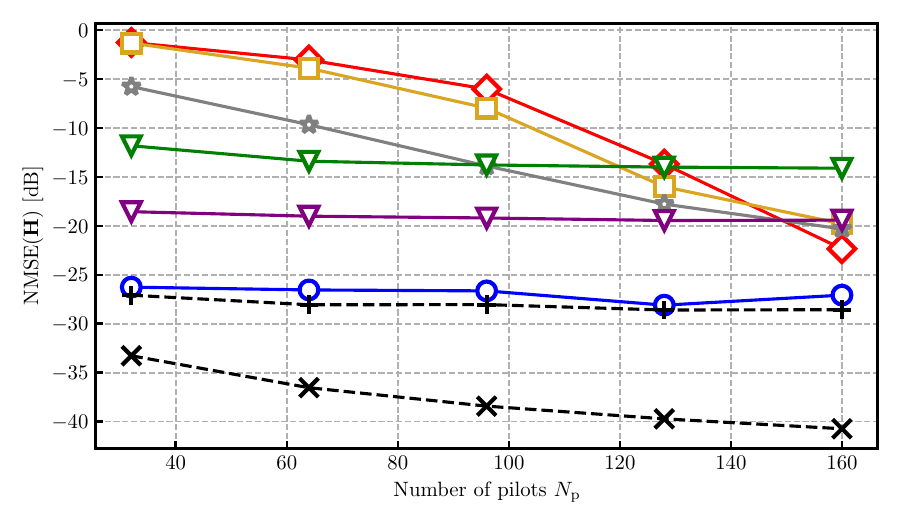}
            \subcaption{ULA case}
            \label{fig:NMSE_vs_Np_ULA}
        \end{minipage}
        \\
        \begin{minipage}{1.0\columnwidth}
            \centering
            \includegraphics[scale=0.5]{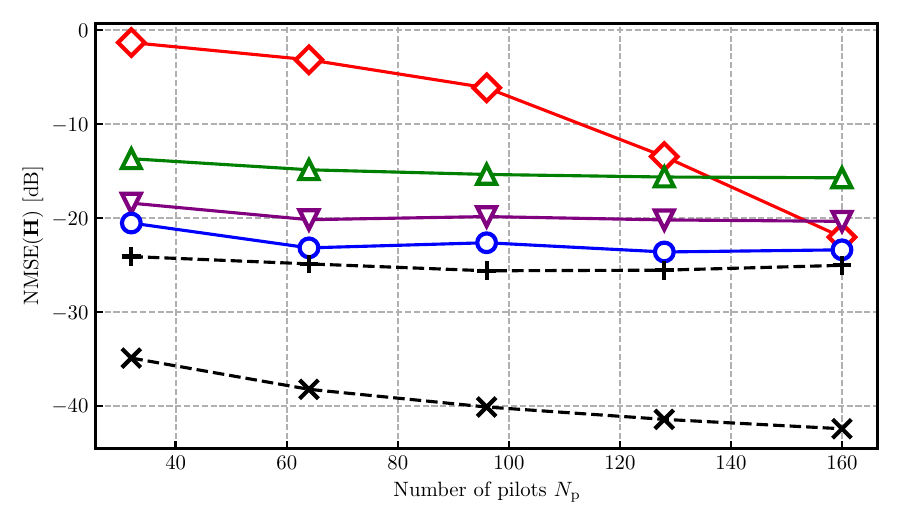}
            \subcaption{UPA case} 
            \label{fig:NMSE_vs_Np_UPA}
        \end{minipage}
        \caption{
        NMSE$(\mathbf{H})$ versus the number of pilot symbols $N_\mathrm{p}$ with $\mathrm{SNR} = 20\ \mathrm{dB}$. The compression ratio $\nu$ varies from 0.25 to 1.25.}
        \label{fig:NMSE_vs_Np}
    \end{figure}

    \begin{figure}[t]
        \begin{minipage}{1.0\columnwidth}
            \centering
            \includegraphics[scale=0.5]{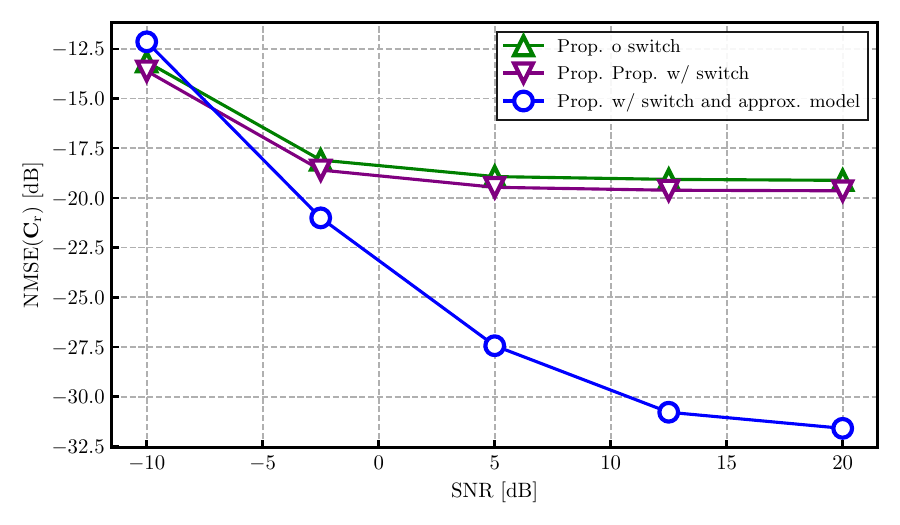}
            \subcaption{ULA case}
            \label{fig:NMSE_vs_Cr_ULA}
        \end{minipage}
        \\
        \begin{minipage}{1.0\columnwidth}
            \centering
            \includegraphics[scale=0.5]{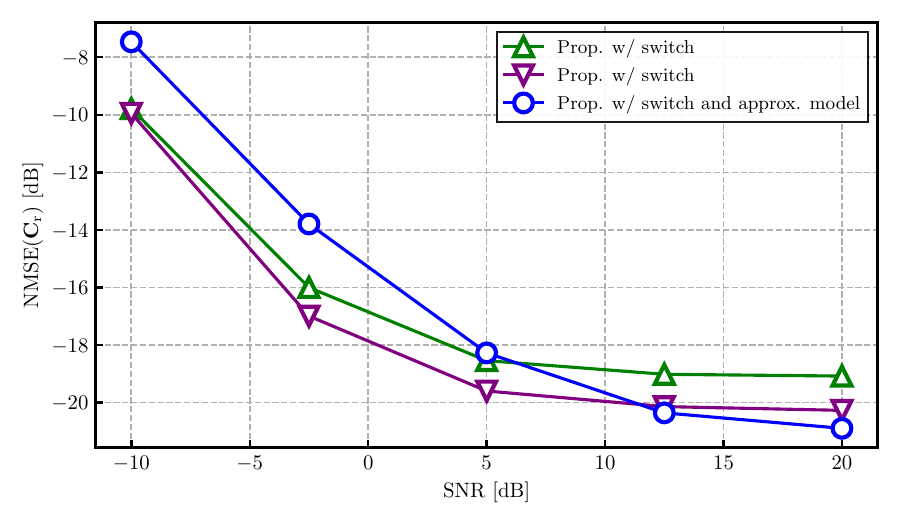}
            \subcaption{UPA case}
            \label{fig:NMSE_vs_Cr_UPA}
        \end{minipage}
        \caption{NMSE$(\mathbf{C}_\mathrm{r})$ versus SNR with $N_\mathrm{p}=50$ (compression ratio $\nu=0.391$), where
        $\mathrm{NMSE}(\mathbf{C}_r) \triangleq \mathbb{E} \left[ \| \mathbf{C}_\mathrm{r} - \hat{\mathbf{C}}_\mathrm{r} \|_\mathrm{F}^2 \big / \| \mathbf{C}_\mathrm{r} \|_\mathrm{F}^2\right] $.
        } 
        \label{fig:NMSE_vs_Cr}
    \end{figure}

    \begin{figure}[t]
        \begin{minipage}{1.0\columnwidth}
            \centering
            \includegraphics[scale=0.25]{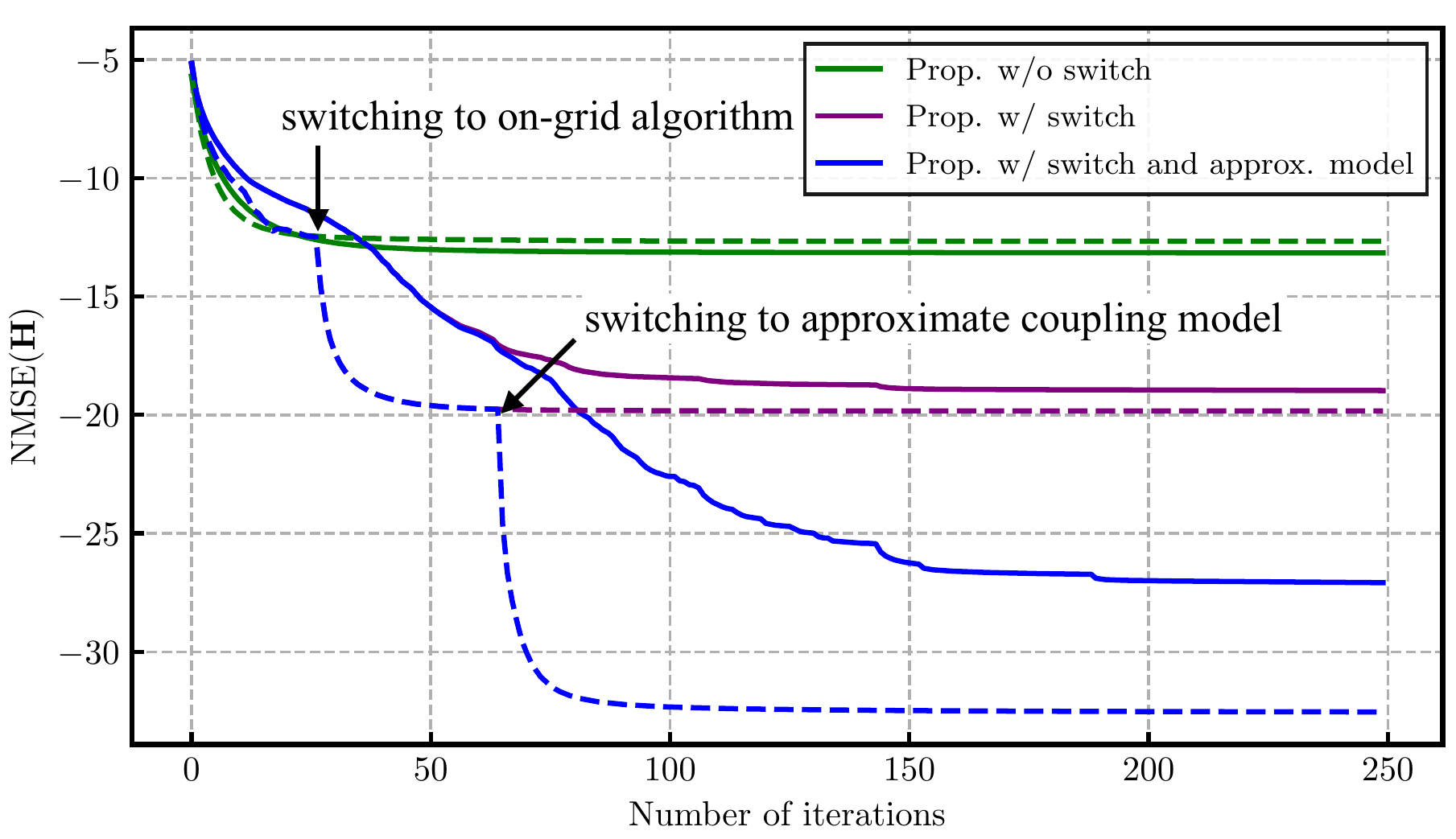}
            \subcaption{ULA case}
            \label{fig:NMSE_vs_Iterations_ULA}
        \end{minipage}
        \\
        \begin{minipage}{1.0\columnwidth}
            \centering
            \includegraphics[scale=0.25]{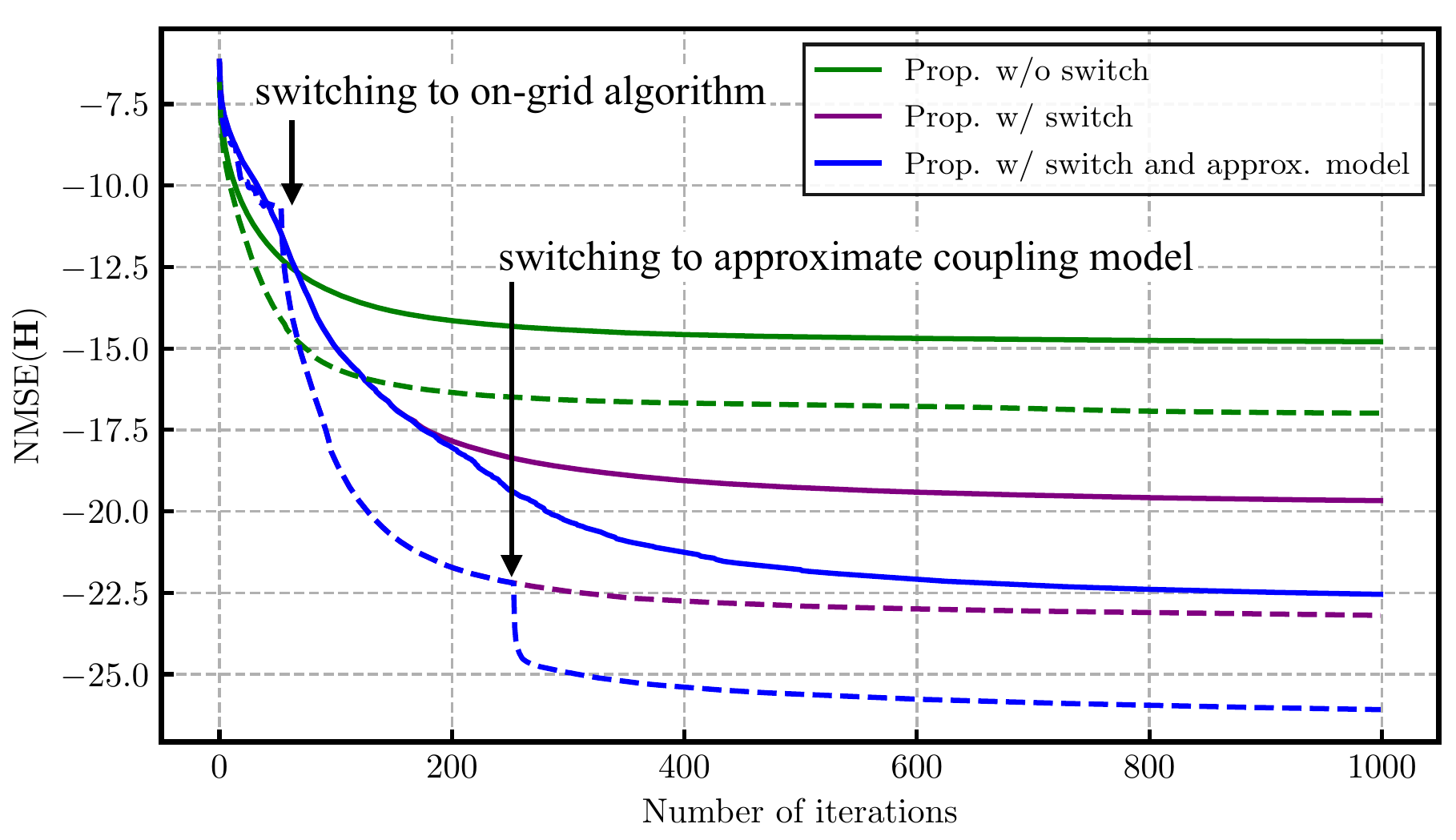}
            \subcaption{UPA case}
            \label{fig:NMSE_vs_Iterations_UPA}
        \end{minipage}
        \caption{NMSE$(\mathbf{H})$ versus the number of algorithmic iterations with $\mathrm{SNR}=20\ \mathrm{dB}$, $M=5$ and $N_\mathrm{p}=50$ (compression ratio is $\nu=0.391$). 
        Dashed lines depict one-shot performance of channel estimation given a single channel realization, whereas solid lines indicate averaged performance calculated by Monte Carlo simulations.} 
        \label{fig:NMSE_vs_Iter}
    \end{figure}

    Fig.~\ref{fig:NMSE_vs_SNR} and Fig.~\ref{fig:NMSE_vs_Np} show the NMSE performance against SNR and the number pilot symbols $N_\mathrm{p}$.
    As presented in these figures, the conventional methods, SOMP w/ CoDL and DA-OMP-BS w/ DLHWBS, fail to improve channel estimation performance, especially in a small number of pilots, because these methods estimate the dictionary matrix, which includes a large number of unknown parameters.
    Although the channel estimation performance improves as the number of pilots increases, it causes significant pilot overhead issues.
    In contrast, the proposed method enhances channel estimation performance compared to conventional methods, even in the small pilot region, because the proposed method decomposes the dictionary matrix into multiple channel components, including array errors, and estimates a small number of parameters.
    Comparing Prop. (w/o switch) and Prop. (w/ switch), it can be seen that the switching mechanism from the on-grid to the off-grid method significantly improves estimation performance, particularly in the high SNR region, owing to the avoidance of convergence to the local optimum in early algorithmic iterations.
    \Kabuto{
    As seen from the result of Prop. (w/o switch) that relies on the Toeplitz structure, the Toeplitz assumption degrades the NMSE performance.
    There is a substantial performance gap compared to the \Kabuto{Perfect calibration}, due to the mismatch in the mutual coupling model under the Toeplitz assumption, which is caused by antenna spacing errors.
    }
    By contrast, the NMSE performance of Prop. (w/ switch and approx. model) improves and approaches \Kabuto{Perfect calibration}, owing to the introduction of the approximate mutual coupling model in \eqref{eq:C_TP_NTP}.
    Although the proposed method approaches the performance of Perfect calibration, its performance saturates in the high SNR region.
    If two closely spaced delays are not resolved, the estimation errors propagate across all subcarriers, since the equivalent path gain $\mathbf{z}_k$ is reconstructed using delays $\bm{\tau}$ shared across all subcarriers, as written in \eqref{eq:zk}.
    Consequently, the performance of the proposed method saturates in the high-SNR region since unresolved delays occur with a certain probability. 
    This leads to a performance gap relative to the Genie-aided LS.
    %

    Fig.~\ref{fig:NMSE_vs_Cr} shows the estimation performance for the mutual coupling matrices $\mathbf{C}_\mathrm{r}$ against SNR.
    Prop. (w/ switch and approx. model) delivers superior performance because a part of the mutual coupling matrix is treated with the exact model, while the other part is modeled using a Toeplitz structure with fewer parameters.
    \Kabuto{
    However, the estimation performance of the UPA case degrades in the low SNR region and the performance gain obtained in the UPA case is smaller than that in the ULA case. 
    This is because the effects of mutual coupling become stronger in the UPA case, as the number of closely spaced antenna elements increases in the $x$–$y$ plane.
    In addition, the number of estimation parameters increases in the UPA case due to the inclusion of two-dimensional parameters. 
    }

    To clarify the performance gains from modifying the coupling model and introducing the switching mechanism, Fig.~\ref{fig:NMSE_vs_Iter} illustrates the NMSE performance over algorithmic iterations with $\mathrm{SNR}=20\ \mathrm{dB}$, $M=5$ and $N_\mathrm{p}=50$.
    In this figure, the dashed lines represent the non-averaged performance for a single channel realization, whereas the solid lines represent the averaged performance obtained through Monte Carlo simulations.
    As seen from Prop. (w/o switch), which use only the off-grid method without the switching mechanism, the performance converges to a low NMSE.
    Employing the off-grid algorithm based on gradient descent during the early iterations might lead to convergence to local optimum with low NMSE, because the estimation accuracy of the array errors is insufficient in the early iterations.
    In contrast, by applying the on-grid method in the early iterations and switching to the off-grid method after convergence, NMSE improvement can be observed from the result of Prop. (w/ switch).
    Furthermore, introducing the approximate mutual coupling model in \eqref{eq:C_TP_NTP} after the convergence of the on-grid method results in performance enhancement, as seen in Prop. (w/ switch and approx. model).
    \Kabuto{
    However, due to the increase in the number of estimation parameters in the UPA case, the convergence speed is slower than in the ULA case, which requires more iterations. 
    Addressing this issue is left as future work.
    }

\vspace{-0.3cm}
\section{Conclusion}
    \label{sec:conclusion}
    In this paper, we proposed a channel estimation method for \Kabuto{wideband hybrid \ac{MIMO} systems} affected by array errors and beam squint effects.
    Due to array deformation from thermal effects and dynamic motion, array errors, such as mutual coupling, gain/phase errors, and antenna spacing errors, change over time. 
    To calibrate these array errors online with small pilot overhead, the proposed method reduces the number of unknown parameters by explicitly decomposing the array response matrices into primary parameters, \Kabuto{including angles, delays, path gains, and array errors.}
    These parameters are estimated iteratively via an alternating minimization technique, where 
    on-grid and off-grid algorithms are switched depending on the rate of change in the objective function, in order to prevent convergence to a local optimum in the early iterations.
    Furthermore, the mutual coupling matrix is approximately decomposed into a Toeplitz part and a non-Toeplitz part.
    The approximate decomposition can improve the channel estimation performance and reduce the computational complexity.
    Simulations demonstrate that the proposed method outperforms the conventional methods with small pilot overhead.

    For future work, more sophisticated estimation methods to improve the convergence speed and enhance the channel estimation performance in the UPA case should be considered.

\vspace{-0.3cm}
\bibliographystyle{IEEEtran}
\bibliography{reference.bib}

\end{document}